\documentclass[12pt]{article}

\usepackage[normalem]{ulem}
\usepackage{hyperref}
\usepackage{graphicx}
\usepackage{caption}
\usepackage{longtable}
\usepackage{subcaption}
\usepackage{amsmath}
\usepackage{textgreek}
\usepackage{cite}
\usepackage{lipsum}
\usepackage{wrapfig}
\usepackage{lineno}
\usepackage{setspace}
\usepackage{geometry}
\usepackage[usenames,dvipsnames]{color}

\newcommand{\Dros}[0]{\emph{Drosophila}}
\newcommand{\Dyak}[0]{\emph{D. yakuba}}
\newcommand{\Dsant}[0]{\emph{D. santomea}}
\newcommand{\Dmel}[0]{\emph{D. melanogaster}}
\newcommand{\Dtei}[0]{\emph{D. teissieri}}
\newcommand{\STome}[0]{S\~{a}o Tom\`{e}}
\newcommand{\FST}{$F_{ST}$}
\newcommand{\deltap}{\textDelta \emph{p}}

\DeclareCaptionType{equ}[][]

\title{ Chromosomal rearrangements and transposable elements in locally adapted island \emph{Drosophila}}
\author{\small Brandon A. Turner*$^{1}$, Theresa R. Erlenbach$^{2}$, Nicholas B. Stewart$^3$,\\ \small Robert W. Reid$^1$, Cathy C. Moore$^1$  and Rebekah L. Rogers$^1$}
\date{}

\begin{document}
\maketitle

\noindent1. Department of Bioinformatics and Genomics, University of North Carolina, Charlotte NC  \\
\noindent2.  Dept of Genetics, University of Georgia, Athens, GA \\
\noindent3.  Dept of Biology, Southern Oregon University, Ashland, OR \\

\clearpage

\doublespacing

\section*{Abstract}

    Chromosomal rearrangements, particularly those mediated by transposable elements (TEs), can drive adaptive evolution by creating chimeric genes, inducing de novo gene formation, or altering gene expression. Here, we investigate rearrangements evolutionary role during habitat shifts in two locally adapted populations, \Dsant{} and \Dyak{}, who have inhabited the island \STome{} for 500,000 and 10,000 years respectively. Using the \Dyak-\Dsant{} species complex, we identified 16,480 rearrangements in the two island populations and the ancestral mainland African population of \Dyak{}. We find a disproportionate association with TEs, with 83.5\% of rearrangements linked to TE insertions or TE-facilitated ectopic recombination. Using significance thresholds based on neutral expectations, we identify 383 and 468 significantly differentiated rearrangements in island \Dyak{} and \Dsant{}, respectively, relative to the mainland population. Of these, 99 and 145 rearrangements also showed significant differential gene expression, highlighting the potential for adaptive solutions from rearrangements and TEs. Within and between island populations, we find significantly different proportions of rearrangements originating from new mutations versus standing variation depending on TE association, potentially suggesting adaptive genetic mechanisms differ based on the timing of habitat shifts. Functional analyses of rearrangements most likely driving local adaptation revealed enrichment for stress response pathways, including UV tolerance and DNA repair, in high-altitude \Dsant{}. These findings suggest that chromosomal rearrangements may act as a source of genetic innovation, and provides insight into evolutionary processes that SNP-based analyses might overlook. \newline
    
    \noindent Keywords: Structural variation, transposable elements, genetic novelty, local adaptation, habitat shifts 
    \newpage

\section*{Significance Statement}
    
    When considering sources of adaptive traits in evolutionary biology, chromosomal rearrangements remain less researched relative to single nucleotide mutations (SNPs). Unlike SNPs, chromosomal rearrangements are associated with transposable element activity, and have different mutational dynamics and fitness effects that may become advantageous as selective pressures shift. This study uses island \Dros{} as a model for local adaptation to uncover how different genetic mechanisms and sources of chromosomal rearrangements may serve as "hopeful monsters" among genetic variation that contribute to adaptive change during habitat shifts.
     
%\section*{Author Summary}
%    We find that structural variation is a source of genetic novelty during local adaptation to island environments. Rearrangements adjacent to UV tolerance genes are among top candidates for local adaptation in high altitude D. santomea. These results offer potential insights about the genetic basis of survival in high UV environments. Locally adapted alleles are associated with significant changes in gene expression, suggesting that regulatory changes are key for innovation during habitat invasion. These results suggest that structural variation can form "hopeful monsters” among genes that are essential for survival as animals invade new environments.
\clearpage
\newpage

%\onehalfspacing
\section*{Introduction}

    Structural variation is a significant source of genetic novelty \cite{conant2008turning, ohno2013evolution, schlotterer2015genes}. Chromosomal rearrangements can duplicate and shuffle DNA segments within the genome \cite{ohno2013evolution, schlotterer2015genes, stewart2019chromosomal}. These rearrangements can create new linkage groups, alter gene expression patterns, and create novel gene sequences \cite{de2009impact, kondrashov2006role, huminiecki2004divergence, harewood2014impact}. Rearrangements may also lead to gene duplication and subsequent neofunctionalization \cite{assis2013neofunctionalization}, as well as the formation of chimeric genes \cite{long1993natural, rogers2012chimeric, zhou2008origin}. While most mutations are neutral or detrimental, rearrangements can result in genetic novelty and potentially contribute to local adaptation. 

    During habitat shifts and shifting environmental conditions, reproductive success within populations is partially driven by adaptive alleles and phenotypes \cite{schoville2012adaptive}. Historically, single nucleotide polymorphisms (SNPs) are considered the primary source of adaptive variation, and their role during habitat shifts and evolution is well established \cite{wellenreuther2019going, morin2004snps}. While the adaptive potential of SNPs is well researched, complex structural variation remains understudied, often due to difficulties in identifying and interpreting these mutations in genetic sequence data \cite{ohno2013evolution, conant2008turning, rogers2017tandem, rogers2012chimeric}. By assessing all sources of adaptive alleles, we can gain insight in to the genetic basis of adaptation, within the context of shifting selective pressures and habitat change.

    \subsection*{Transposable elements and chromosomal rearrangements}

        Transposable elements (TEs) can create genetic novelty \cite{ohno2013evolution, conant2008turning, feschotte2008transposable, oliver2009transposable}. These selfish genetic elements excise and reinsert throughout the genome, which can result in structural variation, such as chromosomal rearrangements \cite{bourque2018ten, stewart2019chromosomal}. While many TE-induced mutations are neutral or detrimental \cite{lynch2011transposon, wei2022dynamics}, TE activity has previously resulted in adaptive phenotypes \cite{casacuberta2013impact, gonzalez2010genome, aminetzach2005pesticide}.

        TEs facilitate chromosomal rearrangements through two main mechanisms: indirectly via homologous recombination or directly through alternative transposition \cite{bourque2018ten, lim1994gross, lister1989molecular}. Indirectly, TEs with similar or identical sequences align during recombination, despite being located at distant genetic positions. This non-homologous, or ectopic, recombination can result in the deletion or duplication of DNA sequence adjacent to TE sites \cite{gray2000takes}. Directly, TEs can cause chromosomal changes via insertion and altered transposition, which can result in the duplication or deletions of entire genes \cite{gray2000takes}. 
        
        Unlike SNPs, which accumulate slowly over generations \cite{ho2021engines}, TEs can proliferate in bursts \cite{cridland2013abundance}, especially under environmental stress \cite{casacuberta2013impact, aminetzach2005pesticide}. Consequently, the proliferation of TEs can result in a burst of new alleles by rapidly shuffling gene sequence adjacent to TE sites \cite{feschotte2008transposable, oliver2009transposable}. This resulting structural variation can serve ass a substrate that facilitates adaptive responses during habitat shifts \cite{casacuberta2013impact, aminetzach2005pesticide}. TEs may fortuitously produce the rapid genetic changes needed for local adaptation during habitat shifts more rapidly than the clock like emergence of adaptive SNPs \cite{ho2021engines}.

    \subsection*{The role of new mutations and standing variation in adaptation}

        In the face of shifting selective pressures, populations will require adaptive solutions. These solutions can arise from either new mutations or standing variation \cite{barrett2008adaptation}. New mutations can modify existing traits or, in some cases, give rise to entirely novel phenotypes, while standing variation acts as a “reservoir” of genetic diversity containing phenotypes that could become advantageous as selective pressures change \cite{barrett2008adaptation, hermisson2005soft}.
        
        Both sources of genetic variation are subject to unique evolutionary constraints. When new mutations occur, they are at risk of being lost due to the random fluctuations of allele frequencies caused by genetic drift \cite{hermisson2005soft, ohta1992nearly, lynch1999perspective}. This stochastic loss is often described as a 'sieve,' where, generally, alleles with smaller fitness effects are more likely to pass through the sieve and be lost from the population \cite{orr2001haldane, hermisson2005soft}. However, alleles that persist despite stochastic loss can remain at low frequency in the population as standing variation, and may become beneficial if selective pressures shift \cite{orr2001haldane, hermisson2005soft}. Standing variation offers immediate adaptive potential, but its effectiveness is limited by the available variation and could be surpassed by new, more adaptive mutations \cite{hermisson2005soft}.

        In contrast to standing variation, new mutations are constrained by mutation rates and the risk of stochastic loss \cite{hermisson2005soft, ohta1992nearly, lynch1999perspective}. This is especially true for new mutations with smaller fitness effects, which are more likely to face stochastic loss via drift \cite{hermisson2005soft, ohta1992nearly, lynch1999perspective}. However, adaptation from new mutations can be as quick as adaptation from standing variation for mutations with large fitness effects, where a new phenotype could heavily outperform existing phenotypes \cite{hermisson2005soft, barrett2008adaptation}. While the long wait times associated with the emergence of new SNPs can hinder rapid adaptation from new mutations \cite{ho2021engines}, other forms of mutations and mutagens, such as chromosomal rearrangements and transposable elements, may offer a mechanism to circumvent these delays, if multiple new insertions in "bursts" of TE activity over relatively few generations \cite{gillespie1994causes, casacuberta2013impact, gonzalez2010genome}.

        Traditionally standing variation is seen as more immediately advantageous during habitat shifts due to its availability \cite{barrett2008adaptation, hermisson2005soft}. However, differences in mutation dynamics for structural variation may contradict molecular clock assumptions of these models. Assessing the “arms race” between the adaptive potential of structural variation from both new mutations and standing variation is crucial for a comprehensive understanding of how species adapt during fluctuating selective pressure and habitat shifts.

    \subsection*{Assessing local adaptation with the \Dsant{}-\Dyak{} species complex}

        Shifting environments and habitat changes present an opportunity to study how populations adapt over time. These changes cause unique selective pressures that can drive evolutionary processes, offering insight into the genetic mechanisms that can contribute local adaptation, genetic differentiation, and species divergence \cite{boyer2021adaptation, rosalino2014adaptation}. One system that has experienced a recent habitat shifts is the \textit{D. santomea-D. yakuba} species complex, residing on the island of \STome{} \cite{coyne1989patterns, bachtrog2006extensive, lachaise2000evolutionary, llopart2005anomalous}. \Dsant{} and \Dyak{} are two sister species of \Dros{} that arrived on \STome{} in separate invasions—\Dsant{} approximately 500,000 years ago \cite{lachaise2000evolutionary} and \Dyak{} around 10,000 years ago \cite{coyne2002sexual, obbard2012estimating, cariou2001divergence}. These two species occupy habitats at distinct elevations on the island, with \Dsant{} at higher elevations, and are absent from the lower regions of the island, where \Dyak{} is abundant \cite{llopart2005anomalous}.

        The invasion of the island \Dyak{} population is more recent, such that a speciation event has not yet occurred from the mainland population, as is the case for \Dsant{} \cite{lachaise2000evolutionary}. Sufficient time has passed since these island invasions to allow for observable genetic differentiation between the island and mainland populations \cite{llopart2005anomalous, comeault2016correlated}. However, the timescale has been short enough that population genetic signals are still largely influenced by standing variation from the ancestral population, rather than being completely obscured by new mutations \cite{bachtrog2006extensive}. These two island invasions thus offer independent opportunities to study local adaptation from the same mainland population, but on different timescales.

        This system benefits from a well-characterized mainland ancestral population, which has been extensively studied for structural variation \cite{llopart2002genetics, rogers2014landscape, andolfatto2011effective}. The mainland \Dyak{} population contains a large number of chromosomal rearrangements \cite{clark2007evolution, bhutkar2008chromosomal, ranz2003sex}, which in previous work have been linked to the creation of \emph{de novo} exons, potentially providing a source of new gene formation \cite{stewart2019chromosomal}. Advances in sequencing technology have enhanced our ability to detect structural variation, offering an opportunity to use this \Dros{} system to assess the adaptive potential of chromosomal rearrangements during habitat shifts \cite{li2011statistical, rogers2014landscape, rogers2015chromosomal}.

        To leverage the strengths of the \Dsant{}-\Dyak{} system, we investigate whether chromosomal rearrangements, including those mediated by transposable elements, are more likely to contribute to adaptive changes in island \Dros{} populations on \STome{}. Our objective is to assess whether local adaptation is facilitated by chromosomal rearrangements, and to determine if rearrangements present as standing variation in the ancestral mainland population differ in dynamics and evolutionary outcomes from those that emerged as new mutations during habitat shifts on the island. To achieve this, we generated population genetic panels for \Dyak{} and \Dsant{} from \STome{}, and performed genome-wide scans to identify rearrangements, calculate allele frequencies, and assess population differentiation. We focus on significantly differentiated rearrangements and employ population genetic scans to identify those showing strong signatures of selection. Furthermore, by assessing these rearrangements with RNA expression data, we connect identified genomic patterns to phenotypic changes. To distinguish the effects of these rearrangements from those of population demography and random genetic drift, we employ simulations. Ultimately, we aim to characterize the role of rearrangements and TEs in evolution and adaptation during habitat shifts. We hypothesize that chromosomal rearrangements, including those induced by TEs, represent a rare source of innovation that can drive adaptive cellular changes.

\section*{Results}

    % summary/conceptual figure for results
    \begin{figure}
        \centering
        \subfloat[I]{\label{habitat_map}\frame{\includegraphics[width=0.85\linewidth]{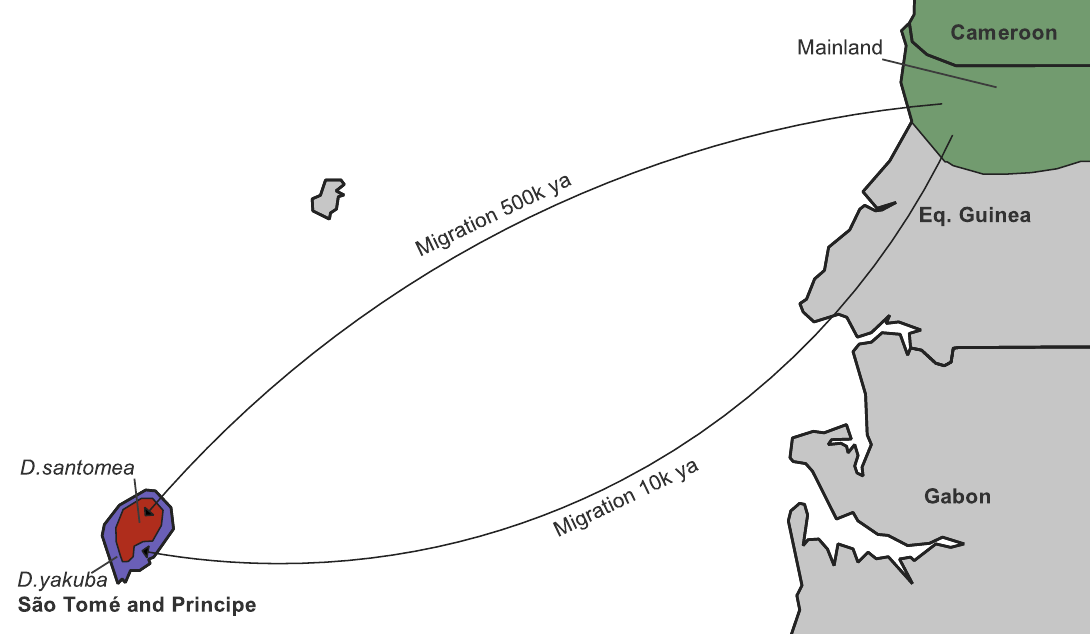}}} \\ 
        \subfloat[II]{\label{read_pairs}\frame{\includegraphics[width=0.9\linewidth]{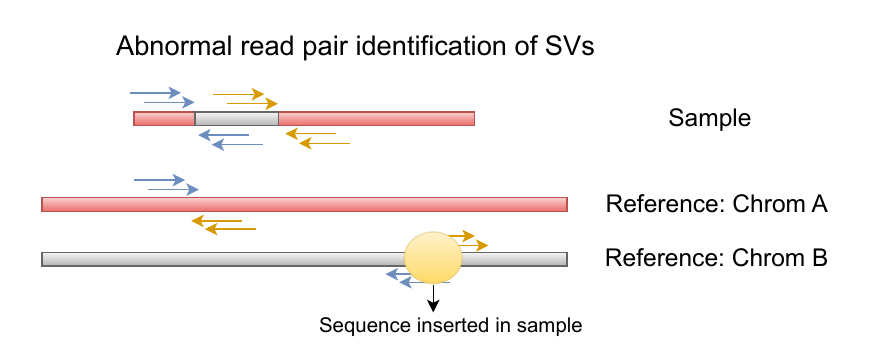}}} \\ 
        \subfloat[III]{\label{highlight_genes}\frame{\includegraphics[width=0.98\linewidth]{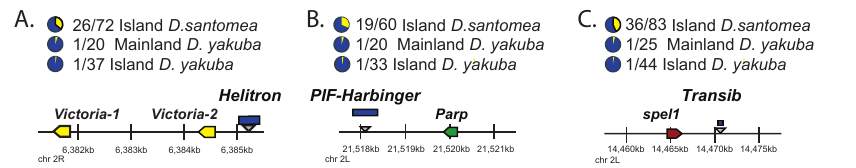}}} \\ 
        \caption{I) Map depicting \Dyak{} habitat shifts from mainland Africa to \STome{} at two different time points. II) Identification of chromosomal rearrangements using abnormally mapping read pairs. III) Three examples of rearrangements with allele frequencies by population. Rearrangements overlapping (or within regulatory regions) of genes (here \emph{Parp}, \emph{Victoria}, and \emph{spel1} are depicted on the chromosome, and labeled with the TE family if TE association was established within rearrangement breakpoints}
        \label{conceptual_1}
    \end{figure}

    Chromosomal rearrangements were characterized in each population to understand their prevalence within and between island and mainland populations (Figure \ref{habitat_map}). Rearrangements were identified using abnormally mapping read pairs from Illumina paired-end reads (Figure \ref{read_pairs}). Across the island populations of \Dsant{}, \Dyak{}, and the ancestral mainland population of \Dyak{}. In total, 16,480 chromosomal rearrangements were identified: 3,412 in 19 strains of mainland \Dyak{}, 7,093 in 35 strains of island \Dyak{}, and 7,796 in 42 strains of \Dsant{} (Figures \ref{varByChrom}). The number of rearrangements per strain varied significantly, ranging from 82 to 875 (Table \ref{breakpoints_dsan}, \ref{breakpoints_dyak}).

    Given the known relationship between transposable elements (TEs) and chromosomal rearrangements, sequence data within rearrangement breakpoints were analyzed and compared to the Repbase repetitive element database \cite{jurka2005repbase}. Of the total rearrangements, 13,763/16,480 (83.5\%) had at least one breakpoint associated with a TE. These were categorized as either likely TE insertions (9,152 instances, 66.5\%) or TE-facilitated ectopic recombination (4,611 instances, 33.5\%). Distinguishing between these mechanisms is crucial, as TE insertions introduce new sequences into the genome, such as regulatory elements or coding regions from the TE itself, while ectopic recombination rearranges existing genomic regions, potentially altering gene structure or regulation. Although TEs constitute only 5.5\% of the reference genome \cite{merel2020transposable}, they are associated with 83.5\% of the identified rearrangements, suggesting a substantial role for TEs in driving chromosomal rearrangements.

    \subsection*{Genome Wide Analysis of Rearrangements}  

        \subsubsection*{Population Differentiation Between Island and Mainland Populations}

            After identifying the total number of rearrangements, derived allele frequencies were calculated for identified rearrangement sites across the three populations (Figures \ref{dsanSFS}, \ref{oranSFS}, \ref{mainSFS}). The site frequency spectra (SFS) were corrected for uneven sample sizes between populations. Comparisons were made between the distributions of rearrangement allele frequencies and those of neutral SNPs, aiming to identify whether rearrangement frequencies deviate from a neutral model within the genome. Significant deviations from neutral expectations were observed for both island \Dsant{} (Kolmogorov-Smirnov, $D=0.65$, $P = 0.00027$) and island \Dyak{} (Kolmogorov-Smirnov, $D=0.65$, $P = 0.0.00027$).

            To further investigate the differences between rearrangement and neutral SNP frequencies, the tails of the allele frequency distributions for each population were examined for high-frequency variants (allele frequency $AF > 0.9$ after SFS correction). In \Dsant{} and island \Dyak{}, 20.9\% (Figure \ref{SFS}A) and 10.2\% (Figure \ref{SFS}B) of allele frequencies, respectively, were classified as high frequency. In contrast, mainland \Dyak{} had only 2.4\% of alleles under the same measurement. Fixed differences for rearrangements were found in 0.855\% and 0.479\% of \Dsant{} and island \Dyak{}, respectively, compared to 0.326\% and 0.174\% for their identified SNPs, meaning rearrangements were seemingly more likely to permeate an entire population compared to SNPs.

            To explore how these rearrangements contribute to genetic differentiation between the island and mainland populations, \deltap{} and \FST{} were calculated for both rearrangements and neutral SNPs. Neutral SNPs were used to establish significance thresholds for rearrangements showing statistically significant deviations from neutral expectations. This analysis identified 244/16,480 (1.48\%) rearrangements in \Dsant{} and 217/16,480 (1.32\%) in island \Dyak{} with significant differentiation from the mainland, based on differences in allele frequency (\deltap{}) (Figure \ref{dsanVsDyak_expchange}. Only 41/383 (10.7\%) of these significantly differentiated rearrangements were shared between the two island populations, which is surprising given the natural introgression observed across these populations \cite{llopart2005anomalous}. These results suggest that a small but crucial subset of rearrangements is driving genetic divergence between island and mainland populations, likely under selective pressures unique to the island environments.

        \subsubsection*{Rearrangements are Strongly Associated with Transposable Elements}

            Notably, the presence of TE-mediated rearrangements is strongly correlated with higher allele frequency and differentiation, suggesting that TEs may enhance the adaptive potential of these rearrangements by increasing their prevalence in the population (Figure \ref{teOnlyPlots}). When TE-mediated rearrangements are filtered from the data, the absence of rearrangements at moderate and high frequencies is striking. (Figure \ref{teRemPlots}). Furthermore, when comparing the allele frequency distribution of TE-associated rearrangements to neutral SNPs, rearrangements significantly deviate from neutral expectations (Kolmogorov-Smirnov, $D = 0.65$, $P = 0.00027$), suggesting the presence of TEs could increases the likelihood of rearrangements conferring a selective advantage, leading to higher allele frequencies than expected under neutrality.           
                        
            Additionally, rearrangements were disproportionately likely to occur near centromeres, where they may facilitate ectopic recombination rather than simply acting as TE insertions. Specifically, 1,990/4,611 (43.2\%) of these rearrangements were located within 3 MB of the centromere ($\chi^2=820.1243$, $P = 1 \times 10^{-5}$), suggesting proximity to the centromere is related with the prevalence of TE-mediated rearrangements (Figures \ref{deltap_ect_dsan_extra}, \ref{deltap_ect_dyak_extra}).
            
            Previous studies have shown that certain TE families can influence transcriptional regulation, affecting the expression of nearby genes \cite{villanueva2019diverse}. Our results indicate that 8,447/13,763 (78.1\%) of rearrangements were associated with just 5/60 (8.33\%) of identified TE families (Figure \ref{AdaptiveTEs_all}), all of which are classified as Class 2 DNA transposons. Together with the increased occurrence of rearrangements near centromeres, these findings highlight the importance of TEs and structural changes in generating genomic diversity that may contribute to adaptive responses in island populations.

        \subsubsection*{Differential Expression and Rearrangements}

            % Figure 3
            \begin{figure}%[h]\centering
                \subfloat[A]{\label{deltap_sig_3L_sub1A}\includegraphics[page=5,width=.48\linewidth]{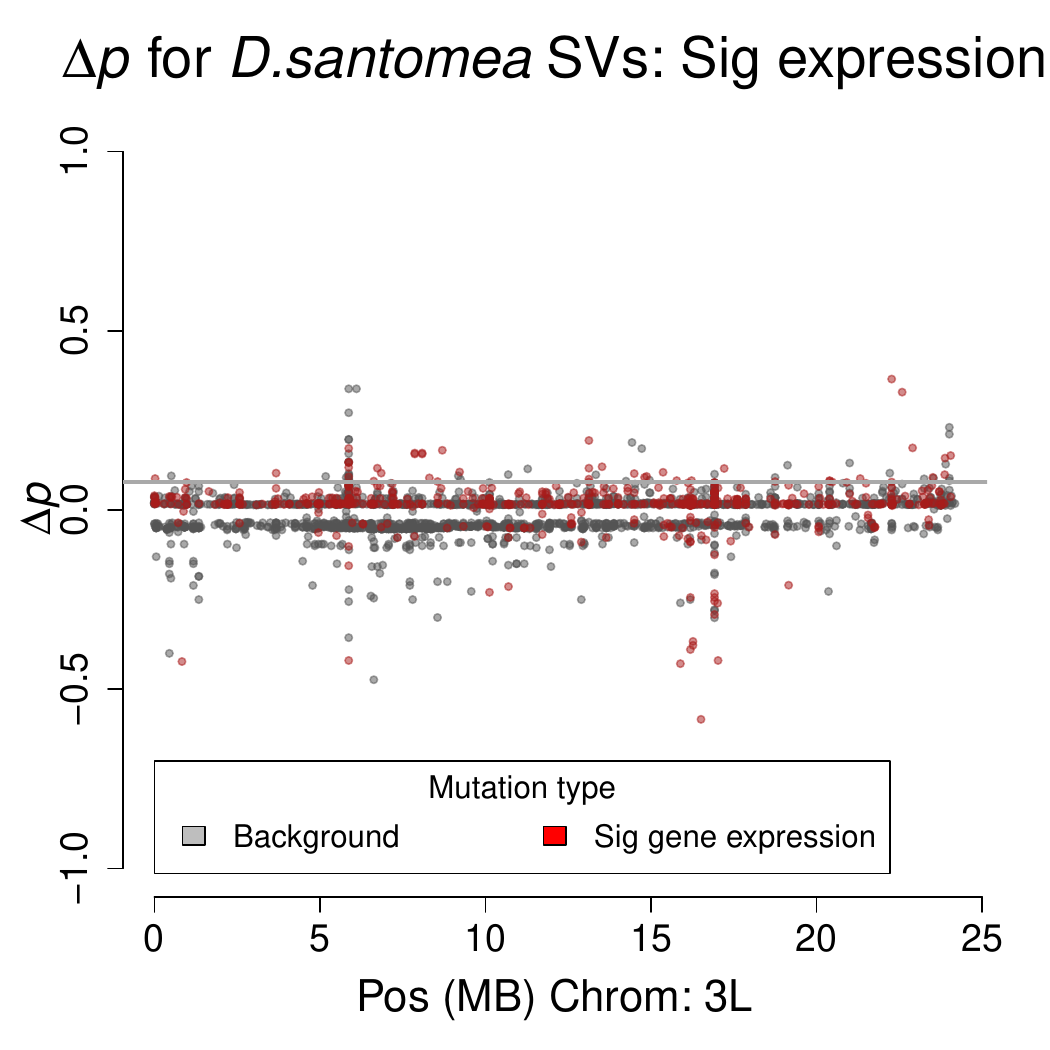}} \hfill
                \subfloat[B]{\label{deltap_sig_X_sub1B}\includegraphics[page=9,width=.48\linewidth]{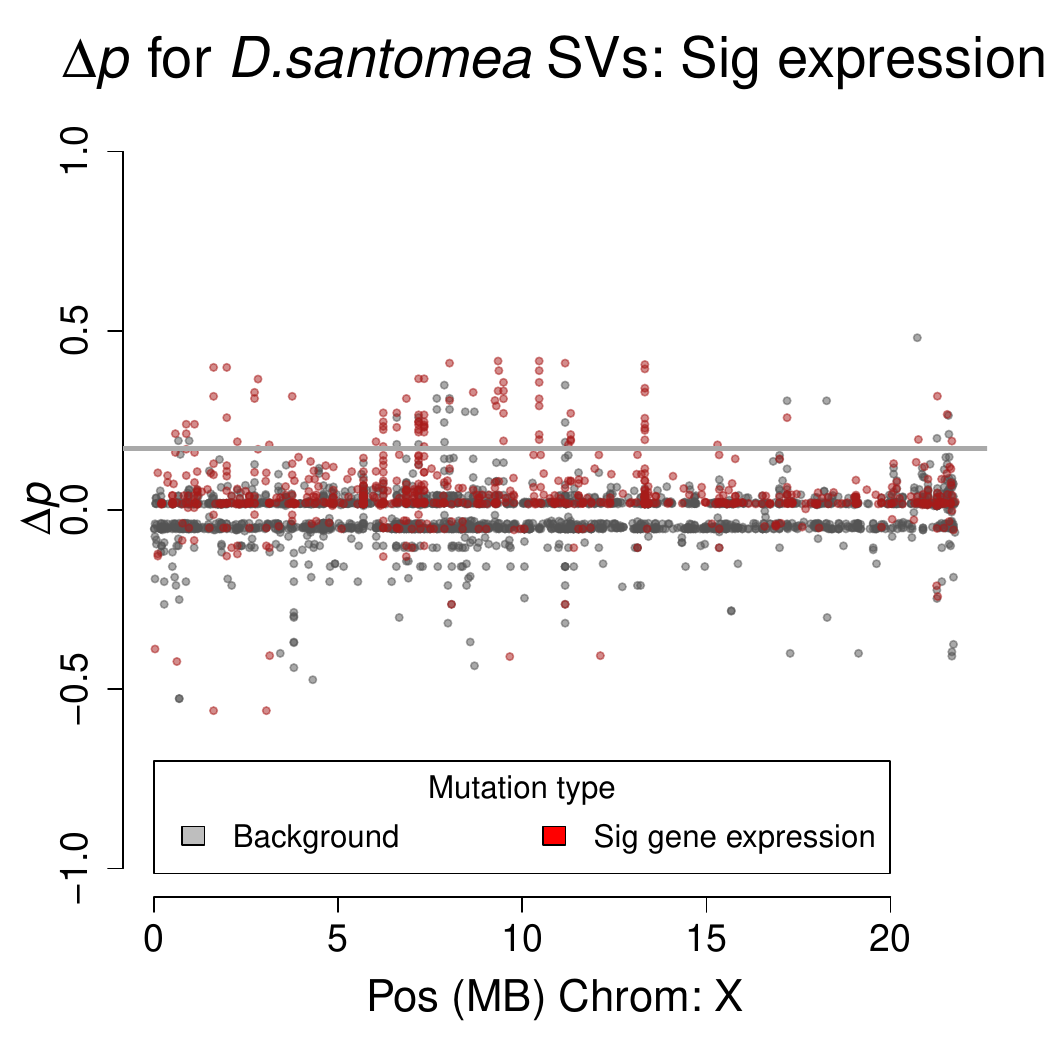}} \\
                \subfloat[C]{\label{deltap_sig_3L_sub1C}\includegraphics[page=6,width=.48\linewidth]{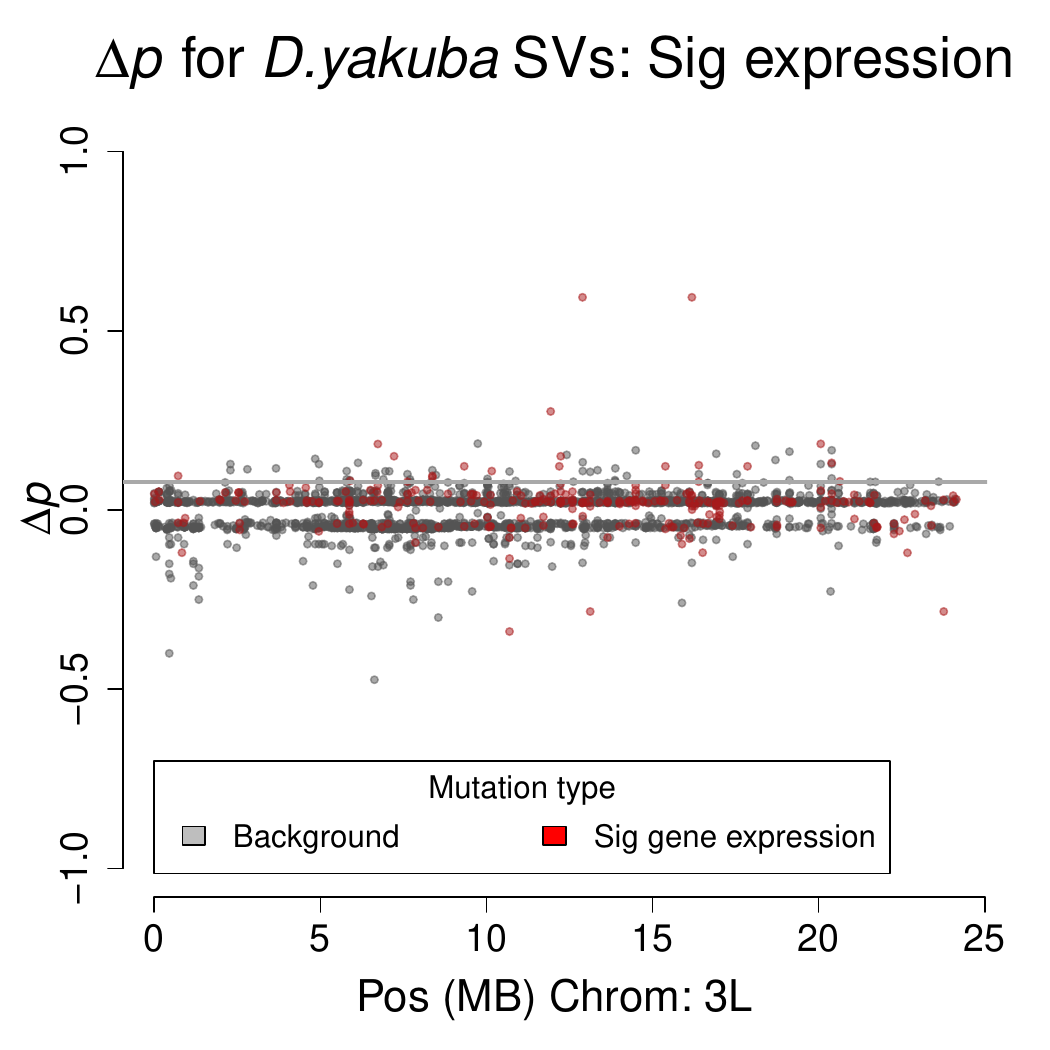}} \hfill
                \subfloat[D]{\label{deltap_sig_X_sub1D}\includegraphics[page=10,width=.48\linewidth]{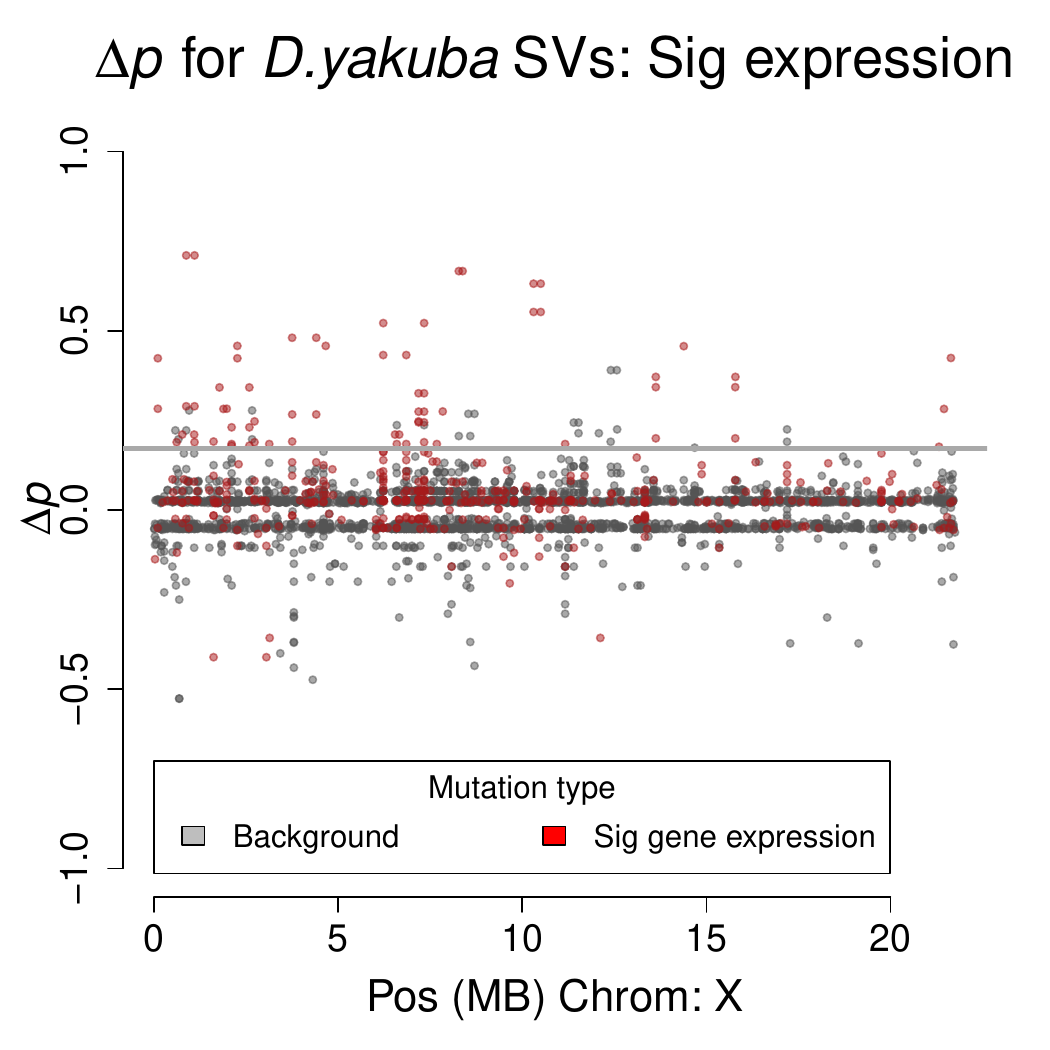}} \\
                \caption{Population differentiation as shown via $\Delta$p A) Between \Dsant{} and mainland ancestral \Dyak{} on autosome 3L.  B)  Between \Dsant{} and mainland ancestral \Dyak{} on the X chromosome  C)  Between Island \Dyak{} and ancestral mainland \Dyak{} on the autosome 3L. D) Between Island \Dyak{} and ancestral mainland \Dyak{} on the X chromosome. The x-axis represents genomic position in megabases, and the y-axis shows the difference in allele frequency of the test population minus the mainland population. Red points indicate rearrangements linked to significant gene expression changes. Differentiation is higher on the X chromosome than on autosomes for both populations, and rearrangements with significant expression changes on the X show strong differentiation, suggesting their importance in local adaptation.}
                \label{dsanVsDyak_expchange}
            \end{figure}

            Differential gene expression near rearrangement breakpoints can modify gene activity, potentially leading to rapid phenotypic changes \cite{de2009impact}. To explore this, we identified 3,008/16,480 (18.3\%) rearrangements that exhibited statistically significant differential gene expression (Fisher's combined $p \leq 0.05$) at adjacent gene sequences within 5 kb of rearrangement breakpoints. In \Dsant{}, his resulted in the identification of 2,032 unique genes that were differentially expressed across at least one tissue sample. These 2,032 genes were expressed a total of 4,572 times across four tissues sampled: testes, ovaries, male soma, and female soma. Specifically, we observed that 1,813 (39.7\%) of these genes were expressed in the testes, 1,205 (26.4\%) in the male soma, 1,361 (29.8\%) in the ovaries, and 193 (4.2\%) in the female soma, with varied expression proportions across different chromosomes (Figures \ref{sigExp_male}, \ref{sigExp_female}).

            In mainland \Dyak{}, 902 unique genes exhibited differential expression associated with 1,947/16,480 (11.8\%) rearrangements across four tissues. Of these genes, 744 (38.2\%) were expressed in the testes, 563 (28.9\%) in the ovaries, 516 (26.5\%) in the male soma, and 124 (6.4\%) in the female soma. For island \Dyak{}, 1,282 unique genes that were differentially expressed, resulting in 2,863 instances of gene expression across the four tissues. Among these, 1,041 (36.4\%) were expressed in the testes, 912 (31.9\%) in the ovaries, 770 (26.8\%) in the male soma, and 140 (4.9\%) in the female soma. 

            To explore links between local adaptation and regulatory changes, we identified rearrangements showing both significant differential gene expression (Fishers adjusted $p$) and significant population differentiation (\deltap). In \Dsant{}, we found that 145/244 (59.4\%) of these rearrangements were linked to both significant expression changes and population differentiation. (Figure \ref{dsanVsDyak_expchange}, \ref{deltap_ge_dyak_extra}). Similarly, 99/217 (45.6\%) of island \Dyak{} rearrangements showed this (Figures \ref{dsanVsDyak_expchange}, \ref{deltap_ge_dyak_extra}). Rearrangements associated with gene expression changes are likely contributing to stronger divergence between island and mainland populations, and potentially can contribute to local adaptation. 
            
            These results indicate that rearrangements are linked to changes in gene expression, suggesting that the evolutionary dynamics of rearrangements may play a critical role in adaptive responses to the unique island environments.

        \subsubsection*{New Mutations, Standing Variation, and Signatures of Selection}

            Building on our previous analyses, we examined whether island rearrangements predominantly arise from new mutations or represent standing variation carried over from the ancestral mainland population. Among the 7,796 rearrangements identified between \Dsant{} and mainland \Dyak{}, we find 7,177 (92.06\%) not observed in the mainland population, likely representing new rearrangements. In contrast, only 619 (7.94\%) rearrangements were identified in the ancestral mainland population, and are likely standing variation. A similar pattern was observed between island and mainland \Dyak{}, with 6,492 (91.5\%) novel rearrangements and 601 (8.50\%) likely representing standing variation. Although new mutations are more prevalent, rearrangements stemming from standing variation show a higher likelihood of further differentiation, with increase in \deltap{}.

            To further investigate how the origin of mutations influences the adaptive potential of rearrangements, we used the integrated haplotype score (iHS) as a secondary test for selection. iHS is particularly sensitive to selective and partial sweeps in regions with polymorphic differences, making it well-suited for detecting selection acting on recent mutations. This method has a known bias toward identifying selection on new mutations over standing variation, which allows us to test whether selective pressures are more likely acting on newly arising rearrangements.    
            
            In \Dsant{}, we observed 2,910 rearrangements with significant iHS values. Of these significant rearrangements, 2,341/2,910 (80.4\%) are likely new mutations, as they have iHS values outside the 95\% confidence interval (CI) established using the distribution of iHS for neutral SNPs. The remaining 569/2,667 (19.6\%) are absent from the ancestral mainland population, and are putatively from standing variation. A similar trend was observed in island \Dyak{}, where 3,707 rearrangements showed significant iHS values, with 1,853/3,707 (82.1\%) likely new mutations and 405/3,707 (17.9\%) from standing variation.  The distribution of iHS values when comparing putative new mutations and standing variation rearrangements is significantly different (Kolmogorov-Smirnov test, $D= 0.092405$, $P = 0.001555$ (Figure \ref{vioPlots}), suggesting that new mutations are more likely to show signatures of recent selection.

            Notably, using iHS identified more statistically significant variants compared to \deltap{} analysis alone. Unlike \deltap{}, which detects changes in allele frequencies that favor higher frequencies, iHS identifies regions with extreme haplotype structures, which can indicate selective sweeps. This allows iHS to capture signals of both positive and partial selection, revealing adaptive changes that may not be detected by \deltap{} alone. By combining these methods, we gain a more comprehensive view of selection, capturing rearrangements that may drive adaptation through different mechanisms. In \Dsant{}, 112/468 (23.9\%) rearrangements that were significant for \deltap{} also showed significant iHS values, while in island \Dyak{}, this overlap was 225/383 (58.7\%) (Figure \ref{upsetihsPlot_dsan}, \ref{upsetihsPlot_oran}). Despite these differences, the overall proportion of significant variants arising from new mutations remained consistent across methods.

    \subsection*{Simulations of Population Demography, Neutrality, and Selection}

        While our results indicate that selection acts on these rearrangements, it is crucial to rule out the influence of other evolutionary forces, such as genetic drift or bottlenecks, that could explain the observed patterns. To address this, we conducted population simulations to compare our empirical results with those from neutral and selection based expectations.  
        
        We first simulated neutral allele frequencies based on known population demography of the island and mainland populations, to establish a baseline for comparison (\ref{sims}. Our results showed no significant difference between the simulated distribution of allele frequencies between simulated neutral SNPs, and those of our empirical within genome neutral SNPs. This suggests that the observed patterns in our data are unlikely to be explained by neutral evolution alone. 

        Next, we simulated population bottlenecks by reducing population size to 20\% of the original, to test whether bottleneck effects could account for the patterns observed in rearrangement frequencies. Even under this extreme scenario, the simulated frequency distributions were not significantly different from those of empirical neutral SNPs (Kolmogorov-Smirnov, $D=0.3$, $P = 0.3356$) (\ref{sims_btl}), indicating that genetic drift, even during population size reductions, has not driven the observed patterns.

        Finally, we examined how selection might influence allele frequencies by simulating mutations with varying selection coefficients (\emph{s}). We found that mutations under stronger selection ($s=0.1$) were far more likely to lead to partial or complete sweeps through a population, compared to scenarios with neutral or weak selection ($s=0.01$). Specifically, scenarios with $s=0.1$ resulted in a 500\% increase in the prevalence of partial or complete sweeps compared to those with weak selection. These findings support the idea that the rearrangement frequencies observed in island populations are driven by selective pressures rather than neutral processes.
       
        Collectively, these simulations bolster our confidence that the patterns observed in rearrangement frequencies are driven by selection in island environments rather than by neutral processes. By ruling out genetic drift, population bottlenecks, and other neutral evolutionary forces, we can more confidently attribute the adaptive variation to selective pressures acting on these rearrangements.

    \subsection*{Candidates of Local Adaptation in Island Populations}

        The identification of rearrangements most likely to be adaptive, or candidate rearrangements, was conducted without \emph{a priori} criteria. These rearrangements were selected based on statistically significant population differentiation, diversity statistics, differential gene expression, and linkage patterns (Figures \ref{upset_dsan}, \ref{upset_oran}, \ref{upsetihsPlot_dsan}, \ref{upsetihsPlot_oran}), with predicted gene function assessed after identification.

        To identify rearrangements that may contribute to local adaptation, those showing both significant differentiation and differential gene expression were further examined. By determining whether these candidate rearrangements arise from new mutations or standing variation, and analyzing their association with signatures of selection or TEs, insights can be gained into how populations adapt during habitat shifts over different timescales. Furthermore, the established relationship between rearrangements and transposable elements (TEs) suggests that analyzing a combination of these features could reveal specific mechanisms that are more likely to contribute to local adaptation.

\begin{figure}
            \centering
            \subfloat[A]{\label{chi_dsan_within}\includegraphics[page=3,width=0.92\linewidth]{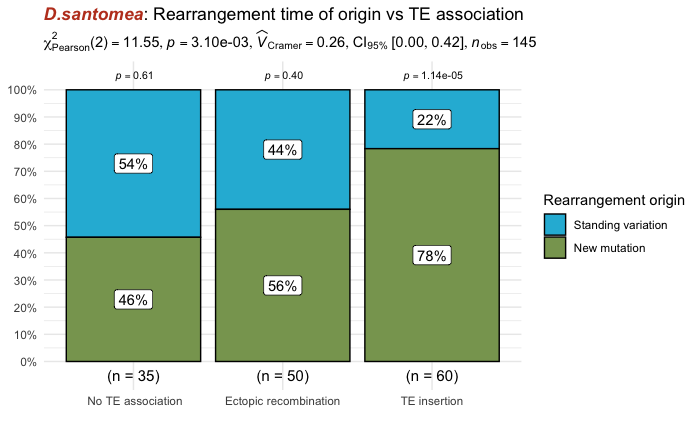}} \\ 
            \subfloat[B]{\label{chi_dyak_within}\includegraphics[page=3,width=0.92\linewidth]{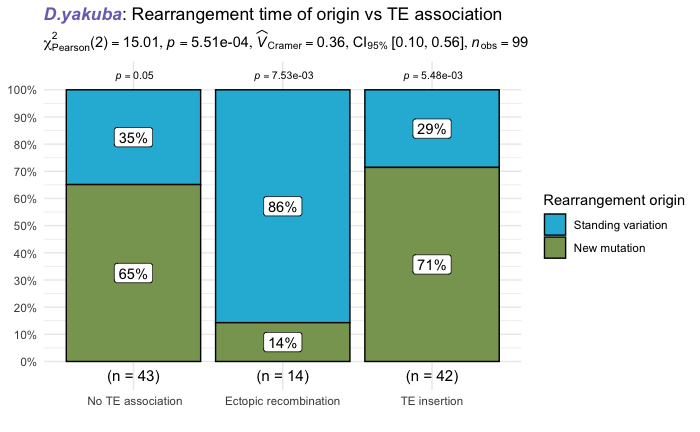}} \\ 
            \caption{Within-population comparisons of rearrangements by TE association (TE insertion, ectopic recombination, or no TE association) and time of origin (standing variation in blue, new mutations in green). Both island populations show statistically significant differences. A) In \Dsant{}, TE insertions differ significantly depending on time of origin. B) Island \Dyak{} has far fewer new mutations associated with ectopic recombination.}
            \label{chi_within}
        \end{figure}

        In \Dsant{}, 145/16,480 (0.88\%) rearrangements exhibited both significant differentiation and differential expression (Figure \ref{upset_dsan}). Of these, 91/145 (62.8\%) are new mutations, while 54/145 (37.2\%) were found in both \Dsant{} and the mainland population. Interestingly, among the candidate rearrangements classified as new mutations, a stronger association with TE insertions was observed, (47/91; 51.6\%) compared to those facilitated by ectopic recombination (28/91; 30.7\%) or those without TE association (16/91, 17.7\%) (Figure \ref{chi_dsan_within}). In contrast, rearrangements from standing variation showed a different distribution: TE insertions (13/54; 24.1\%), ectopic recombination (22/54; 40.7\%), and no TE association (19/54; 35.2\%). 

        % chi square within
        
        This clear distinction between new mutations and standing variation suggests that new TE insertions may play a significant role in the adaptive potential of newly arising rearrangements. The differences in TE association between new mutations and standing variation are statistically significant within \Dsant{} ($\chi^2=11.5548$, $p < 3.09\times10^{-3}$) In island \Dyak{}, a small subset of candidate rearrangements also showed significant differentiation and differential expression, with 99/16,480 (0.60\%) meeting these criteria. Among these, 60/99 (60.6\%) were new mutations, and 39/99 (39.4\%) were derived from standing variation. 
    
        Similar to \Dsant{}, new mutations in island \Dyak{} were strongly associated with TE insertions (30/60; 50.0\%). However, there was a notable difference from \Dsant{} in the role of new mutations associated with ectopic recombination, where only 2/60 (3.3\%) rearrangements (Figure \ref{chi_dyak_within}), and the remaining 28/60 (46.7\%) having no TE association. Rearrangements from standing variation in island \Dyak{} showed a more balanced distribution of associations: TE insertions (12/39; 30.8\%), ectopic recombination (12/39; 30.8\%), and no TE association (15/39; 38.4\%). The association between rearrangement type (new mutations vs. standing variation) and TE involvement (TE insertions, ectopic recombination, or no TE association) was statistically significant within island \Dyak{} ($\chi^2=15.0081$, $p < 5.51\times10^{-4}$), indicating that the distribution of TE mechanisms differs depending on whether the rearrangements arose from new mutations or standing variation.
        
\begin{figure}
            \centering
            \subfloat[A]{\label{chi_sv_between}\includegraphics[width=0.85\linewidth]{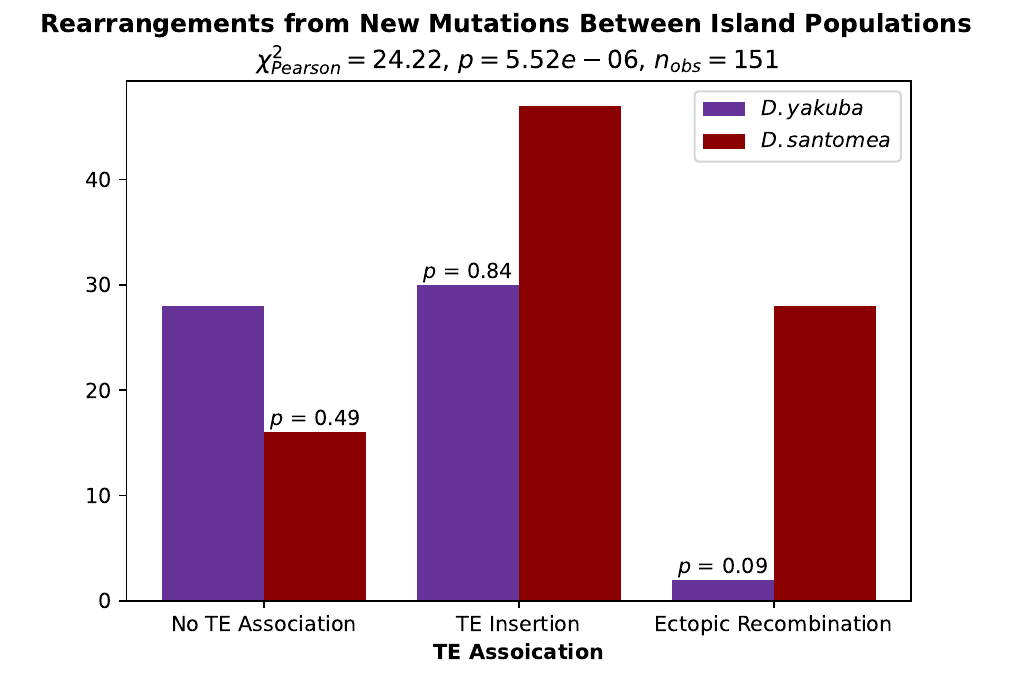}} \\ 
            \subfloat[B]{\label{chi_nm_between}\includegraphics[width=0.85\linewidth]{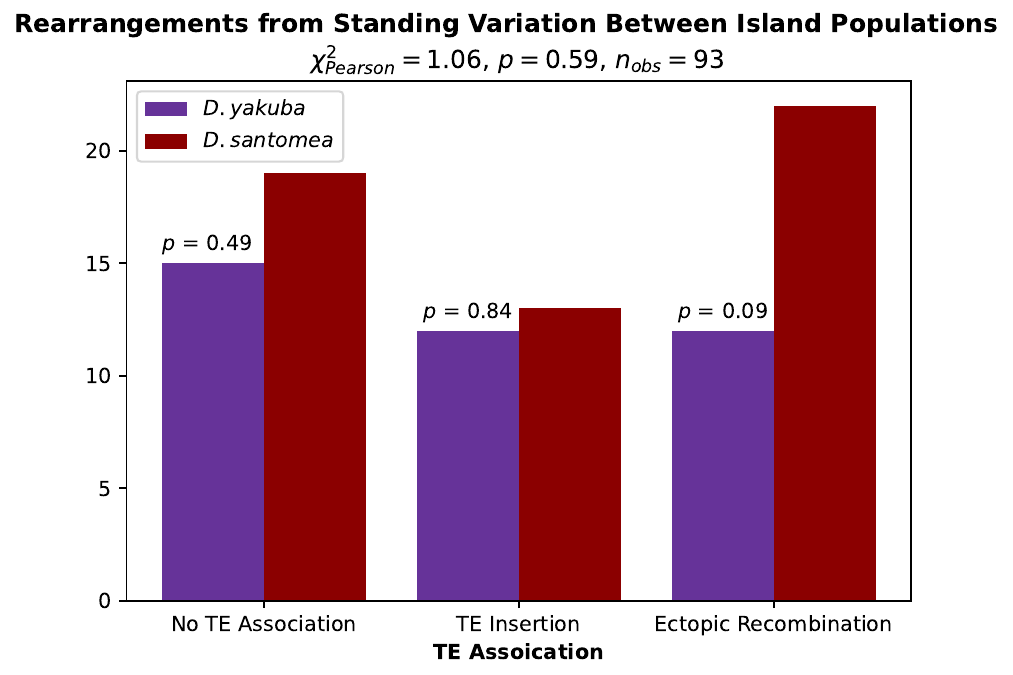}} \\ 
            \caption{Comparison of rearrangements by TE association (TE insertion, ectopic recombination, or no TE association) and population (\Dsant{} in purple, \Dyak{} in red). A) New mutations show significant differences in TE categories, with rearrangements facilitating ectopic recombination being nearly absent in island \Dyak{}. B)  Standing variation shows no significant differences in TE categories between populations.}
            \label{chi_between}
        \end{figure}

        The comparison of rearrangements between island populations, rather than within, for new mutations or standing variation did not show significant differences ($\chi^2=0.3693$, $p=0.543$) (Figure \ref{chi_between}). However, when rearrangements were further separated by TE association, new mutations showed significant differences in TE association between the two island populations ($\chi^2=24.2157$, $p < 1.00\times10^{-5}$) (Figure \ref{chi_between}). Ectopic recombination was notably rare in island \Dyak{}, perhaps suggesting that the time elapsed since a habitat shift influences the mechanisms driving rearrangement formation. 

        \newpage
        % chisq between
        \newpage

        For instance, in \Dsant{}, which has been isolated on \STome{} for a longer period, ectopic recombination events are more frequent and may have had more opportunity to persist within the population. In contrast, \Dyak{} experienced a relatively more recent shift, which could limit the time available for ectopic recombination-based variation to establish and spread. This suggests that rearrangement mechanisms themselves, influenced by colonization timelines, may impact the adaptive potential of new mutations in island populations.

    \subsection*{Functional Relevance and Adaptive Significance of Island Rearrangements}
    
        Functional annotation analysis can inform us of enriched gene functions associated with allf 16,480 identified putative rearrangements identified in all populations. Enriched functions may indicate phenotypes that contribute to local adaptation. When assessing functional annotation clusters for all identified rearrangements, 127 total functional clusters were identified in \Dsant{} and 129 in island \Dyak{}, though no significant functional enrichment was found in either.
            
        In contrast, among the 145 candidate rearrangements in \Dsant{} that are both significantly differentiated and differentially expressed, one significant functional enrichment cluster was identified ($ES \geq 3.46; p < 0.05$). This cluster was composed of differentially expressed genes related to heat shock, stress, and DNA repair. No significant functional enrichment was observed in \Dyak{} for the 99 candidate rearrangements with significant differentiation and differential expression. Genes within this enriched functional cluster are associated with just 0.88\% of all rearrangements.

        The candidate rearrangements linked to the genes within this significant functional cluster consistently exhibited multiple statistically significant measures. For example, four separate rearrangements were identified within 5kb of the gene orthologous to \emph{eIF2alpha}, with one rearrangement displaying an allele frequency of 0.441 in \Dsant{}, strong differentiation from the mainland (\deltap{} of 0.389 and an \FST{} of 0.104), and signatures of positive selection via an extended cluster of statistically significant iHS values, or approximately $iHS < -2.0$ at rearrangement breakpoints. This cluster of low iHS values implies extended haplotypes carrying a derived allele. 

        %%% Table for fusion/express genes
        \begin{table}[ht!]
            \begin{center}
                \caption{Genes associated with rearrangements found in the \Dsant{} significantly enriched function cluster, and their allele frequency in each population.}
                \label{gene_freq_table}
                \begin{tabular}{c|c|c|c} % <-- Alignments: 1st column left, 2nd middle and 3rd right, with vertical lines in between
                \textbf{Region/Ortholog} & \textbf{\emph{D.san} freq} & \textbf{Mainland \emph{D.yak} freq} & \textbf{Island \emph{D.yak} freq}\\
                \hline
                \emph{eIF2alpha}& 34/77 & 1/19 & 2/36\\
                \emph{spel1}& 36/83 & 1/25 &1/44\\
                \emph{Victoria} & 26/72 & 1/20 & 1/37\\
                \emph{Rtel1} & 35/81 & 1/15 & 0/36\\
                \emph{Hsp70Aa} & 21/73 & 0/23 & 1/46
                \end{tabular}
            \end{center}
        \end{table}

        For several candidate rearrangements, the precise cellular and phenotypic impacts in island environments remain less clear. For instance, a candidate rearrangement was observed near a gene orthologous to \emph{pnut}, which encodes proteins involved in cytokinesis and cellular organization \cite{neufeld1994drosophila}. This rearrangement was found at a frequency of 0.205 in \Dsant{} but was absent in \Dyak{} (Figure \ref{nonrandom_highlight}). Another example is a rearrangement at moderate frequency near a gene orthologous to \emph{Smr}, a transcriptional corepressor influencing \Dros{} developmental genes, including \emph{Notch} and ecdysone signaling \cite{heck2011transcriptional}. This rearrangement was present in 0.361\% of \Dsant{} and 0.05\% of mainland \Dyak{}, indicating a potential partial sweep in island \Dsant{}. Additionally, two candidate genes were identified (\ref{random_highlight}) without genes within 10 kb, both showing significant population differentiation (\deltap{} = 0.4651 and \deltap{} = 0.3485) (Table \ref{gene_freq_table}).

        Many of these candidate rearrangements exhibit signatures of selection, with statistically significant indicators such as allele frequency shifts, high population differentiation, extended runs of homozygosity (iHS), and differential gene expression. By evaluating these metrics, delineating candidate rearrangements by TE association and their origin as new mutations or standing variation, we can identify the genetic mechanisms or features most likely to contribute to adaptive cellular changes, even in the absence of functional enrichment. While they may not show significant enrichment, genes associated with candidate rearrangements displaying multiple indicators of selective pressure should not be overlooked. Rearrangements at moderate frequency related to developmental processes in \Dros{} may point to contributions to adaptive phenotypes that are not yet fully understood.

\section*{Discussion}

    \subsection*{Chromosomal Rearrangements in Island \Dros{} Respond To Habitat Change}

        The genetic basis of adaptation during habitat shifts is thought to arise from either pre-existing genetic variation (standing variation) or novel genetic variation (new mutations) \cite{hermisson2005soft, barrett2008adaptation}. Chromosomal rearrangements, often facilitated by transposable elements (TEs) \cite{cridland2013abundance}, are increasingly recognized as key contributors to local adaptation \cite{casacuberta2013impact, gonzalez2010genome, aminetzach2005pesticide}, facilitating “bursts” of genetic novelty under environmental stress \cite{casacuberta2013impact, cridland2013abundance}. These structural mutations can rapidly reshape genome architecture, influencing gene regulation and phenotypic traits \cite{de2009impact, kondrashov2006role, huminiecki2004divergence, harewood2014impact}. The \Dyak{}-\Dsant{} species complex offers an excellent model for studying the role of chromosomal rearrangements during habitat shifts, with two distinct species adapted to the same island: \Dsant{} at high altitude and island \Dyak{} in the lowlands \cite{coyne1989patterns, bachtrog2006extensive, lachaise2000evolutionary, llopart2005anomalous} (Figure \ref{james_exposed}). While genome structure changes remain less explored in evolutionary genetics compared to single nucleotide polymorphisms (SNPs), our findings demonstrate that structural variations can profoundly alter the genetic landscape, potentially driving local adaptation during selective shifts.

        The adaptive potential of structural variation has been documented across a wide range of taxa, from plants to vertebrates \cite{ho2021engines, wellenreuther2019going}. However, distinguishing random genomic changes, such as those driven by genetic drift or population bottlenecks, from those shaped by novel selective pressures during habitat shifts is crucial \cite{bouzat2010conservation}. Demographic events, like bottlenecks followed by population expansion, can reduce the efficacy of negative selection, thereby influencing the allele frequency and genetic divergence of both SNPs and structural variants \cite{bouzat2010conservation, monroe2022mutation, barghi2020polygenic}. Under such conditions, rearrangements would be expected to exhibit allele frequency patterns similar to neutral SNPs \cite{garrigan2010measuring}. In contrast, we observe a significant enrichment of moderate and high-frequency alleles in island populations (Figure \ref{dsanSFS}, \ref{oranSFS}), a pattern far less pronounced in the mainland population (Figure \ref{mainSFS}). This supports the idea that the partial sweep of these structural variants is not stochastic but instead driven by selective pressures, potentially contributing to adaptation to the distinct ecological challenges of \STome. Furthermore, these moderate and high-frequency alleles typically exhibit strong divergence from the mainland population. Using Bonferroni corrected significance thresholds from neutral SNPs, we find that 1.48\% and 1.32\% of rearrangements in \Dsant{} and island \Dyak{}, respectively, show significant differentiation from the mainland. 

        The parallel analysis of rearrangements in two independently derived island populations, each with a shared ancestral source, provides a robust framework for understanding how chromosomal rearrangements contribute to adaptive divergence (Figure \ref{conceptual_1}). Notably, only 7.3\% of these differentiated rearrangements are shared between the two island populations, highlighting their largely independent evolutionary trajectories despite exposure to some shared selective pressures on \STome. This limited overlap in variation suggests that the time since colonization shapes the relative influence of different sources and mechanisms of structural variation. Specifically, evaluating whether rearrangements are associated with transposable elements, standing variation, or new mutations can offer critical insights into which mechanisms disproportionately drive adaptation during habitat shifts.

    \subsection*{TE-mediated rearrangements drive population divergence from the ancestral mainland}    
        
        Transposable elements (TEs) facilitate the formation of chromosomal rearrangements and are strongly associated with increased prevalence, frequency, and genetic differentiation in island populations \cite{ohno2013evolution, conant2008turning, feschotte2008transposable, oliver2009transposable}. TE insertions and TE-facilitated ectopic recombination can introduce new genetic material, inducing large structural changes such as duplications or translocations that alter genomic structure \cite{stewart2019chromosomal, montgomery1991chromosome, bourque2018ten}. Previous studies have shown that TEs can activate in response to stress, introducing variation that remodels gene expression under specific environmental conditions or developmental processes \cite{laudencia2012genotype, cridland2013abundance, casacuberta2013impact}. This TE-mediated structural variation can rapidly alter the genome during habitat shifts \cite{casacuberta2013impact, gonzalez2010genome, aminetzach2005pesticide}, potentially bypassing the lengthy adaptation times typically associated with SNP-based changes \cite{hermisson2005soft}.
        
        TE associated rearrangements behave significantly differently from both non TE associated rearrangements and neutral SNPs within and between populations. Analysis of sequence data at rearrangement breakpoints shows that 83.5\% of the rearrangements are associated with TEs, despite TEs constituting only 5.5\% of the reference genome \cite{merel2020transposable}. Among TE-associated rearrangements, 66.5\% arise from TE insertions, while 33.5\% result from TE-mediated ectopic recombination. TE-associated rearrangements contribute substantially to moderate- and high-frequency alleles in island populations (Figures \ref{teRemPlots}, \ref{teOnlyPlots}) compared to non-TE-associated rearrangements or neutral SNPs. Furthermore, TEs are strongly associated with significant population differentiation and differential gene expression, aligning with findings that TEs drive regulatory divergence \cite{wei2022dynamics, villanueva2019diverse}. The most common TE families in our study, including Helitron, Transib, and DNAREP1, are all Class 2 DNA transposons (Figure \ref{AdaptiveTEs_all}) and are the most frequently associated with rearrangements that show significant differentiation and gene expression differences (Figure \ref{AdaptiveTEs_sig}).        
        
        Collectively, these findings indicate that chromosomal rearrangements in the island populations are not randomly maintained but are shaped by selective pressures that favor genetic variants contributing to local adaptation. While the occurrence of mutations, including those facilitated by TEs, is random, their retention and prevalence in the population appear to be influenced by selective pressures. The association between TE activity and moderate to high frequency rearrangements supports the idea that TEs may introduce structural variants that may offer adaptive advantages, particularly in environments where rapid evolutionary responses are necessary.

    \subsection*{Effects of Chromosomal Rearrangements on Gene Expression Patterns}

        Chromosomal rearrangements can influence gene expression and drive phenotypic diversity, particularly in response to environmental pressures \cite{de2009impact}. These structural changes can reposition genes relative to regulatory elements, disrupt gene function, or even create entirely new regulatory landscapes \cite{de2009impact, kondrashov2006role, huminiecki2004divergence, harewood2014impact}. The impact of rearrangements on gene regulation has been documented across diverse taxa, highlighting their potential as key contributors to adaptive evolution \cite{kondrashov2006role, huminiecki2004divergence, harewood2014impact}. Transposable elements (TEs), which often mediate the formation of rearrangements, can amplify these effects by introducing new promoters, enhancers, or other regulatory sequences into novel genomic contexts \cite{lynch2011transposon, feschotte2008transposable}. 

        In the context of our study, chromosomal rearrangements often exhibit strong links to differential gene expression, particularly in reproductive tissues. Differential expressions associated with rearrangements varied significantly between island and mainland populations: 12.1\% in \Dsant{}, 11.8\% in island \Dyak{}, and 5.7\% in mainland \Dyak{}. Reproductive tissues, including testes and ovaries, showed the strongest association with rearrangement-linked differential expression. These findings align with broader theories suggesting that TEs and rearrangements may play a role in rapidly modulating gene expression, particularly in tissues critical to fitness and reproductive success.

        The timeline of colonization on \STome{} further contextualizes these patterns. In \Dsant{}, 59.4\% of rearrangements associated with significant differentiation from the mainland also showed significant expression changes, while 45.6\% of island \Dyak{} rearrangements exhibited this pattern. Despite this difference, both species exhibit substantial differentiation at rearrangement sites compared to their mainland counterparts, suggesting that novel genetic material introduced through rearrangements plays a crucial role in regulatory changes during habitat shifts. These patterns are consistent with theories that emphasize TEs as drivers of rapid genomic restructuring in response to environmental pressures \cite{gonzalez2010genome, casacuberta2013impact}.

    \subsection*{Adaptive Potential of New Mutations And Standing Variation \newline through TE-Mediated Rearrangements}
    
        Our findings indicate that chromosomal rearrangements associated with transposable elements (TEs) contribute to significant differentiation and gene expression changes, with various levels of adaptive potential arising from both new mutations and standing variation. New mutations were more likely to exhibit significant differentiation and differential expression. However, standing variation remains a notable contributor to adaptation, with significant rearrangements from standing variation appearing at rates higher than expected compared to proportions from neutral expectations. Theory supports this trend, as pre-existing alleles benefit from a “head start” over new mutations in the selection process during habitat shifts \cite{lynch2011transposon, hermisson2005soft, barrett2008adaptation}. Despite being less prevalent, the utility of rearrangements from standing variation is clear. New mutations at low frequency are likely to be lost due to stochastic processes, even when they confer a selective advantage, making standing variation a critical reservoir for rapid adaptive responses \cite{hermisson2005soft}. Standing variation may enable rearrangements to provide immediate adaptive advantages, while new mutations offer a burst of potentially adaptive genetic variation.

        These findings raise two key questions for future research: First, are TE insertions and TE-mediated ectopic recombination equally likely to produce rearrangements that are also associated with significant differential gene expression? Second, do adaptive rearrangements arise primarily from new mutations within the island populations, or from standing variation present at the time of colonization? Addressing these questions will enhance our understanding of the relationship between structural variation, TEs, and regulatory evolution, shedding light on how species adapt to new environments and how genomic changes accumulate and diversify over time.
        
        Novel TE insertions in island populations appear to play a particularly strong role in contributing rearrangements with significant measures in \Dsant{} and island \Dyak{}. In both populations, significant rearrangements are disproportionately associated with novel TE insertions compared to those from standing variation (Figure \ref{chi_within}), suggesting that TE insertions may increase the prevalence of beneficial rearrangements. Seemingly, novel TE insertions appear to escape removal at a higher rate than other rearrangement types, supporting the theory of TE-mediated adaptation via the restructuring of the genome after habitat shifts [cite]. Conversely, in island \Dyak{}, only 3.3\% of significant new rearrangements are associated with ectopic recombination, in contrast to 30.7\% in \Dsant{} (Figure \ref{chi_dyak_within}). This difference suggests that TE-mediated rearrangements may play distinct roles in local adaptation based on both the age of the mutation and the mechanism resulting in a chromosomal rearrangement.

        The disparity in ectopic recombination association between island \Dsant{} and island \Dyak{} may reflects their distinct timescales of adaptation. In \Dsant{}, which has been isolated for approximately 500,000 years, TE-mediated ectopic recombination has likely had sufficient time to establish structural changes in low-recombination regions, particularly near centromeres (Figure \ref{deltap_ect_dsan_extra}) \cite{carneiro2009recombination, haenel2018meta}. Rearrangements in these regions can introduce impactful mutations by altering essential genes or regulatory elements, potentially leading to reduced fitness or genomic instability. However, when rearrangements in conserved regions prove beneficial, they may confer significant adaptive advantages due to their ability to introduce novel genetic solutions. Such rearrangements could allow for adaptation at a small subset of loci, despite the inherent risks of deleterious effects. This pattern is less evident in island \Dyak{}, possibly due to its more recent colonization approximately 10,000 years ago. In island \Dyak{}, the adaptive contributions of TE-mediated rearrangements come almost entirely from new insertions rather than TE-facilitated ectopic recombination, implying that the establishment of adaptive variation is related to the time since a habitat shift.

    \subsection*{Functional Implications of TE-Mediated Rearrangements in Island Populations}

        Connecting rearrangements with statistically significant signals of selection to the genes they interact with provides insights into the evolutionary mechanisms underlying adaptive changes in island populations. Although island \Dyak{} showed no significant functional enrichment, \Dsant{} exhibited enrichment for genes within the heat shock pathway, including those involved in DNA repair and the co regulation of UV stress response (Table \ref{gene_freq_table}). Multiple rearrangements associated with \emph{eIF2alpha} and \emph{Parp} show statistically significant signatures of selection in \Dsant{}, and the the co-regulation of DNA repair and UV stress between these genes is well documented \cite{fortuna2021ddx17}. These rearrangements may reflect strong selective pressures specific to the high-elevation environments inhabited by \Dsant{}.

        Populations living at higher altitudes, like \Dsant{}, face increased UV exposure, which may impose unique selective pressures on genomic integrity and stress response pathways \cite{coyne2002sexual, storz2021high}. Despite these pressures, \Dsant{} exhibits a surprisingly pale phenotype compared to \Dyak{}, a longstanding puzzle given the typical advantage of darker pigmentation under high UV conditions \cite{llopart2002genetics, matute2009little}. Several significant rearrangements in \Dsant{} are located near genes implicated in DNA repair and UV response in \Dmel{} and other model organisms (Table \ref{gene_freq_table}) \cite{matute2013influence, sekelsky2017dna}. These findings may provide a framework for understanding how chromosomal rearrangements partially contribute to the complex puzzle of stress response and UV tolerance, which likely involves a coordinated network of regulatory and structural changes.

        Our recent empirical work supports this hypothesis. UV exposure experiments revealed sex-specific effects in \Dsant{}, with island females displaying greater UV tolerance than their mainland counterparts \cite{titus2023sex}. Elevated allele frequencies of standing variation at UV resistance loci, particularly those linked to stress response genes, provide further evidence that pre-existing genetic variation plays a critical role in facilitating \Dsant{} adaptation to increased UV exposure on \STome{} \cite{titus2023sex, llopart2002genetics, matute2013influence}. The genetic basis for UV tolerance and DNA repair is inherently complex, relying on contributions from multiple genes and pathways. Rearrangements affecting a small but crucial subset of these genes may significantly influence adaptive traits, offering a potential explanation for the observed patterns of UV tolerance on \STome{}.

        TE activity plays a pivotal role in accelerating adaptive shifts by introducing structural changes that reduce the wait times typically required for deterministic sweeps at SNPs \cite{hermisson2005soft, gillespie1994causes}. Our findings suggest that novel TE insertions have dynamically reshaped the genome by translocating DNA segments, driving differentiation between island and mainland populations, and altering gene expression patterns \cite{kondrashov2006role, harewood2014impact}. This dynamic restructuring provides a versatile genetic substrate for rapid adaptation, especially in response to novel environmental challenges over short evolutionary timescales. While UV tolerance is possibly a polygenic trait involving multiple interacting pathways, structural variation offers diverse genetic routes to maintaining DNA integrity under high UV radiation \cite{brodsky2000drosophila}. Investigating this often-overlooked source of genetic variation may reveal critical insights into the genetic basis of UV tolerance at high altitudes, addressing long-standing questions about the origins and maintenance of this adaptive phenotype \cite{bastide2014pigmentation, currall2013mechanisms}.

\section*{Materials and Methods}

    \subsection*{Whole genome sequencing and SNP calling}

        % extractions
        We generated whole genome sequencing for single \Dros {} from 42 strains of \Dsant {} from \STome, 35 strains of \Dyak{} from the island of \STome, and 19 strains of \Dyak{} from the central African mainland (Cameroon and Kenya). DNA was extracted from flies flash frozen in liquid nitrogen following QIAamp Mini Kit (Qiagen) protocol without using RNase A. The resulting DNA samples were quantified (Qubit dsDNA HS assay kit, ThermoFisher Scientific), assessed for quality (Nanodrop ND-2000; A260/A280$>$1.8), and stored at -20°C. 

        % libraries
        Illumina TruSeq Nano DNA libraries were prepared manually following the manufacturer’s protocol (TruSeq Nano DNA, RevD; Illumina). Briefly, samples were normalized to 100ng DNA and sheared by sonication with Covaris M220 (microTUBE 50; Sage Science). The samples were end repaired, purified with Ampure XP beads (Agencourt; Beckman Coulter), adaptors adenylated, and Unique Dual Indices (Table) ligated. Adaptor enrichment was performed using eight cycles of PCR. Following Ampure XP bead cleanup, fragment sizes for all libraries were measured using Agilent Bioanalyzer 2100 (HS DNA Assay; Applied Biosystems). The libraries were diluted 1:10 000 and 1:20 000 and quantified in triplicate using the KAPA Library Quantification Kit (Kapa Biosystems). Equimolar samples were pooled and the libraries were size selected targeting 400-700bp range to remove adaptor monomers and dimers using Pippen Prep DNA Size Selection system (1.5\% Agarose Gel Cassette \#CDF1510; Sage Sciences). Library pools (24 samples per lane) were run on an llumina HiSeq 4000 platform using the 150bp paired end (PE) Cluster Kit.

        % alignment, sorting
        We aligned short read sequences to the \Dyak{} reference genome $Prin.Dyak.Tai18E2$ and \emph{Wolbachia} endoparasite sequence NC\_002978.6 using bwa aln (version 0.7.17) \cite{li2009fast}, and resolved paired end mappings using bwa sampe. We sorted alignments by position using samtools sort (version 1.17) \cite{danecek2021twelve}. Sequence depth on sorted bam files was calculated using samtools depth -aa. Using GATK \cite{mckenna2010genome} we called the SNPs for each strain with the default parameters. 

        % long read validation 
        HiFi PacBio sequences were used to confirm the Illumina structural variant calls. We aling and call structural variants using Minimap2 \cite{li2018minimap2}, Sniffles \cite{sedlazeck2018accurate}, and PacBip's pbsv \cite{wengercomprehensive}. All structural variant calls were done with alignments created by minimp2, except in the case of pbsv which requires alignments from pbmm2 \cite{wengercomprehensive}, a Minimap2 front-end for PacBio native data formats.

    \subsection*{Identification of Chromosomal Rearrangements and Frequency}

        % abnormally mapping
        We identified abnormally mapping read pairs on different chromosomes and long-spanning read pairs greater than 100 kb apart as signals of putative chromosomal rearrangements (Figure \ref{conceptual_1}). Mutations with more than three supporting read pairs were included as candidate chromosomal rearrangements. Rearrangement calls were then clustered across samples using an in-house Python script. Rearrangements were assumed to have both breakpoints (putative start and stop positions) within 325 base pairs of each other to align with the sequencing library insert size. Once clustered, each group of mutations was represented as a single putative rearrangement, with the minimum and maximum positions of all breakpoints defining the cluster boundaries.

        For each rearrangement, we calculated the derived allele frequency for the island and mainland populations. Rearrangement frequencies were determined by clustering across samples using the in-house Python script. This script sorted rearrangements by chromosome and position, grouping them into bins with sizes determined by the library insert size. Adjacent bins were cross-compared to ensure edge cases were not excluded due to sorting limitations. Frequencies were corrected for residual heterozygosity resulting from inbreeding resistance at inversions, using methods similar to those in prior work \cite{rogers2015tandem}, and were polarized based on the ancestral state of each rearrangement (both methods describe further below).

        Using the allele frequency data for each variant, separated by each population, we generated site frequency spectra (SFS) to account for uneven sample sizes across the genome. Allele frequencies in each subpopulation were projected to a uniform sample size of $n=19$ using a hypergeometric transformation to evaluate SFSs under unequal population sizes \cite{stewart2019chromosomal, nielsen2005molecular}.

    \subsection*{False Positive and Negative Corrections}

        To identify heterozygous regions (later used to correct allele frequencies), we used a Hidden Markov Model (HMM) to parse SNP heterozygosity for each strain using the R package "HMM"  (version 1.0.1). After initially running the HMM with incorrect parameters we observe no changes of state, we used the Baum-Welsh algorithm to estimate the parameters of the HMM. With the correct parameters we can rerun the HMM and determine haplotypes for SNPs and calculate their frequencies. In \Dyak, some strains had reference strain contamination, resulting in no heterozygous or homozygous non-reference SNPs. In \Dsant{} no indication of references contamination was observed. We used a three-state HMM for \Dyak{} and an two-state HMM for \Dsant{} to identify reference contamination, inbred haplotypes, and non-inbred haplotypes. Transitions probabilities were set to $10^{-10}$. Emission probabilities were set as heterozygosity for inbred regions would be 0, and heterozygosity in non-inbred regions was $\theta=0.01$, with a lower threshold for probabilities on off-diagonals of $\epsilon=0.00005$ to prevent chilling effects of zero probability. Haplotype calls from the HMM were used to generate correct site frequency spectra for SNPs and structural variants given the variable number of chromosomes sampled across different parts of the genome. 

        % hetero correction
        Within inbred regions, we assume mutations are homozygous, and assign a genotype of 1 structural variant on 1 sampled chromosome. Within heterozygous haplotypes, read pair information alone cannot detect ploidy. For outbred regions with two chromosomes with different ancestry, we used coverage changes to distinguish hemizygous and homozygous mutations. We compared the average coverage of observed mutations to the average coverage across the entire chromosome in a strain looking for an observed coverage between $1.25 <= x <= 1.75$ or $1.75 < x <= 2.25$. Regions whose haplotypes were approximately 1.5x the average coverage of the chromosome are expected to be heterozygous, and regions who have approximately 2x the average are expected to be homozygous \cite{rogers2014landscape}. False negative genotypes are possible when read pair support is insufficient to identify rearrangements de novo. For each mutation, if another strain showed 1.5x or 2x coverage changes and lesser support of 1 or 2 read pairs, allele frequencies were adjusted to avoid low frequency reads, similarly to prior work on gene duplications and rearrangements \cite{rogers2014landscape, stewart2019chromosomal}.

    \subsection*{Polarization of the Ancestral State}

        Genotyping identifies mutations in populations that differ from the reference sequence but does not inherently determine whether a mutation is ancestral or novel. To establish the ancestral state of rearrangements, we compared each variant to the reference genome of \emph{D. teissieri}. This comparison allowed us to determine whether mutations were novel or if the ancestral state had been rearranged in the \Dyak{} reference strain. To polarize mutations, we employed BLASTn \cite{camacho2009blast+} with an E-value threshold of $10^{-20}$ to compare the \Dyak{} reference sequence within +/- 1kb of each rearrangement breakpoint against the \emph{D. teissieri} genome (Prin.Dtei.1.1).

        A mutation was considered ancestral if both sequences mapped to the same location in \emph{D. teissieri} with an alignment length greater than 150 bp, greater than 95\% sequence identity, and less than 10\% overlap in their respective alignments within \emph{D. teissieri}. For samples identified as containing the ancestral state found in \Dtei{}, allele frequencies ($p$) from genotyping were reversed to $(1-p)$ to reflect the reference genome as representing the novel mutation.  

    \subsection*{Associating Rearrangements with Transposable Elements}

        To identify transposable elements (TEs) associated with rearrangements, we used BLAST \cite{camacho2009blast+} to compare variant sequences against the Repbase database \cite{jurka2005repbase}. Variants that mapped to TEs with an E-value threshold of $10^{-20}$ were marked as TEs in our analysis. By performing BLAST on both breakpoints for each rearrangement, we determined whether TEs were present on one or both sides of the rearrangement breakpoints. To verify these mappings, we assessed the coverage of the regions and compared it to the average coverage for the respective chromosome and strain. This ensured accurate identification of TEs associated with rearrangements.

    \subsection*{Population Differentiation}

        Differentiation between populations was assessed using two metrics: \deltap{} and \FST{}. \deltap{} captures directional differences in allele frequencies and is particularly effective for moderate timescales, while \FST{} quantifies population differentiation without directional context and may be more sensitive to changes at low allele frequencies. Both metrics were calculated to identify structural variants with increased frequencies in island environments.

        To establish thresholds for significant differentiation, the distribution of neutral SNP, specifically those located between base pairs 8 and 30 of first introns \cite{halligan2006ubiquitous}, was used as a neutral reference. Thresholds for \deltap{} were determined per chromosome using a Bonferroni corrected P-value of 0.05 (Table \ref{threshTable}).

        Significance  population differentiation was determined by calculating a Bonferroni corrected 95\% confidence interval for \deltap{}, stratified by chromosome to account for chromosomal variation in its distribution. Mutations exceeding the upper bound of the confidence interval (2L: 0.0936917; 2R: 0.09405691; 3L: 0.08218144; 3R: 0.08330457; X: 0.1462702) were classified as significantly differentiated.

        While \deltap{} can sufficiently describe population differentiation, some alleles present on the mainland will naturally be missed in finite sample sizes. This means that the impact of new mutations versus standing variation has the potential to be misunderstood. To characterize the prevalence of missing alleles, we can estimate the expectation of the allele frequencies under a Bayesian framework. We find that for alleles unsampled from mainland populations given a sample size of n=20 the average allele frequency is p=0.002. Thus we conclude that alleles currently identified in our screens as potentially new mutations are at very low frequency, potentially below the level required to establish deterministic sweeps in natural populations \cite{gillespie1994causes, hermisson2005soft}. Regardless, we address the possibility of misattributed new mutations by employing the iHS statistic. By using LD patterns and runs of homozygosity, we can detect if the surrounding region is being misattributed as a new mutation.

    \subsection*{Selective Pressure Inferred From Extended Haplotype Homozygosity}

        Haplotype-based statistics, such as iHS (integrated haplotype score), identify genomic regions under selective pressure by detecting unusually long runs of homozygosity associated with elevated population frequencies. This method is particularly effective for pinpointing loci that have undergone recent directional selection \cite{qanbari2011application}. Additionally, iHS helps differentiate between mutations originating as standing variation and those resulting from new mutations. By leveraging linkage disequilibrium (LD) patterns, iHS is sensitive to new mutations within a population and can provide insights into their contribution to local adaptation.
        
        To determine regions that have recently come under positive selective pressure, we preformed another genome wide scan using iHS. LD patterns surroundings rearrangements that are new mutations will marked differences when compared to those from standing variation. We did this using the R package rehh version 3.2.2 \cite{rehh}. By providing a VCF of each sample, rehh is capable of performing genome wide scans of haplotype homozygosity and linkage to determine outliers. These outliers are established in an identical manner as the \deltap{} thresholds. Using a subset of SNPs from the mainland \Dyak{}, we determine what values are represent a significant departure from a neutral distribution of iHS values. rehh uses phased haplotypes as input. To account for chromosomal rearrangements inside of each samples VCF, we take the midpoint of each of the rearrangement breakpoints and the genotype information for each rearrangement to the unphased VCF. We then phase each VCF using Beagle version 5.4 \cite{ayres2012beagle}. Once the VCFs with rearrangements are phased, rehh can perform genome wide scans and calculate iHS values for each provided site.
                
        %The sensitivity of iHS to new mutations is further supported by the fact that the \Dyak{} population have more significant rearrangements than \Dsant{} when using iHS as a metric of selective pressure. As the island \Dyak{} are relatively new to the island environment compared to \Dsant{}, this increased proportion of significant rearrangements fits with our theory of a TE mediated burst of variation. It is possible that a burst of variation is the cause of the amount of significant rearrangements determined from iHS. While variables statistically significant between \deltap{} and iHS are not 1:1, the proportion of each features is not significantly different, with the most common significant variant still being rearrangements that are new mutations in both island populations of \Dros{} (\ref{upsetihsPlot_dsan}, \ref{upsetihsPlot_oran}).        

    \subsection*{Gene Ontology and Enrichment}

        To assess the functional relevance of rearrangements, we examined a 5 kb window upstream and downstream of each mutation to identify genes either overlapping or within regulatory distance. Variants matching \Dyak{} gene coordinates were mapped to their orthologs in \emph{D. melanogaster} using FlyBase \cite{drysdale2008flybase}. Genes without identified orthologs were excluded from further analysis. Functional annotations of the orthologous genes were analyzed using DAVID\cite{sherman2009systematic} with the "low" stringency setting to identify enriched gene ontology (GO) terms.

    \subsection*{Tissue dissection and RNA sequencing}

        % RNA extractions
        To determine how rearrangements might change gene expression profiles or create new genes, we collected RNAseq data for a subset of 14 strains. Because new gene formation commonly occurs at genes with testes-specific expression \cite{zhou2008origin, rogers2014revised, betran2002retroposed, bachtrog2010dosage} we collected gene expression data for gonads and soma of adult male and female flies. For each strain, we collected virgin males and females within 2 hours of eclosion and placed into separate vials to age for 5-7 days. Once adults, we placed individual males and females on a glass slide with Ringer's solution for gonad dissection. We removed testes and the accessory gland from males and the ovaries from females. Gonads and the carcass (rest of the body minus the gonads) were placed into separate Eppendorf tubes and flash frozen immediately in liquid nitrogen. Five biological replicates were used for each tissue.
    
        RNA was extracted from fly tissue frozen in liquid nitrogen following Zymo DirectZol RNA Microprep (Zymo Research) without DNAase treatment. The resulting RNA samples were quantified (Qubit RNA HS assay kit, ThermoFisher Scientific), assessed for quality (Nanodrop ND-2000; A260/A280$>$2.0), and stored at -80°C.

        Poly (A) enriched strand-specific Illumina TruSeq libraries were manually prepared following the manufacturer’s protocol (TruSeq Stranded mRNA LS, RevD; Illumina). Briefly, samples were normalized to 100ng RNA and poly (A) containing mRNA molecules purified using poly (T) oligo attached magnetic beads. The poly (A) molecules were chemically fragmented for 8 minutes and primed for cDNA synthesis. Reverse transcription was performed using SuperScript IV enzyme (Invitrogen), samples purified with Ampure XP beads (Agencourt; Beckman Coulter), adaptors adenylated, and Unique Dual Indices (Table) ligated. Adaptor enrichment was performed using 15 cycles of PCR. Following Ampure XP bead cleanup, fragment sizes for all libraries were measured using Agilent Bioanalyzer 2100 (HS DNA Assay; Applied Biosystems). The libraries were diluted 1:10 000 and 1:20 000 and quantified in triplicate using the KAPA Library Quantification Kit (Kapa Biosystems). Equimolar samples were pooled (20 samples per lane) and run on an Illumina HiSeq 4000 platform using the 150bp paired end (PE) Cluster Kit.

    \subsection*{Differential expression analysis}

         To determine if strains contain structural rearrangements, we performed a Tophat fusion search version 2.1.0 to find split reads on paired-end RNA sequencing data \cite{kim2011tophat}. This program has previously been validated to identify chimeric constructs using split-read mapping and abnormal read pairs in RNASeq data at loci with genomic DNA signals indicating rearrangements \cite{rogers2017tandem, rogers2015chromosomal, stewart2019chromosomal}.

        We then identified structural variants that were differentially expressed using the program CuffDiff \cite{trapnell2012differential}. To determine whether gene expression changes and fusion transcripts are associated with local adaptation, these were also examined for signatures of selection. We separated structural variant data into male carcass, female carcass, male gonads, and female gonads to determine if these differed across the tissue sample. To identify statistically significant expression changes for rearrangements within 5kb of a gene, we calculated Fisher's Adjusted P-values \cite{tsuyuzaki2013metaseq} using the p-values for expression in each strain that contained that structural variant. To determine whether structural variants as a class are more likely to induce changes in gene expression than we would expect for the genome at large, we performed 10,000 bootstrap replicates with random sampling.

    \subsection*{Demographic modeling of mainland and island \Dros{}}

        % demography simulations
        To simulate island demography we use msprime version 1.1.1 \cite{baumdicker2022efficient}. The demography parameters are assigned based on common values for divergence, arrival time, population size, mutation rate, and recombination rate from literature for each of the 3 \Dros{} populations. We split the populations based on these parameters using msprime, and run ancestry and mutations simulations. We run these simulations for a variety of situations, where we run simulations to test neutrality, bottlenecks ranging from 50\% to 20\% reduction of initial population size, before rebounding in 10 generations. We also run simulations with varying levels of selective pressure on inserted mutations, with values of $s=0.001, 0.01, 0.1$. Msprime is not currently capable of handling simulations involving new mutations, as it uses backwards simulation and the coalescent theory to add mutations to ancestral trees, rather than other popular forward simulation programs.

    \subsection*{Acknowledgements}

        This work was supported by the National Institute of General Medical Sciences at the National Institutes of Health [R35-GM133376 to R.L.R]; T.M was funded in part by the National Institute of General Medical Sciences at the National Institutes of Health [DEB-1737824 to Kelly Dyer]; University of Georgia Training Grant [T32GM007103 to T.M]; and the University of North Carolina, Charlotte [startup funding to R.L.R]. The authors thank Daniel Matute for collecting, maintaining, and sharing \emph{Drosophila} stocks. As requested, we disclose that AI was used for proofreading and minor edits to enhance the clarity and flow of the manuscript. AI was not involved in the design, analysis, or interpretation of results. All scientific content, analyses, figures, tables, and conclusions are the sole work of the authors.

\section*{Data Availability}

    All sequence data are available under SRA PRJNA764098, PRJNA764689, PRJNA764691, PRJNA764693,PRJNA764695, PRJNA764098, PRJNA269314. Supplementary Data are available at\newline \url{https://www.dropbox.com/sh/4fkf4fojcbbnzil/AABpp5TGeefHjaQk1YqI9bPKa?dl=0}.

\clearpage

\bibliographystyle{vancouver}
\bibliography{SvAdapt_Turner.bib}

\begin{thebibliography}{10}

\bibitem{conant2008turning}
Conant GC, Wolfe KH.
\newblock Turning a hobby into a job: how duplicated genes find new functions.
\newblock {Nature Reviews Genetics}. 2008;9(12):938-50.

\bibitem{ohno2013evolution}
Ohno S.
\newblock Evolution by gene duplication.
\newblock Springer Science \& Business Media; 2013.

\bibitem{schlotterer2015genes}
Schl{\"o}tterer C.
\newblock Genes from scratch--the evolutionary fate of \emph{de novo} genes.
\newblock {Trends in Genetics}. 2015;31(4):215-9.

\bibitem{stewart2019chromosomal}
Stewart NB, Rogers RL.
\newblock Chromosomal rearrangements as a source of new gene formation in \emph{{D}rosophila yakuba}.
\newblock {PLoS Genetics}. 2019;15(9):e1008314.

\bibitem{de2009impact}
De S, Teichmann SA, Babu MM.
\newblock The impact of genomic neighborhood on the evolution of human and chimpanzee transcriptome.
\newblock {Genome Research}. 2009;19(5):785-94.

\bibitem{kondrashov2006role}
Kondrashov FA, Kondrashov AS.
\newblock Role of selection in fixation of gene duplications.
\newblock {Journal of Theoretical Biology}. 2006;239(2):141-51.

\bibitem{huminiecki2004divergence}
Huminiecki L, Wolfe KH.
\newblock Divergence of spatial gene expression profiles following species-specific gene duplications in human and mouse.
\newblock {Genome Research}. 2004;14(10a):1870-9.

\bibitem{harewood2014impact}
Harewood L, Fraser P.
\newblock The impact of chromosomal rearrangements on regulation of gene expression.
\newblock {Human Molecular Genetics}. 2014;23(R1):R76-82.

\bibitem{assis2013neofunctionalization}
Assis R, Bachtrog D.
\newblock Neofunctionalization of young duplicate genes in \emph{{D}rosophila}.
\newblock {Proceedings of the National Academy of Sciences}. 2013;110(43):17409-14.

\bibitem{long1993natural}
Long M, Langley CH.
\newblock Natural selection and the origin of \emph{jingwei}, a chimeric processed functional gene in \emph{{D}rosophila}.
\newblock {Science}. 1993;260(5104):91-5.

\bibitem{rogers2012chimeric}
Rogers RL, Hartl DL.
\newblock Chimeric genes as a source of rapid evolution in \emph{{D}rosophila melanogaster}.
\newblock Molecular Biology and Evolution. 2012;29(2):517-29.

\bibitem{zhou2008origin}
Zhou Q, Zhang G, Zhang Y, Xu S, Zhao R, Zhan Z, et~al.
\newblock On the origin of new genes in \emph{{D}rosophila}.
\newblock {Genome Research}. 2008;18(9):1446-55.

\bibitem{schoville2012adaptive}
Schoville SD, Bonin A, Fran{\c{c}}ois O, Lobreaux S, Melodelima C, Manel S.
\newblock Adaptive genetic variation on the landscape: methods and cases.
\newblock {Annual Review of Ecology, Evolution, and Systematics}. 2012;43:23-43.

\bibitem{wellenreuther2019going}
Wellenreuther M, M{\'e}rot C, Berdan E, Bernatchez L.
\newblock Going beyond SNPs: The role of structural genomic variants in adaptive evolution and species diversification.
\newblock Molecular Ecology. 2019;28(6).

\bibitem{morin2004snps}
Morin PA, Luikart G, Wayne RK, Group SW, et~al.
\newblock SNPs in ecology, evolution and conservation.
\newblock Trends in Ecology \& Evolution. 2004;19(4):208-16.

\bibitem{rogers2017tandem}
Rogers RL, Shao L, Thornton KR.
\newblock Tandem duplications lead to novel expression patterns through exon shuffling in \emph{{D}rosophila yakuba}.
\newblock {PLoS Genetics}. 2017;13(5):e1006795.

\bibitem{feschotte2008transposable}
Feschotte C.
\newblock Transposable elements and the evolution of regulatory networks.
\newblock {Nature Reviews Genetics}. 2008;9(5):397-405.

\bibitem{oliver2009transposable}
Oliver KR, Greene WK.
\newblock Transposable elements: powerful facilitators of evolution.
\newblock Bioessays. 2009;31(7):703-14.

\bibitem{bourque2018ten}
Bourque G, Burns KH, Gehring M, Gorbunova V, Seluanov A, Hammell M, et~al.
\newblock Ten things you should know about transposable elements.
\newblock Genome Biology. 2018;19:1-12.

\bibitem{lynch2011transposon}
Lynch VJ, Leclerc RD, May G, Wagner GP.
\newblock Transposon-mediated rewiring of gene regulatory networks contributed to the evolution of pregnancy in mammals.
\newblock {Nature Genetics}. 2011;43(11):1154-9.

\bibitem{wei2022dynamics}
Wei KHC, Mai D, Chatla K, Bachtrog D.
\newblock Dynamics and impacts of transposable element proliferation in the \emph{Drosophila nasuta} species group radiation.
\newblock Molecular Biology and Evolution. 2022;39(5):msac080.

\bibitem{casacuberta2013impact}
Casacuberta E, Gonz{\'a}lez J.
\newblock The impact of transposable elements in environmental adaptation.
\newblock Molecular Ecology. 2013;22(6):1503-17.

\bibitem{gonzalez2010genome}
Gonz{\'a}lez J, Karasov TL, Messer PW, Petrov DA.
\newblock Genome-wide patterns of adaptation to temperate environments associated with transposable elements in \emph{Drosophila}.
\newblock PLoS Genetics. 2010;6(4):e1000905.

\bibitem{aminetzach2005pesticide}
Aminetzach YT, Macpherson JM, Petrov DA.
\newblock Pesticide resistance via transposition-mediated adaptive gene truncation in \emph{{D}rosophila}.
\newblock {Science}. 2005;309(5735):764-7.

\bibitem{lim1994gross}
Lim JK, Simmons MJ.
\newblock Gross chromosome rearrangements mediated by transposable elements in \emph{Drosophila melanogaster}.
\newblock Bioessays. 1994;16(4):269-75.

\bibitem{lister1989molecular}
Lister C, Martin C.
\newblock Molecular analysis of a transposon-induced deletion of the \emph{nivea} locus in \emph{Antirrhinum majus}.
\newblock Genetics. 1989;123(2):417-25.

\bibitem{gray2000takes}
Gray YH.
\newblock It takes two transposons to tango: transposable-element-mediated chromosomal rearrangements.
\newblock Trends in Genetics. 2000;16(10):461-8.

\bibitem{ho2021engines}
Ho EK, Bellis ES, Calkins J, Adrion JR, Latta~IV LC, Schaack S.
\newblock Engines of change: transposable element mutation rates are high and variable within \emph{Daphnia magna}.
\newblock PLoS Genetics. 2021;17(11):e1009827.

\bibitem{cridland2013abundance}
Cridland JM, Macdonald SJ, Long AD, Thornton KR.
\newblock Abundance and distribution of transposable elements in two \emph{{D}rosophila} {QTL} mapping resources.
\newblock {Molecular Biology and Evolution}. 2013;30(10):2311-27.

\bibitem{barrett2008adaptation}
Barrett RD, Schluter D.
\newblock Adaptation from standing genetic variation.
\newblock {Trends in Ecology \& Evolution}. 2008;23(1):38-44.

\bibitem{hermisson2005soft}
Hermisson J, Pennings PS.
\newblock Soft sweeps: molecular population genetics of adaptation from standing genetic variation.
\newblock {Genetics}. 2005;169(4):2335-52.

\bibitem{ohta1992nearly}
Ohta T.
\newblock The nearly neutral theory of molecular evolution.
\newblock {Annual Review of Ecology and Systematics}. 1992;23(1):263-86.

\bibitem{lynch1999perspective}
Lynch M, Blanchard J, Houle D, Kibota T, Schultz S, Vassilieva L, et~al.
\newblock Perspective: spontaneous deleterious mutation.
\newblock {Evolution}. 1999;53(3):645-63.

\bibitem{orr2001haldane}
Orr HA, Betancourt AJ.
\newblock Haldane's sieve and adaptation from the standing genetic variation.
\newblock Genetics. 2001;157(2):875-84.

\bibitem{gillespie1994causes}
Gillespie JH.
\newblock The causes of molecular evolution. vol.~2.
\newblock Oxford University Press On Demand; 1994.

\bibitem{boyer2021adaptation}
Boyer S, H{\'e}rissant L, Sherlock G.
\newblock Adaptation is influenced by the complexity of environmental change during evolution in a dynamic environment.
\newblock PLoS Genetics. 2021;17(1):e1009314.

\bibitem{rosalino2014adaptation}
Rosalino LM, Verdade LM, Lyra-Jorge MC.
\newblock Adaptation and evolution in changing environments.
\newblock Applied Ecology and Human Dimensions in Biological Conservation. 2014:53-71.

\bibitem{coyne1989patterns}
Coyne JA, Orr HA.
\newblock Patterns of speciation in \emph{{D}rosophila}.
\newblock {Evolution}. 1989;43(2):362-81.

\bibitem{bachtrog2006extensive}
Bachtrog D, Thornton K, Clark A, Andolfatto P.
\newblock Extensive introgression of mitochondrial {DNA} relative to nuclear genes in the \emph{{D}rosophila yakuba} species group.
\newblock {Evolution}. 2006;60(2):292-302.

\bibitem{lachaise2000evolutionary}
Lachaise D, Harry M, Solignac M, Lemeunier F, Benassi V, Cariou ML.
\newblock Evolutionary novelties in islands: \emph{{D}rosophila santomea}, a new \emph{melanogaster} sister species from Sao Tome.
\newblock {Proceedings of the Royal Society of London Series B: Biological Sciences}. 2000;267(1452):1487-95.

\bibitem{llopart2005anomalous}
Llopart A, Lachaise D, Coyne JA.
\newblock An anomalous hybrid zone in \emph{{D}rosophila}.
\newblock {Evolution}. 2005;59(12):2602-7.

\bibitem{coyne2002sexual}
Coyne JA, Kim SY, Chang AS, Lachaise D, Elwyn S.
\newblock Sexual isolation between two sibling species with overlapping ranges: \emph{{D}rosophila santomea} and \emph{{D}rosophila yakuba}.
\newblock {Evolution}. 2002;56(12):2424-34.

\bibitem{obbard2012estimating}
Obbard DJ, Maclennan J, Kim KW, Rambaut A, O’Grady PM, Jiggins FM.
\newblock Estimating divergence dates and substitution rates in the \emph{{D}rosophila} phylogeny.
\newblock {Molecular Biology and Evolution}. 2012;29(11):3459-73.

\bibitem{cariou2001divergence}
Cariou ML, Silvain JF, Daubin V, Da~Lage JL, Lachaise D.
\newblock Divergence between \emph{{D}rosophila santomea} and allopatric or sympatric populations of \emph{{D}. yakuba} using paralogous amylase genes and migration scenarios along the Cameroon volcanic line.
\newblock {Molecular Ecology}. 2001;10(3):649-60.

\bibitem{comeault2016correlated}
Comeault AA, Venkat A, Matute DR.
\newblock Correlated evolution of male and female reproductive traits drive a cascading effect of reinforcement in \emph{{D}rosophila yakuba}.
\newblock {Proceedings of the Royal Society B: Biological Sciences}. 2016;283(1835):20160730.

\bibitem{llopart2002genetics}
Llopart A, Elwyn S, Lachaise D, Coyne JA.
\newblock Genetics of a difference in pigmentation between \emph{{D}rosophila yakuba} and \emph{{D}rosophila santomea}.
\newblock {Evolution}. 2002;56(11):2262-77.

\bibitem{rogers2014landscape}
Rogers RL, Cridland JM, Shao L, Hu TT, Andolfatto P, Thornton KR.
\newblock Landscape of standing variation for tandem duplications in \emph{{D}rosophila yakuba} and \emph{{D}rosophila simulans}.
\newblock {Molecular Biology and Evolution}. 2014;31(7):1750-66.

\bibitem{andolfatto2011effective}
Andolfatto P, Wong KM, Bachtrog D.
\newblock Effective population size and the efficacy of selection on the {X} chromosomes of two closely related \emph{{D}rosophila} species.
\newblock {Genome Biology and Evolution}. 2011;3:114-28.

\bibitem{clark2007evolution}
Clark AG, Eisen MB, Smith DR, Bergman CM, Oliver B, Markow TA, et~al.
\newblock Evolution of genes and genomes on the \emph{{D}rosophila} phylogeny.
\newblock {Nature}. 2007;450(7167):203-18.

\bibitem{bhutkar2008chromosomal}
Bhutkar A, Schaeffer SW, Russo SM, Xu M, Smith TF, Gelbart WM.
\newblock Chromosomal rearrangement inferred from comparisons of 12 \emph{{D}rosophila} genomes.
\newblock {Genetics}. 2008;179(3):1657-80.

\bibitem{ranz2003sex}
Ranz JM, Castillo-Davis CI, Meiklejohn CD, Hartl DL.
\newblock Sex-dependent gene expression and evolution of the \emph{{D}rosophila} transcriptome.
\newblock {Science}. 2003;300(5626):1742-5.

\bibitem{li2011statistical}
Li H.
\newblock A statistical framework for {SNP} calling, mutation discovery, association mapping and population genetical parameter estimation from sequencing data.
\newblock {Bioinformatics}. 2011;27(21):2987-93.

\bibitem{rogers2015chromosomal}
Rogers RL.
\newblock Chromosomal rearrangements as barriers to genetic homogenization between archaic and modern humans.
\newblock {Molecular Biology and Evolution}. 2015;32(12):3064-78.

\bibitem{jurka2005repbase}
Jurka J, Kapitonov VV, Pavlicek A, Klonowski P, Kohany O, Walichiewicz J.
\newblock Repbase Update, a database of eukaryotic repetitive elements.
\newblock {Cytogenetic and Genome Research}. 2005;110(1-4):462-7.

\bibitem{merel2020transposable}
M{\'e}rel V, Boulesteix M, Fablet M, Vieira C.
\newblock Transposable elements in \emph{{D}rosophila}.
\newblock {Mobile DNA}. 2020;11(1):1-20.

\bibitem{villanueva2019diverse}
Villanueva-Ca{\~n}as JL, Horvath V, Aguilera L, Gonz{\'a}lez J.
\newblock Diverse families of transposable elements affect the transcriptional regulation of stress-response genes in \emph{{D}rosophila melanogaster}.
\newblock {Nucleic Acids Research}. 2019;47(13):6842-57.

\bibitem{neufeld1994drosophila}
Neufeld TP, Rubin GM.
\newblock The \emph{{D}rosophila} \emph{peanut} gene is required for cytokinesis and encodes a protein similar to yeast putative bud neck filament proteins.
\newblock {Cell}. 1994;77(3):371-9.

\bibitem{heck2011transcriptional}
Heck BW, Zhang B, Tong X, Pan Z, Deng WM, Tsai CC.
\newblock The transcriptional corepressor \emph{SMRTER} influences both \emph{Notch} and ecdysone signaling during \emph{{D}rosophila} development.
\newblock {Biology Open}. 2011;1(3):182-96.

\bibitem{bouzat2010conservation}
Bouzat JL.
\newblock Conservation genetics of population bottlenecks: the role of chance, selection, and history.
\newblock Conservation Genetics. 2010;11:463-78.

\bibitem{monroe2022mutation}
Monroe JG, Srikant T, Carbonell-Bejerano P, Becker C, Lensink M, Exposito-Alonso M, et~al.
\newblock Mutation bias reflects natural selection in \emph{Arabidopsis thaliana}.
\newblock Nature. 2022;602(7895):101-5.

\bibitem{barghi2020polygenic}
Barghi N, Hermisson J, Schl{\"o}tterer C.
\newblock Polygenic adaptation: a unifying framework to understand positive selection.
\newblock Nature Reviews Genetics. 2020;21(12):769-81.

\bibitem{garrigan2010measuring}
Garrigan D, Lewontin R, Wakeley J.
\newblock Measuring the sensitivity of single-locus “neutrality tests” using a direct perturbation approach.
\newblock Molecular Biology and Evolution. 2010;27(1):73-89.

\bibitem{montgomery1991chromosome}
Montgomery E, Huang S, Langley C, Judd B.
\newblock Chromosome rearrangement by ectopic recombination in \emph{{D}rosophila melanogaster}: genome structure and evolution.
\newblock {Genetics}. 1991;129(4):1085-98.

\bibitem{laudencia2012genotype}
Laudencia-Chingcuanco D, Fowler DB.
\newblock Genotype-dependent burst of transposable element expression in crowns of hexaploid wheat (\emph{{T}riticum aestivum L.}) during cold acclimation.
\newblock {Comparative and Functional Genomics}. 2012;2012.

\bibitem{carneiro2009recombination}
Carneiro M, Ferrand N, Nachman MW.
\newblock Recombination and speciation: loci near centromeres are more differentiated than loci near telomeres between subspecies of the European rabbit (\emph{Oryctolagus cuniculus}).
\newblock Genetics. 2009;181(2):593-606.

\bibitem{haenel2018meta}
Haenel Q, Laurentino TG, Roesti M, Berner D.
\newblock Meta-analysis of chromosome-scale crossover rate variation in eukaryotes and its significance to evolutionary genomics.
\newblock Molecular Ecology. 2018;27(11):2477-97.

\bibitem{fortuna2021ddx17}
Fortuna TR, Kour S, Anderson EN, Ward C, Rajasundaram D, Donnelly CJ, et~al.
\newblock \emph{DDX17} is involved in DNA damage repair and modifies \emph{FUS} toxicity in an RGG-domain dependent manner.
\newblock Acta Neuropathologica. 2021;142:515-36.

\bibitem{storz2021high}
Storz JF.
\newblock High-altitude adaptation: mechanistic insights from integrated genomics and physiology.
\newblock Molecular Biology and Evolution. 2021;38(7):2677-91.

\bibitem{matute2009little}
Matute DR, Butler IA, Coyne JA.
\newblock Little effect of the tan locus on pigmentation in female hybrids between \emph{{D}rosophila santomea} and \emph{{D} melanogaster}.
\newblock {Cell}. 2009;139(6):1180-8.

\bibitem{matute2013influence}
Matute DR, Harris A.
\newblock The influence of abdominal pigmentation on desiccation and ultraviolet resistance in two species of \emph{Drosophila}.
\newblock Evolution. 2013;67(8):2451-60.

\bibitem{sekelsky2017dna}
Sekelsky J.
\newblock DNA repair in \emph{Drosophila}: mutagens, models, and missing genes.
\newblock Genetics. 2017;205(2):471-90.

\bibitem{titus2023sex}
Titus-McQuillan JE, Turner BA, Rogers RL.
\newblock Sex-specific ultraviolet radiation tolerance across \emph{Drosophila}.
\newblock arXiv preprint arXiv:231001743. 2023.

\bibitem{brodsky2000drosophila}
Brodsky MH, Nordstrom W, Tsang G, Kwan E, Rubin GM, Abrams JM.
\newblock \emph{{D}rosophila} \emph{p53} binds a damage response element at the \emph{reaper} locus.
\newblock {Cell}. 2000;101(1):103-13.

\bibitem{bastide2014pigmentation}
Bastide H, Yassin A, Johanning EJ, Pool JE.
\newblock Pigmentation in \emph{{D}rosophila melanogaste} reaches its maximum in {E}thiopia and correlates most strongly with ultra-violet radiation in sub-{S}aharan {A}frica.
\newblock {BMC Evolutionary Biology}. 2014;14(1):1-14.

\bibitem{currall2013mechanisms}
Currall BB, Chiangmai C, Talkowski ME, Morton CC.
\newblock Mechanisms for structural variation in the human genome.
\newblock Current Genetic Medicine Reports. 2013;1:81-90.

\bibitem{li2009fast}
Li H, Durbin R.
\newblock Fast and accurate short read alignment with Burrows--Wheeler transform.
\newblock Bioinformatics. 2009;25(14):1754-60.

\bibitem{danecek2021twelve}
Danecek P, Bonfield JK, Liddle J, Marshall J, Ohan V, Pollard MO, et~al.
\newblock Twelve years of SAMtools and BCFtools.
\newblock Gigascience. 2021;10(2):giab008.

\bibitem{mckenna2010genome}
McKenna A, Hanna M, Banks E, Sivachenko A, Cibulskis K, Kernytsky A, et~al.
\newblock The Genome Analysis Toolkit: a MapReduce framework for analyzing next-generation {DNA} sequencing data.
\newblock {Genome Research}. 2010;20(9):1297-303.

\bibitem{li2018minimap2}
Li H.
\newblock Minimap2: pairwise alignment for nucleotide sequences.
\newblock {Bioinformatics}. 2018;34(18):3094-100.

\bibitem{sedlazeck2018accurate}
Sedlazeck FJ, Rescheneder P, Smolka M, Fang H, Nattestad M, Von~Haeseler A, et~al.
\newblock Accurate detection of complex structural variations using single-molecule sequencing.
\newblock {Nature Methods}. 2018;15(6):461-8.

\bibitem{wengercomprehensive}
Wenger AM.
\newblock Comprehensive structural and copy-number variant detection with long reads. 2019.

\bibitem{rogers2015tandem}
Rogers RL, Cridland JM, Shao L, Hu TT, Andolfatto P, Thornton KR.
\newblock Tandem duplications and the limits of natural selection in \emph{{D}rosophila yakuba} and \emph{{D}rosophila simulans}.
\newblock {PLoS One}. 2015;10(7):e0132184.

\bibitem{nielsen2005molecular}
Nielsen R.
\newblock Molecular signatures of natural selection.
\newblock {Annu Rev Genet}. 2005;39:197-218.

\bibitem{camacho2009blast+}
Camacho C, Coulouris G, Avagyan V, Ma N, Papadopoulos J, Bealer K, et~al.
\newblock BLAST+: architecture and applications.
\newblock BMC Bioinformatics. 2009;10:1-9.

\bibitem{halligan2006ubiquitous}
Halligan DL, Keightley PD.
\newblock Ubiquitous selective constraints in the \emph{{D}rosophila} genome revealed by a genome-wide interspecies comparison.
\newblock {Genome Research}. 2006;16(7):875-84.

\bibitem{qanbari2011application}
Qanbari S, Gianola D, Hayes B, Schenkel F, Miller S, Moore S, et~al.
\newblock Application of site and haplotype-frequency based approaches for detecting selection signatures in cattle.
\newblock BMC Genomics. 2011;12(1):1-12.

\bibitem{rehh}
Gautier M, Klassmann A, Vitalis R.
\newblock rehh 2.0: a reimplementation of the R package rehh to detect positive selection from haplotype structure.
\newblock Molecular Ecology Resources. 2017;17(1):78-90.

\bibitem{ayres2012beagle}
Ayres DL, Darling A, Zwickl DJ, Beerli P, Holder MT, Lewis PO, et~al.
\newblock BEAGLE: an application programming interface and high-performance computing library for statistical phylogenetics.
\newblock Systematic Biology. 2012;61(1):170-3.

\bibitem{drysdale2008flybase}
Drysdale R, Consortium F, et~al.
\newblock FlyBase.
\newblock {\emph{Drosophila}}. 2008:45-59.

\bibitem{sherman2009systematic}
Sherman BT, Lempicki RA, et~al.
\newblock Systematic and integrative analysis of large gene lists using DAVID bioinformatics resources.
\newblock {Nature Protocols}. 2009;4(1):44-57.

\bibitem{rogers2014revised}
Rogers RL, Shao L, Sanjak JS, Andolfatto P, Thornton KR.
\newblock Revised annotations, sex-biased expression, and lineage-specific genes in the \emph{{D}rosophila melanogaster} group.
\newblock {G3: Genes, Genomes, Genetics}. 2014;4(12):2345-51.

\bibitem{betran2002retroposed}
Betr{\'a}n E, Thornton K, Long M.
\newblock Retroposed new genes out of the {X} in \emph{{D}rosophila}.
\newblock {Genome Research}. 2002;12(12):1854-9.

\bibitem{bachtrog2010dosage}
Bachtrog D, Toda NR, Lockton S.
\newblock Dosage compensation and demasculinization of {X} chromosomes in \emph{{D}rosophila}.
\newblock {Current Biology}. 2010;20(16):1476-81.

\bibitem{kim2011tophat}
Kim D, Salzberg SL.
\newblock TopHat-Fusion: an algorithm for discovery of novel fusion transcripts.
\newblock {Genome Biology}. 2011;12(8):1-15.

\bibitem{trapnell2012differential}
Trapnell C, Roberts A, Goff L, Pertea G, Kim D, Kelley DR, et~al.
\newblock Differential gene and transcript expression analysis of {RNA}-seq experiments with TopHat and Cufflinks.
\newblock {Nature Protocols}. 2012;7(3):562-78.

\bibitem{tsuyuzaki2013metaseq}
Tsuyuzaki K, Nikaido I.
\newblock metaSeq: Meta-analysis of {RNA}-seq count data.
\newblock {Tokyo University of Science, Tokyo}. 2013.

\bibitem{baumdicker2022efficient}
Baumdicker F, Bisschop G, Goldstein D, Gower G, Ragsdale AP, Tsambos G, et~al.
\newblock Efficient ancestry and mutation simulation with msprime 1.0.
\newblock Genetics. 2022;220(3):iyab229.

\end{thebibliography}

%\printbibliography

%%%%%%%%%%%%%%%%%%%%% SUPPLEMENT   %%%%%%%%%%%%%%%
\section*{Supplementary Information}

\pagenumbering{arabic}
\renewcommand{\thefigure}{S\arabic{figure}}
\renewcommand{\thetable}{S\arabic{table}}
\setcounter{figure}{0}
\setcounter{table}{0}

    \subsection*{Resolving Complex Variation}

        % complex var
        Previous work has shown that PacBio can confirm 100\% of Illumina paired-end reads when using alignment methods in tandem, as the rate varies on the confirmation method (BLAST, MM2, PBSV, etc). (Figure \ref{confirm_rate}). The majority of unconfirmed regions have erratic coverage, implying the identification of complex variation that aligners and assemblers struggle to solve. To investigate further, we conducted targeted \emph{de novo} assembly of these regions and plotted them with Mummer. 

    \subsection*{Complex variation causes genotyping challenges}

        % long read sv calling layout 
        Long read sequencing is proposed as a promising solution to structural variant calling. However, computational and technical limitations remain, especially in the context of population genetics where false positives and false negatives may alter allele frequencies. We explored the utility of HiFi PacBio reads to determine structural variants using 3 pre-existing tools and a BLAST comparison. 

        % long reads method
        By looking at three programs designed to align and call structural variants, Minimap2, Pbsv, and Sniffles, we ascertained that these programs were not confirming our Illumina reads at a sufficient rate individually. Individual confirmation rates ranged from 3.2\%-67.5\% across individual bioinformatics pipelines. \Dros{} is known to have many small structural variants in comparison with mammals. Deletions accumulate rapidly in Drosophila, creating gaps in alignment that make it more difficult to clarify mutation state. We find that complex variation is difficult for existing aligners to solve. To confirm our suspicions we used coverage data and investigated these complex regions where we frequently, and were able to find signs in the coverage that there was likely some kind of variation occurring, whether it be duplications, rearrangements, or other forms of variation. 

\begin{figure}
    \centering
    \subfloat[A]{\label{confirm_B130013}\includegraphics[page=1, width=.33\linewidth]{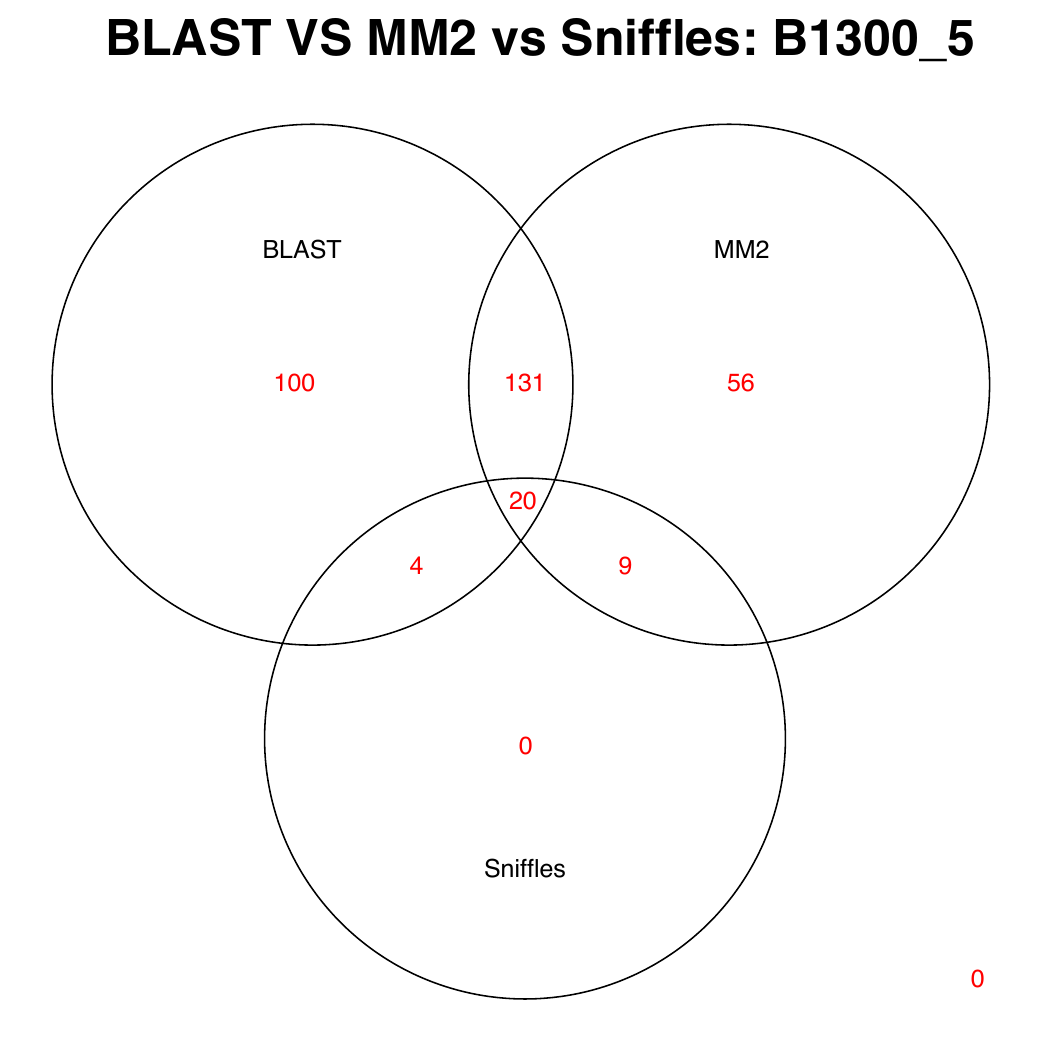}}
    \subfloat[B]{\label{confirm_B13005}\includegraphics[width=.33\linewidth]{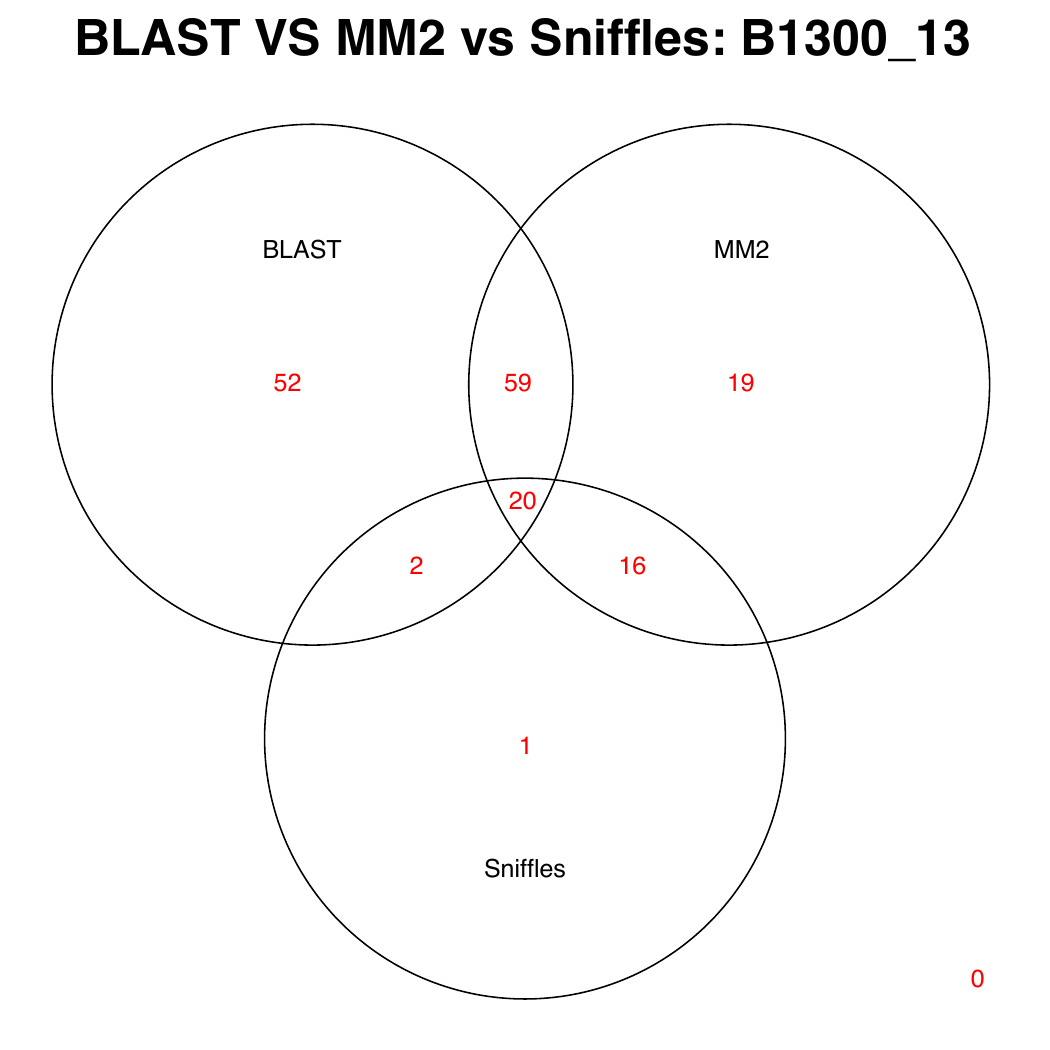}}
    \subfloat[C]{\label{confirm_car}\includegraphics[width=.33\linewidth]{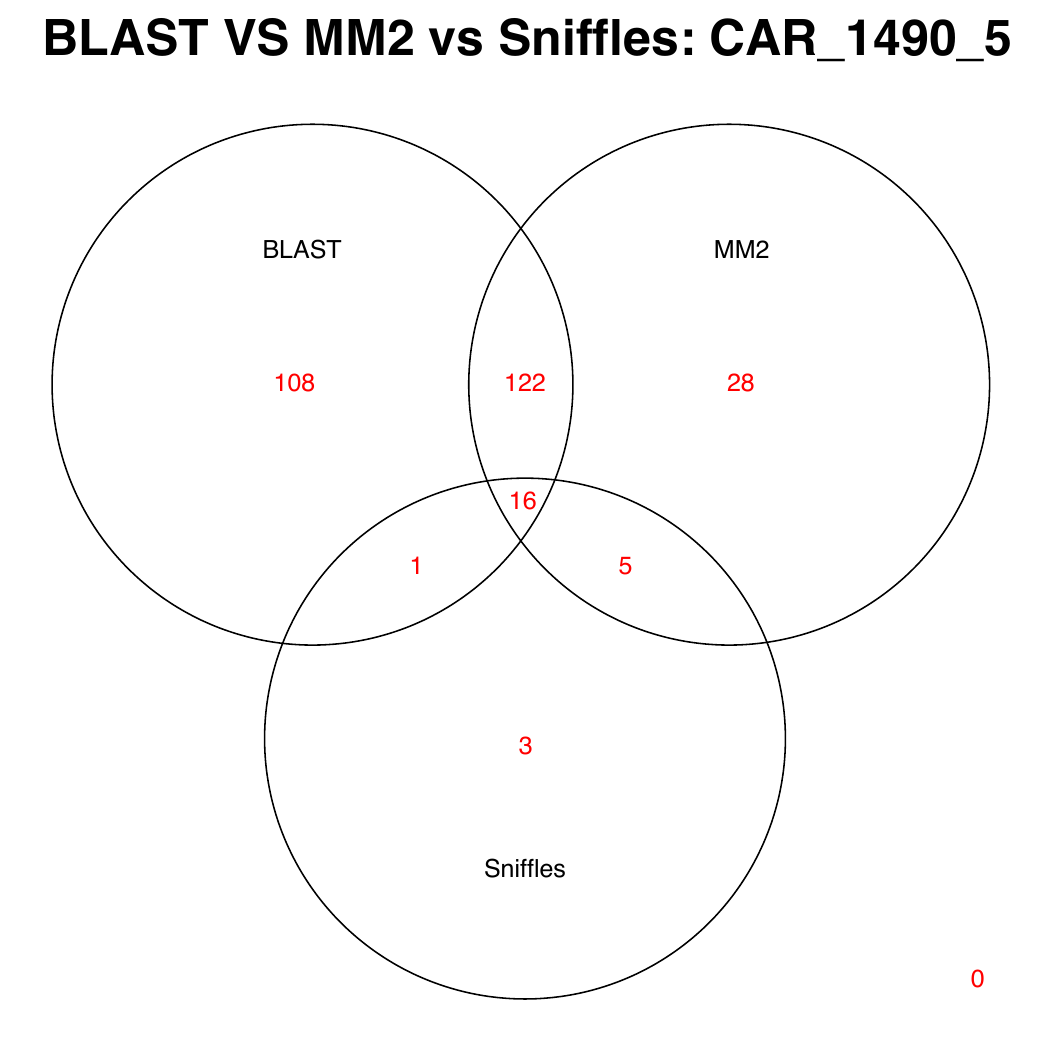}}
    \caption{Confirmation rates for 336 rearrangements in three strains with PacBio HiFi data.   We observe low confirmation rates from Sniffles, which is tuned to identify large rearrangements common in mammals.  Minimap2 alignments confirm a moderate number of rearrangements in each strain.  Using BLAST to search for reads matching to rearrangement breakpoints, we are able to confirm an additional 308 rearrangements.  These differing rates of confirmation suggest that greater methods development is required to identify moderately sized rearrangement sequences in long read data.}
    \label{confirm_rate}
\end{figure}

        % when we use blast to align
        Taking all of these factors into consideration, we aligned sequence data for these regions a BLAST  against and the long-read data. Once we did this we were able to confirm rearrangements at a rate of 79.7\%-87.2\% from strain to strain \ref{confirm_rate}. In aggregate across all 4 bioinformatic pipelines, we confirm 100\% of rearrangement genotypes. BLAST is not designed to align long read sequence data, however with BLAST we are able to confirm a total of 100\% that these regions do share similarity based on nucleotide identity and that the current tools commonly used to call these mutations are not accurately handling all cases of structural variation in \Dros{}. 

        % long read conclusion
        The prospects of structural variant calling in Drosophila with new HiFi sequence technology are promising.  However, greater computational and bioinformatic analysis, likely with species-specific parameter tuning is likely necessary to solve these technical issues in data analysis for complex regions of the genome.

% \clearpage

 %%%%%%%%%%%%%%%%%%%

         % Sig_exp_deltap 
\newpage
\begin{table}[ht!]
    \begin{center}
    \caption{Number of rearrangements breakpoints per chromosome}
    \begin{tabular}{c|c|c|c|c|c}
    \textbf{Population} & \textbf{2L} & \textbf{2R} & \textbf{3L} & \textbf{3R} & \textbf{X}\\
    \hline
        All populations & 6,338 & 5,597 & 5,279 & 6,633 & 4,818\\
        \Dsant{} & 3,301 & 2,792 & 2,475 & 3,037 & 2,152\\
        Island \Dyak{} & 2,600 & 2,340 & 2,155 & 3,017 & 2,020\\              
        Mainland \Dyak{} & 1,176 & 1,047 & 1,105 & 1,225 & 1,186\\
    \end{tabular}

    \label{varByChrom}
    \end{center}
\end{table}

 % Breakpoints
\begin{figure}[htp]
    \centering
    \includegraphics[page=1,width=.4\textwidth]{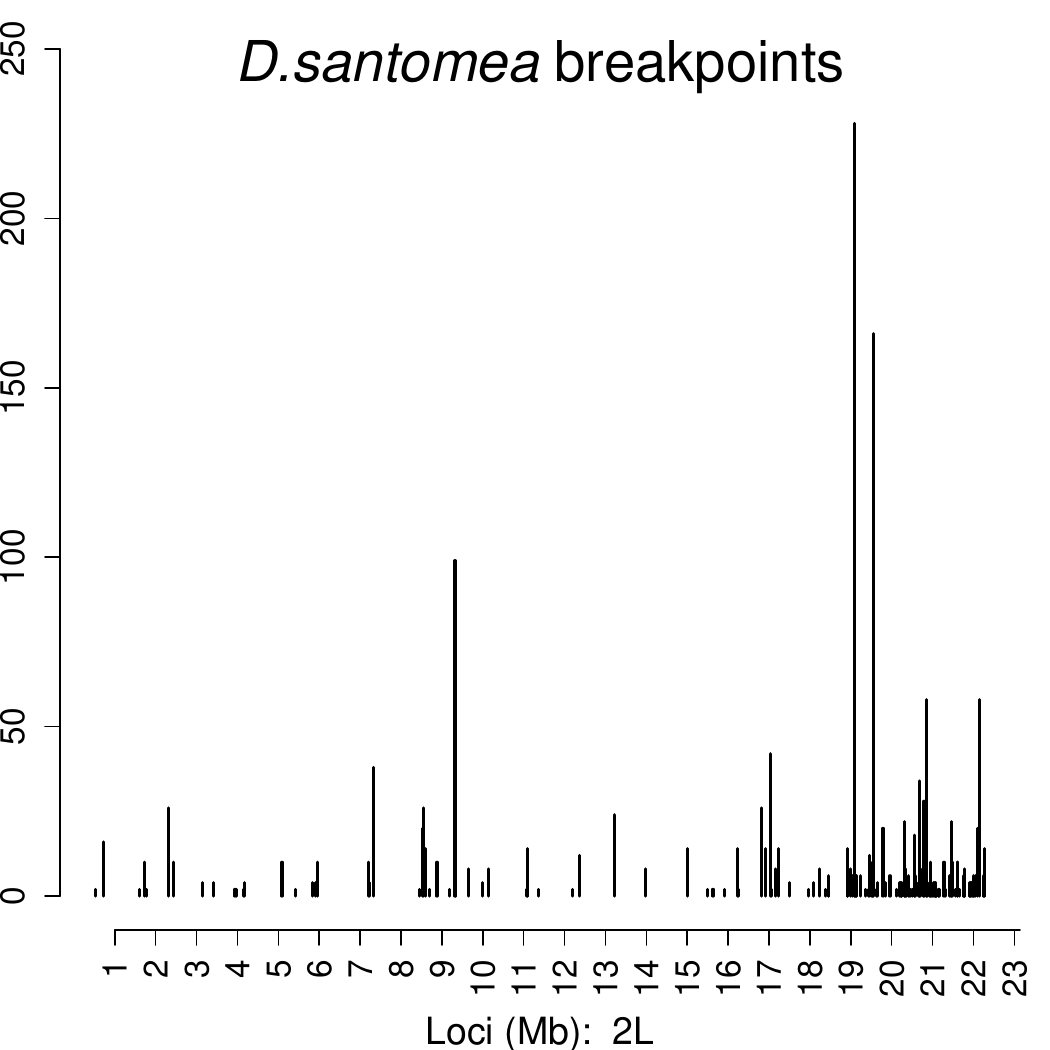}\quad
    \includegraphics[page=2,width=.4\textwidth]{Breakpoints/bp_sant.pdf}\quad
    \includegraphics[page=3,width=.4\textwidth]{Breakpoints/bp_sant.pdf}
    \medskip
    \includegraphics[page=4,width=.4\textwidth]{Breakpoints/bp_sant.pdf}\quad
    \includegraphics[page=5,width=.4\textwidth]{Breakpoints/bp_sant.pdf}
    \caption{Number of rearrangement breakpoints in 10kb windows along each autosome and the X chromosome in \Dsant.  Centromeres on chromosome 2, 3, and the X display hotspots, with a maximum of 250 rearrangements per window. Recombination suppression near the cenetromeres is expected to allow detrimental rearrangements to persist in populations, increasing rearrangement numbers.  }
    \label{breakpoints_dsan}
\end{figure}
\clearpage

% BP yak
\begin{figure}[htp]
    \centering
    \includegraphics[page=1,width=.4\textwidth]{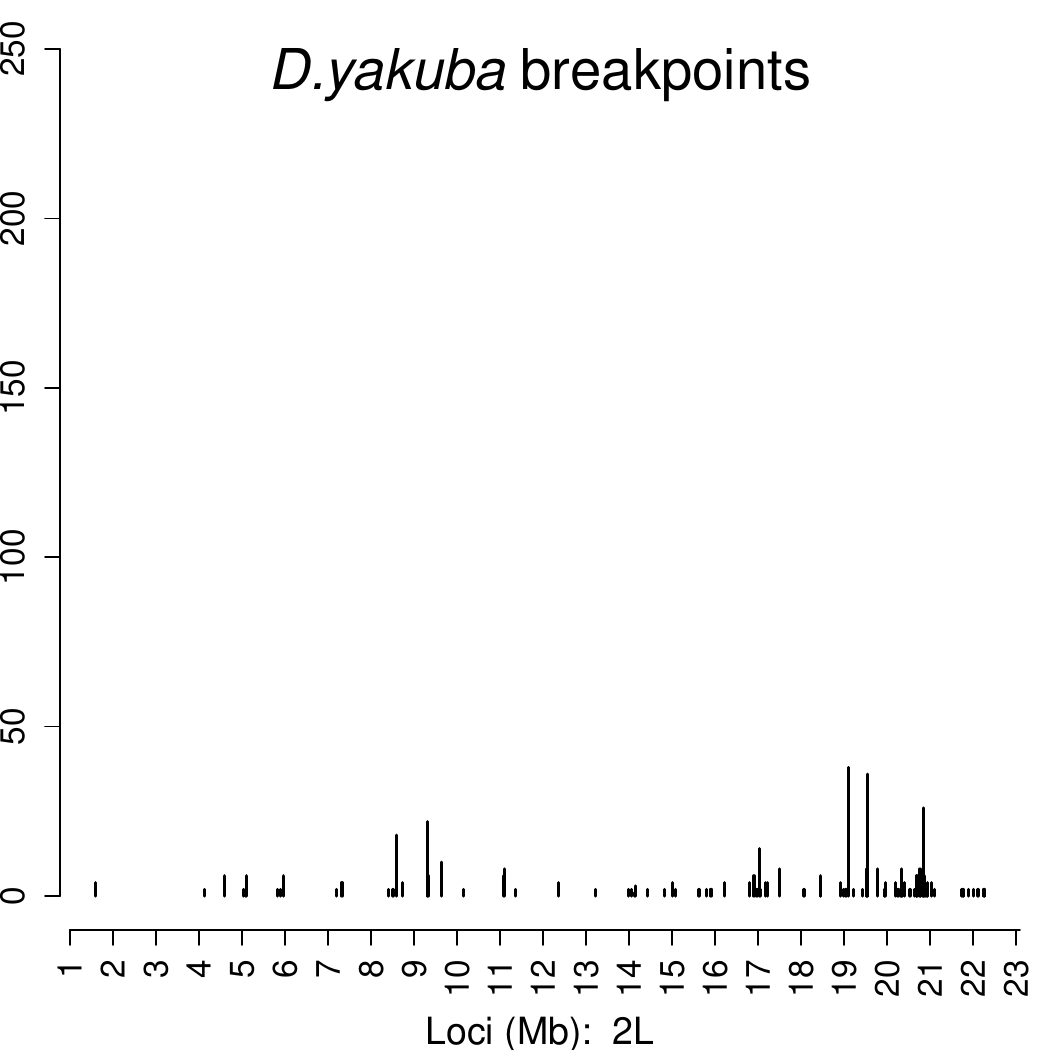}\quad
    \includegraphics[page=2,width=.4\textwidth]{Breakpoints/bp_yak.pdf}\quad
    \includegraphics[page=3,width=.4\textwidth]{Breakpoints/bp_yak.pdf}
    \medskip
    \includegraphics[page=4,width=.4\textwidth]{Breakpoints/bp_yak.pdf}\quad
    \includegraphics[page=5,width=.4\textwidth]{Breakpoints/bp_yak.pdf}
    \caption{Number of rearrangement breakpoints in 10kb windows along each autosome and the X chromosome in \Dyak. Centromeres on chromosome 2, 3, and the X display hotspots, with a maximum of 80 rearrangements per window, less dynamic than compared with more distantly related \Dsant.   }
    \label{breakpoints_dyak}
\end{figure}
\clearpage

% plain sfs
\begin{figure}
    \centering
    \subfloat[A]{\label{dsanSFS}\includegraphics[width=.33\linewidth]{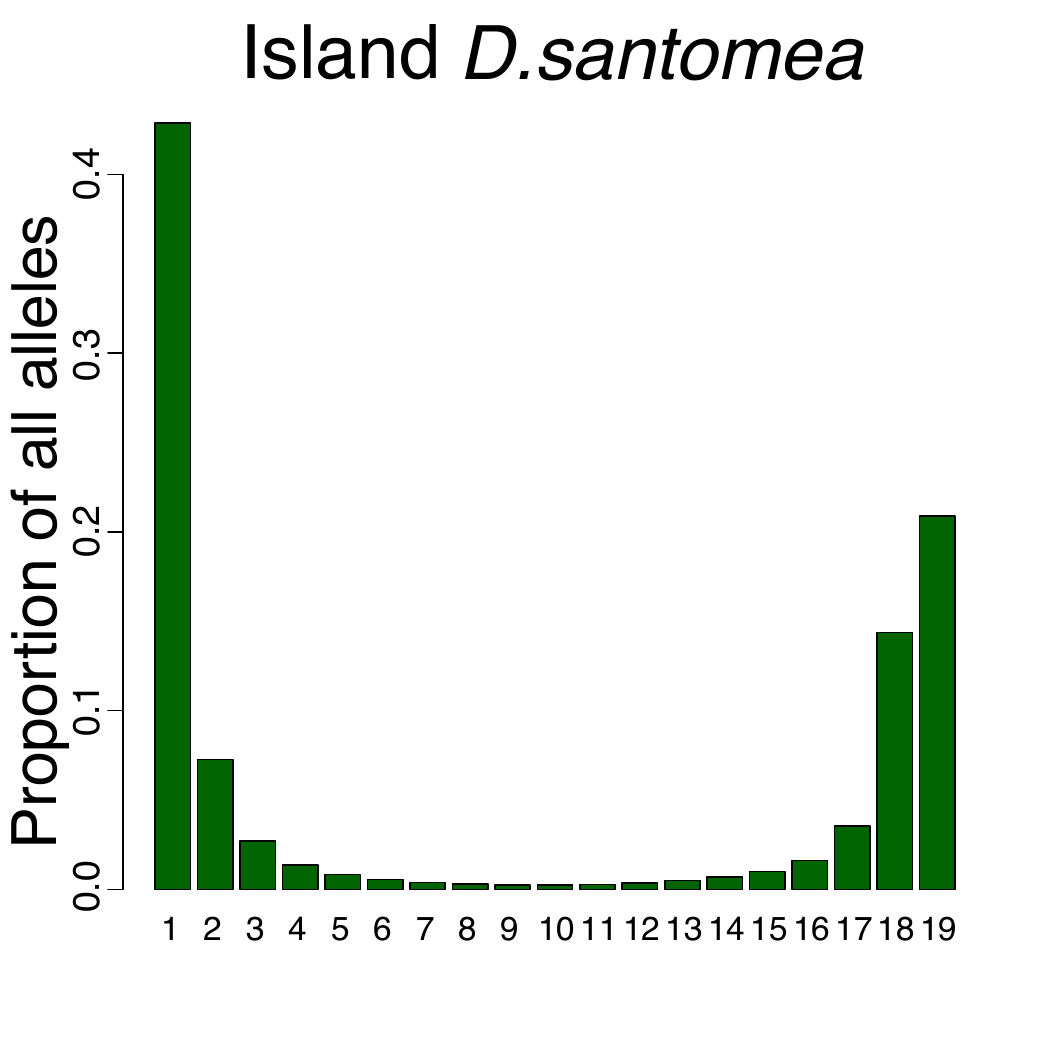}}%\hfill
    \subfloat[B]{\label{oranSFS}\includegraphics[width=.33\linewidth]{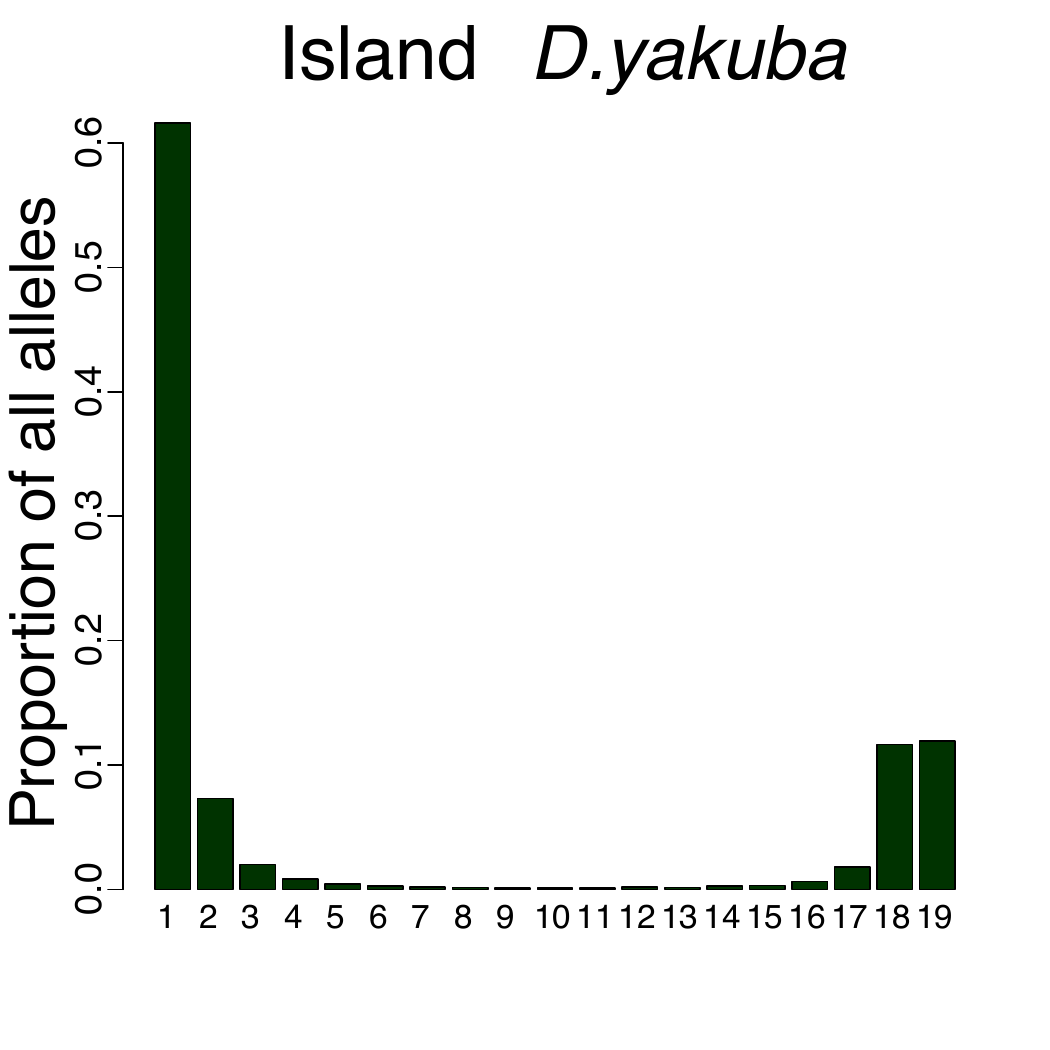}}%\par 
    \subfloat[C]{\label{mainSFS}\includegraphics[width=.33\linewidth]{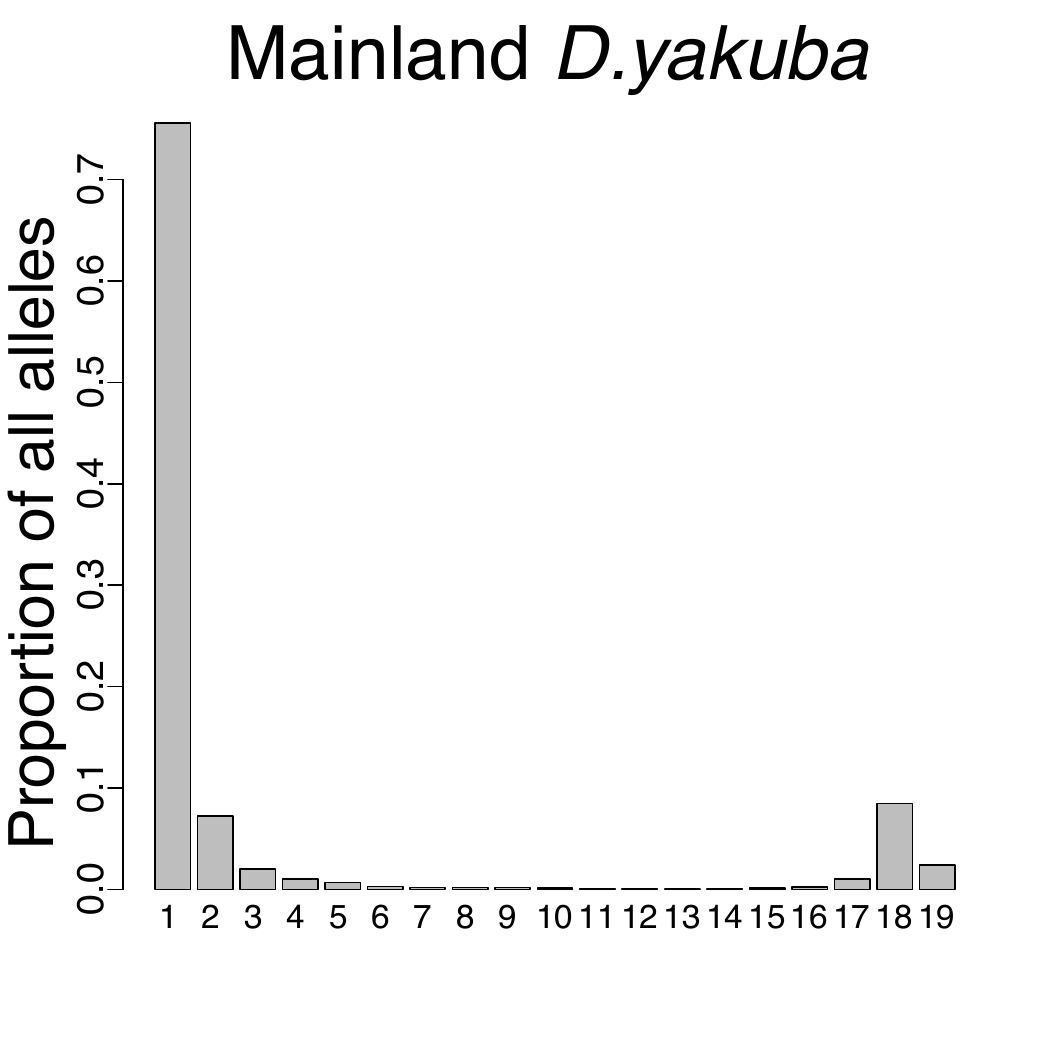}}
    \caption{A) Site frequency spectrum of \Dsant{}.  The increase in high frequency variants is much greater than that of the mainland \Dyak{} B) Site frequency spectrum of the island \Dyak{}. The increase in high frequency variants is slightly greater than that of the mainland \Dyak{}, but not as much as \Dsant{}, likely related to divergence time C) Site frequency spectrum of mainland \Dyak{} lacks the increase in high frequency variants that is present in both of the other sub-populations. }%All SFS are olarized ancestral state with \Dtei{} and projected down to n=19 using the geometric distribution.
    \label{SFS}
\end{figure}

% Extra sfs (te only)
\begin{figure}[htp]
    \centering
    \subfloat[A]{\includegraphics[page=4,width=.4\textwidth]{SFS/r_sfs_teOnly.pdf}}\quad
    \subfloat[B]{\includegraphics[page=1,width=.4\textwidth]{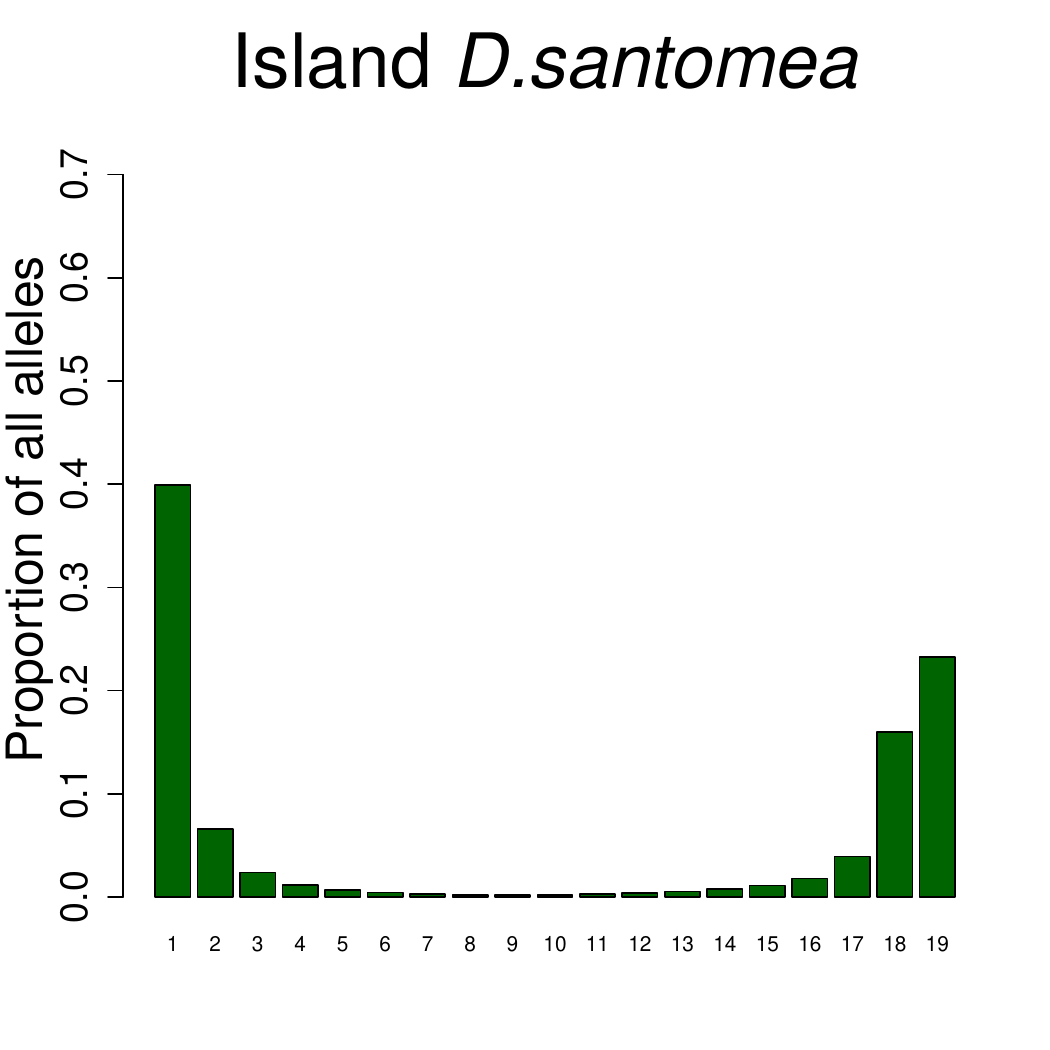}}\quad
    \subfloat[C]{\includegraphics[page=2,width=.4\textwidth]{SFS/r_sfs_teOnly.pdf}}
    \medskip
    \subfloat[D]{\includegraphics[page=3,width=.4\textwidth]{SFS/r_sfs_teOnly.pdf}}\quad
    \caption{Site frequency spectrum once only rearrangements associated with TEs are selected A) Site frequency spectrum of the meta-population of the study. Polarized ancestral state with \Dtei{} and projected down to n=19 using the geometric distribution. The increase in high frequency variants is represented in only some of the sub-populations once they are separated. B) Site frequency spectrum of \Dsant{}. Polarized ancestral state with \Dtei{} and projected down to n=19 using the geometric distribution. The increase in high frequency variants is much greater than that of the mainland \Dyak{} C) Site frequency spectrum of the island \Dyak{}. Polarized ancestral state with \Dtei{} and projected down to n=19 using the geometric distribution. The increase in high frequency variants is slightly greater than that of the mainland \Dyak{}, but not as much as \Dsant{}, likely related to divergence time D) Site frequency spectrum of mainland \Dyak{}. Polarized ancestral state with \Dtei{} and projected down to n=19 using the hypergeometric distribution. Lacks the increase in high frequency variants that is present in both of the other sub-populations}
    \label{teOnlyPlots}
\end{figure}

% Extra sfs (te removed)
% S5
\begin{figure}[htp]
    \centering
    \subfloat[A]{\includegraphics[page=4,width=.4\textwidth]{SFS/r_sfs_teRem.pdf}}\quad
    \subfloat[B]{\includegraphics[page=1,width=.4\textwidth]{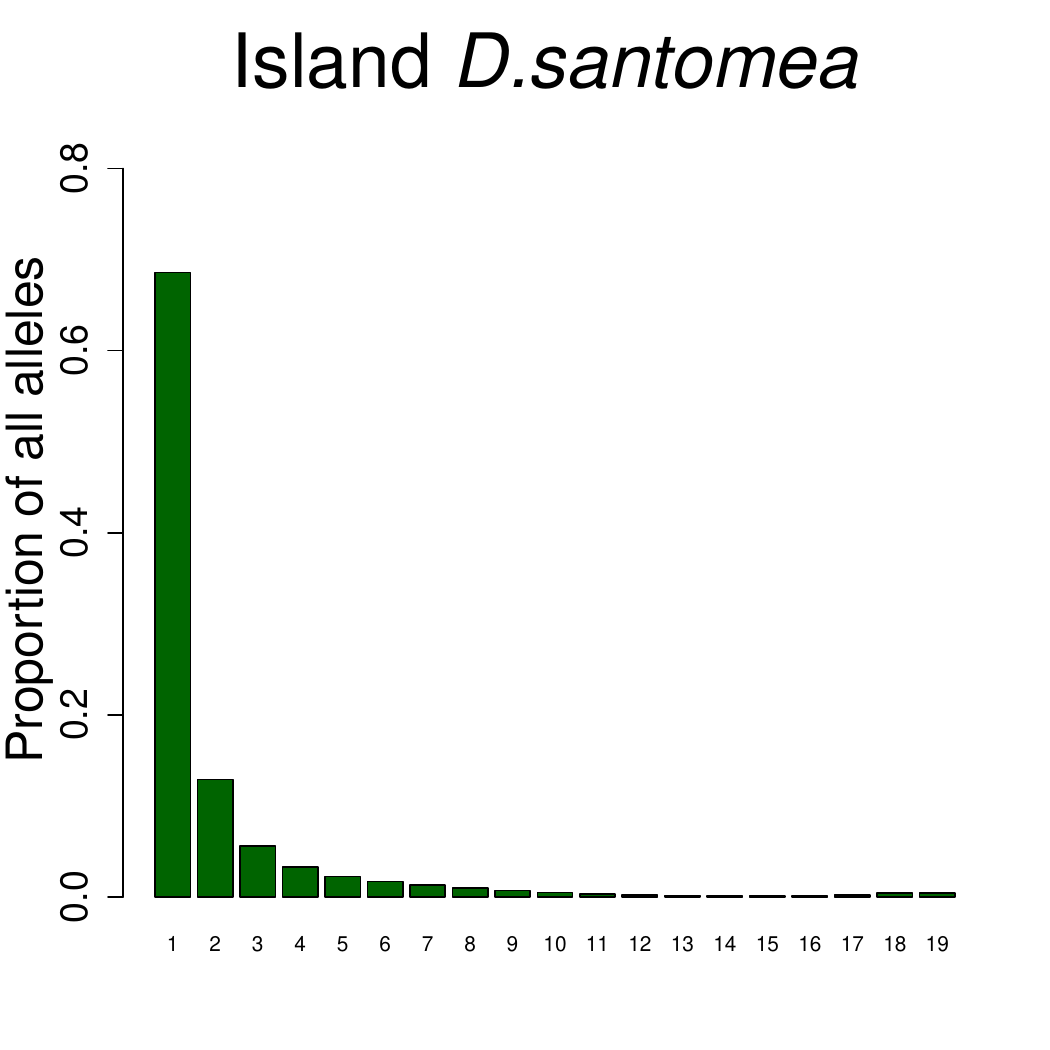}}\quad
    \subfloat[C]{\includegraphics[page=2,width=.4\textwidth]{SFS/r_sfs_teRem.pdf}}
    \medskip
    \subfloat[D]{\includegraphics[page=3,width=.4\textwidth]{SFS/r_sfs_teRem.pdf}\quad}
    \caption{Site frequency spectrum once only rearrangements associated with TEs are removed A) Site frequency spectrum of the meta-population of the study. Polarized ancestral state with \Dtei{} and projected down to n=19 using the geometric distribution. The increase in high frequency variants is represented is missing from all populations. B) Site frequency spectrum of \Dsant{}. Polarized ancestral state with \Dtei{} and projected down to n=19 using the geometric distribution. The increase in high frequency variants is lost once TEs are removed. C) Site frequency spectrum of the island \Dyak{}. Polarized ancestral state with \Dtei{} and projected down to n=19 using the geometric distribution. The increase in high frequency variants is lost once TEs are removed. D) Site frequency spectrum of mainland \Dyak{}. Polarized ancestral state with \Dtei{} and projected down to n=19 using the hypergeometric distribution. The increase in high frequency variants is lost once TEs are removed.}
    \label{teRemPlots}
\end{figure}

% Extra deltap: ect dsan 
\begin{figure}[htp]
    \centering
    \includegraphics[page=2,width=.3\textwidth]{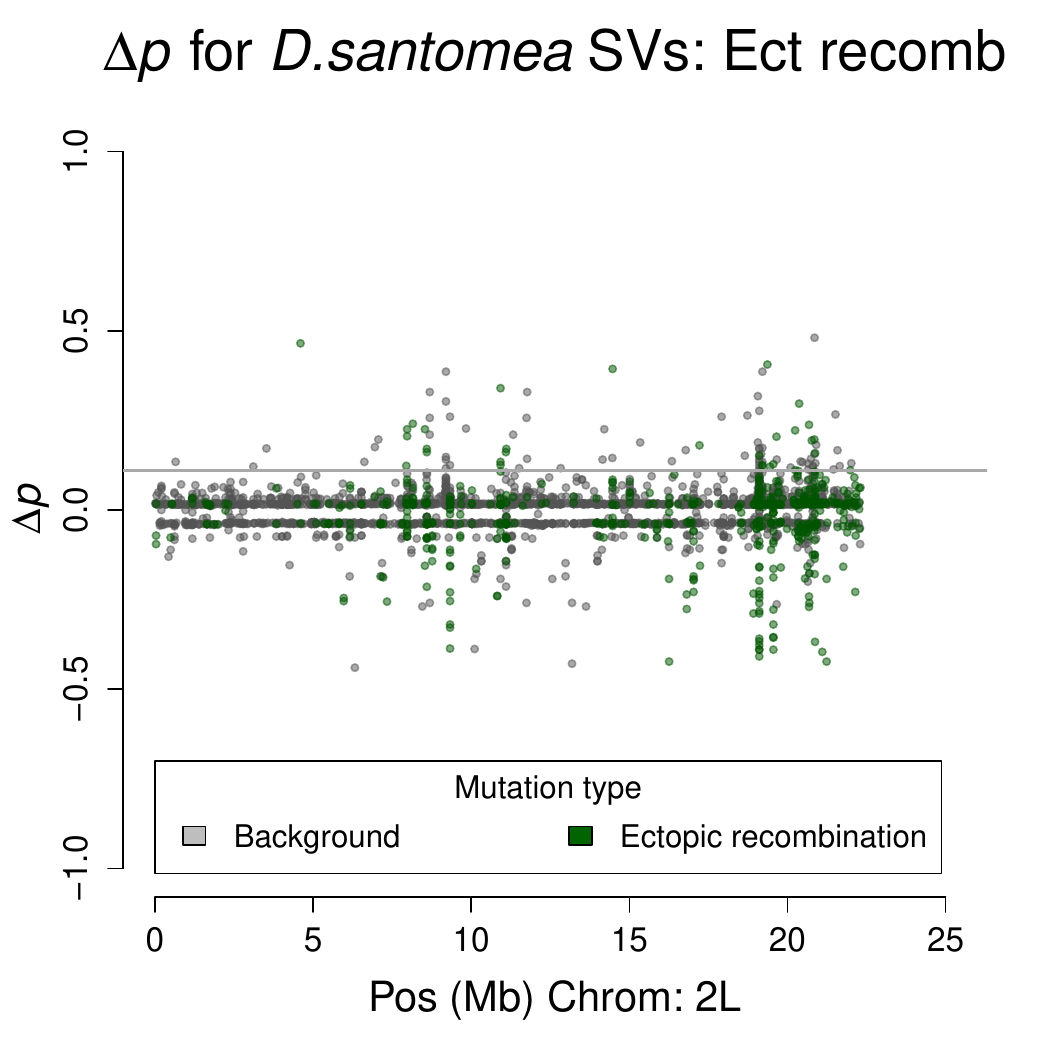}\quad
    \includegraphics[page=6,width=.3\textwidth]{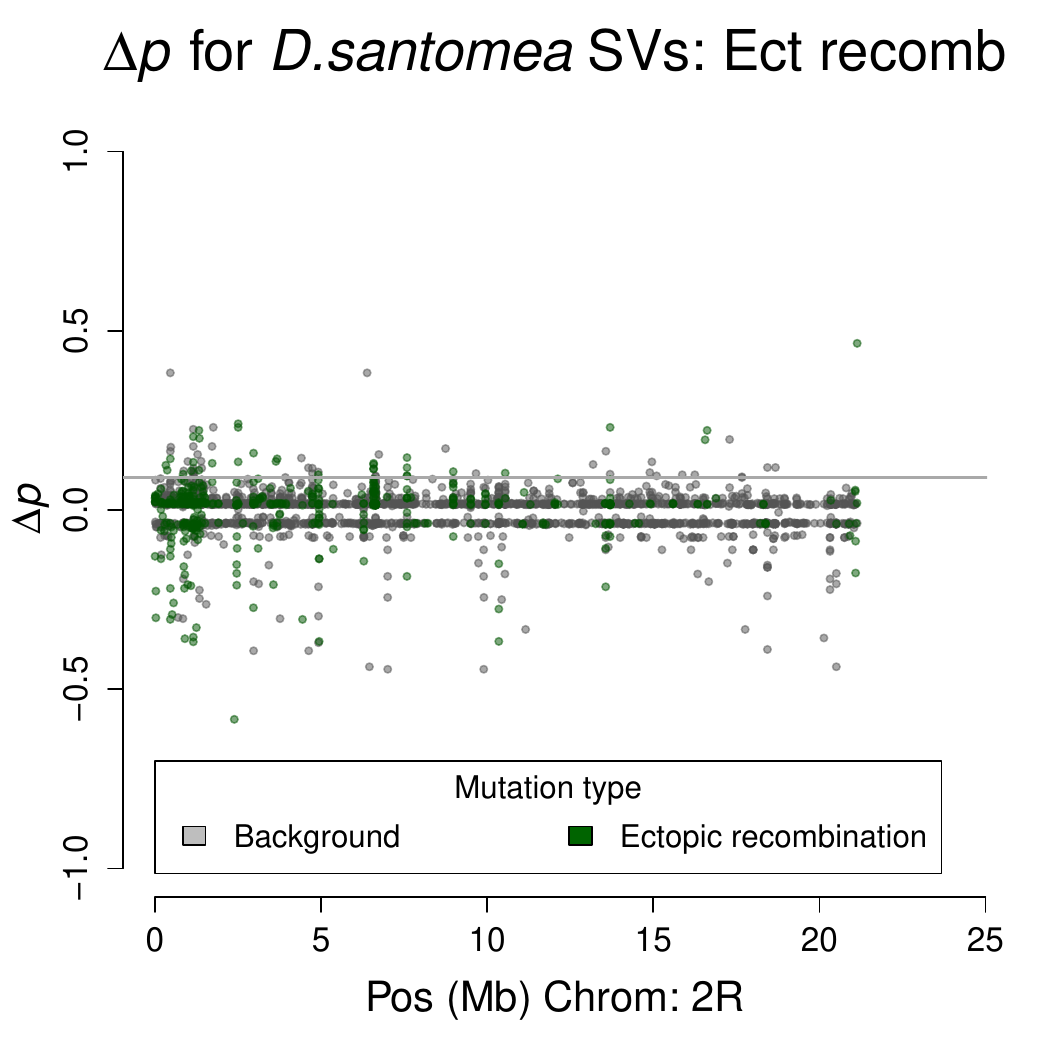}\quad
    \includegraphics[page=10,width=.3\textwidth]{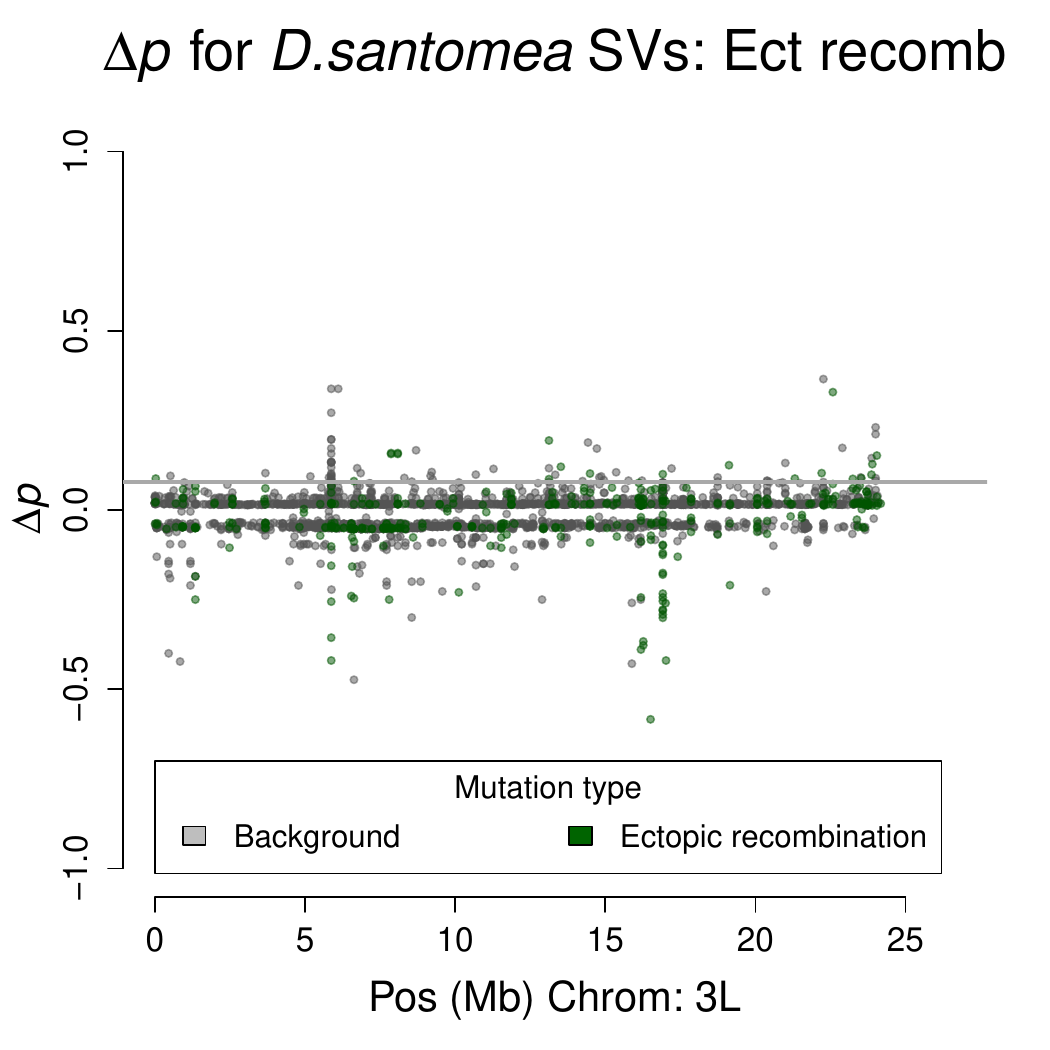}
    \medskip
    \includegraphics[page=14,width=.3\textwidth]{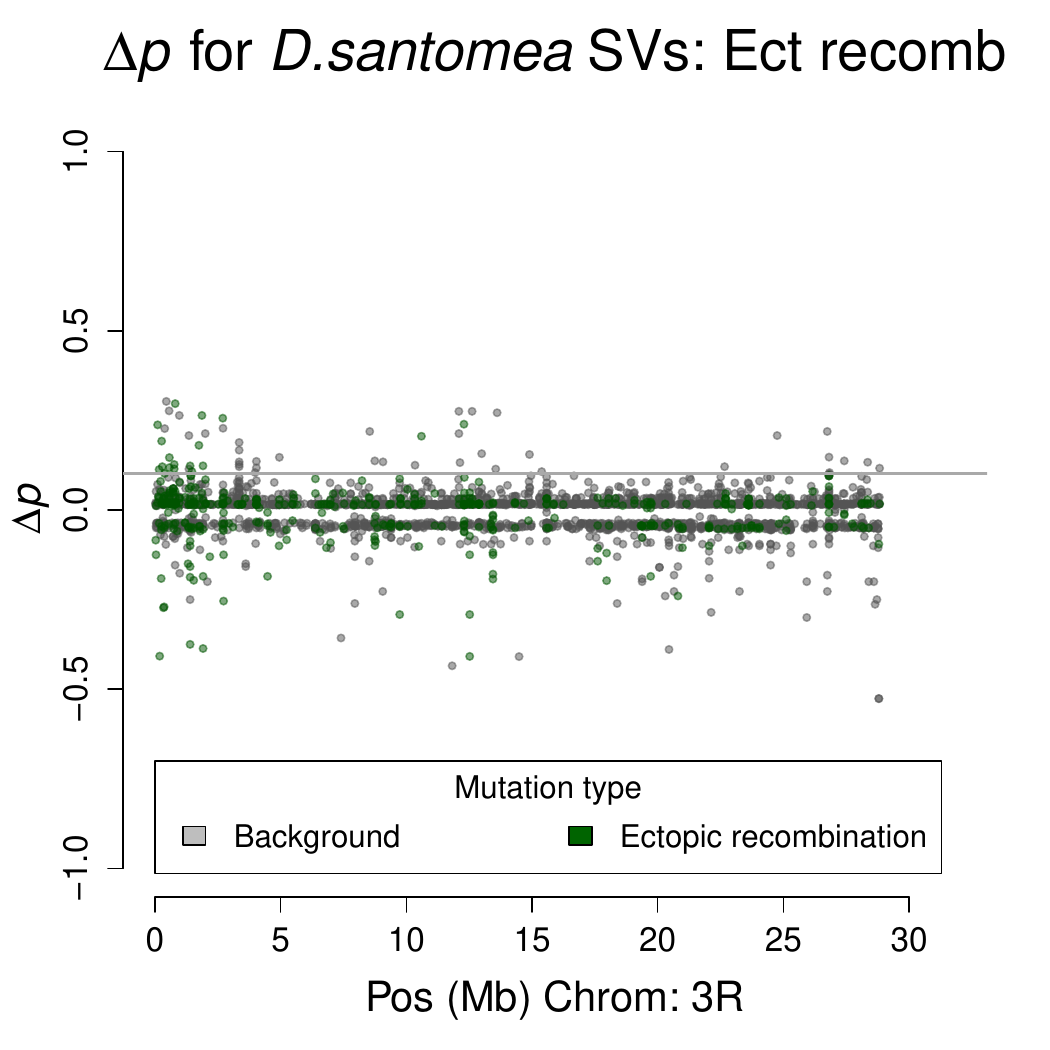}\quad
    \includegraphics[page=18,width=.3\textwidth]{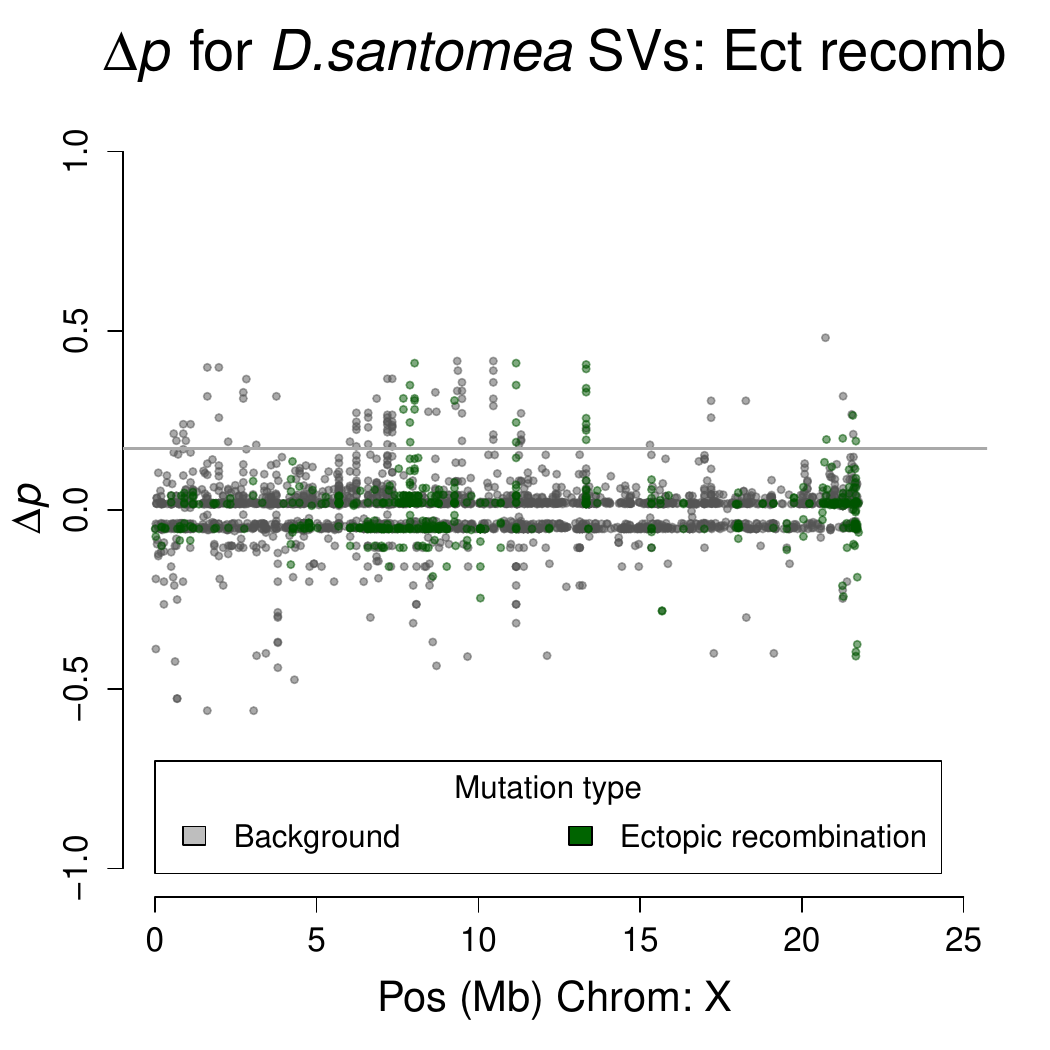}
    \caption{Population differentiation as shown via $\Delta$p between island \Dyak{} and mainland ancestral \Dyak{} on each chromosome. The x-axis shows genomic  position on the chromosome in megabases, and the y-axis is difference in allele frequencies between the island \Dyak{} and the mainland \Dyak{}. Points in dark green highlight the rearrangements associated with facilitating ectopic recombination. We observe greater levels of differentiation on the X compared to the autosomes, as well as a concentration of green points near the centromeres.}
    \label{deltap_ect_dsan_extra}
\end{figure}

% Extra deltap: ect dyak
\begin{figure}[htp]
    \centering
    \includegraphics[page=4,width=.33\textwidth]{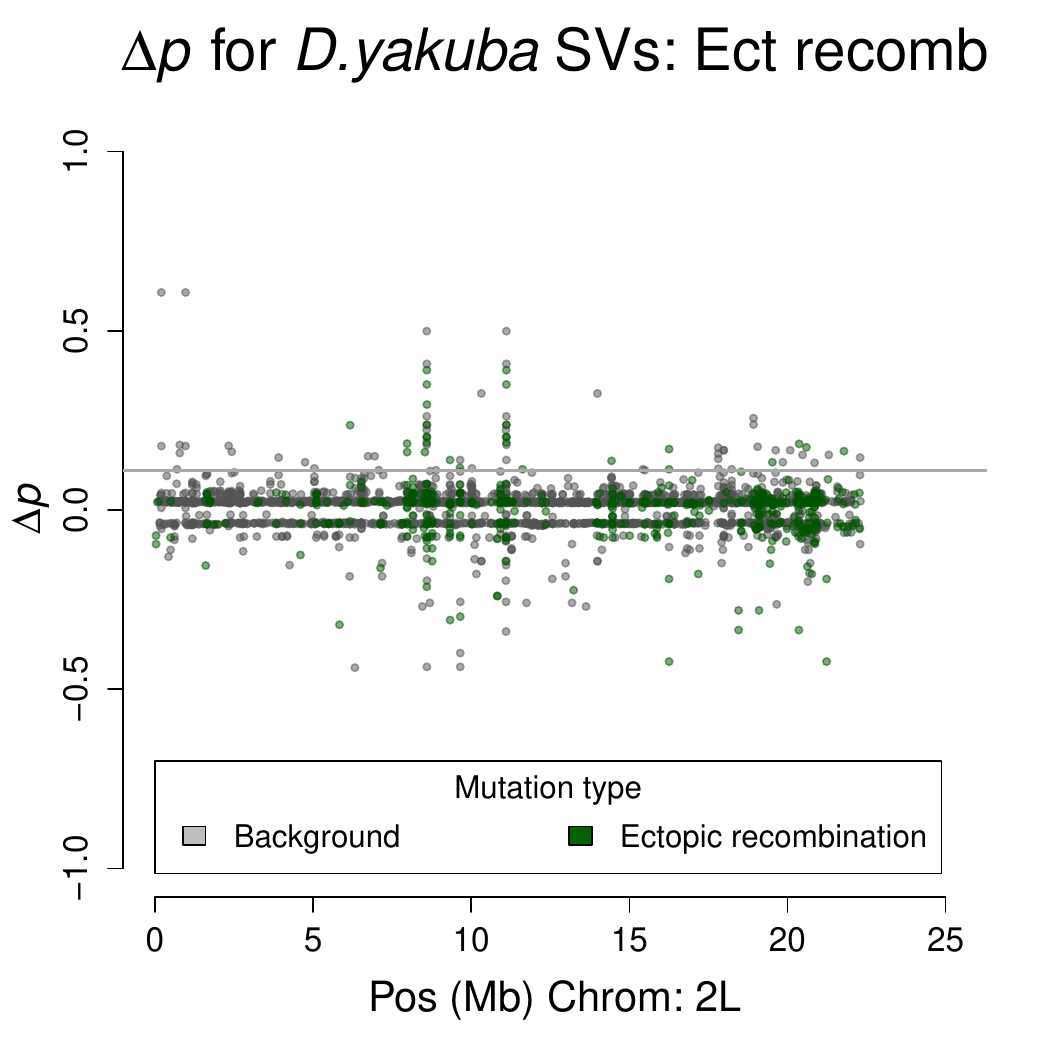}\quad
    \includegraphics[page=8,width=.33\textwidth]{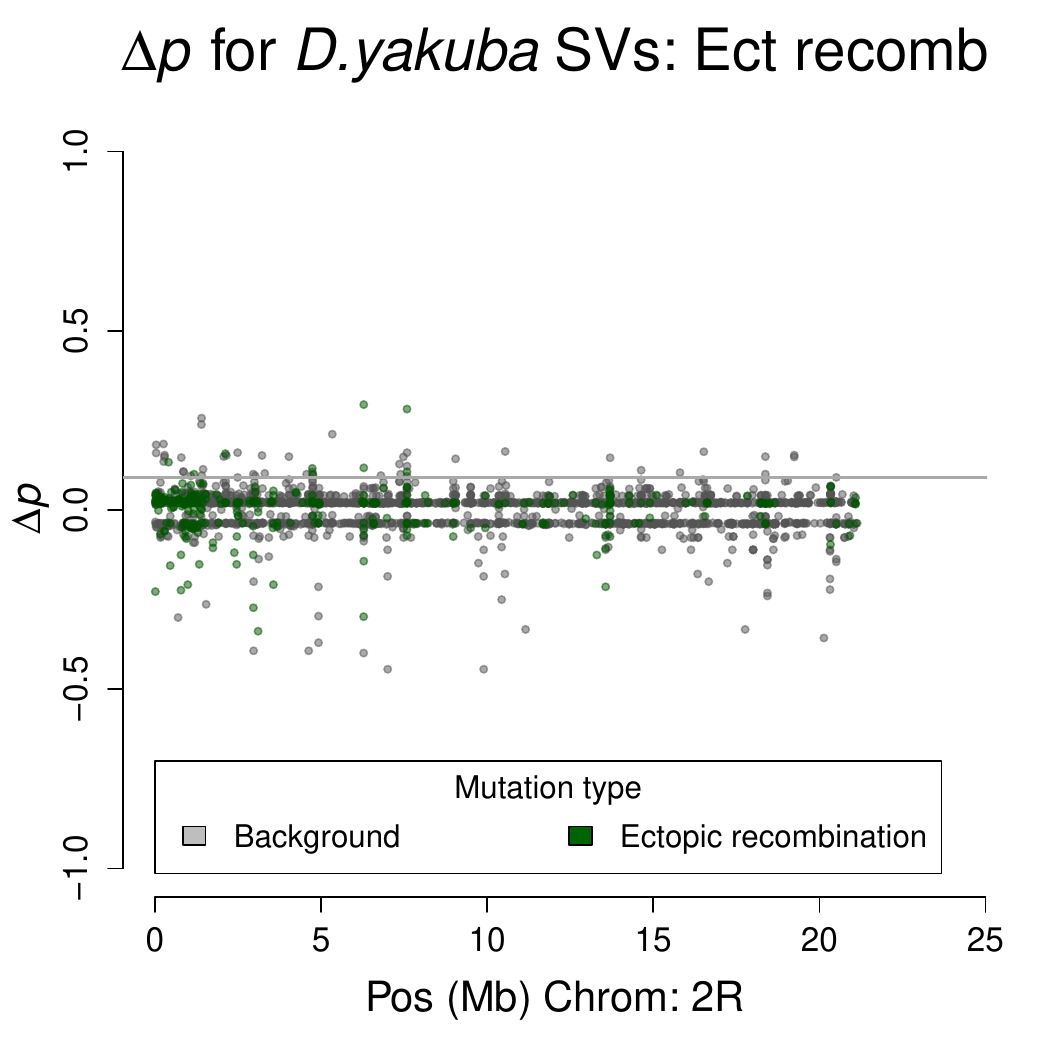}\quad
    \includegraphics[page=12,width=.33\textwidth]{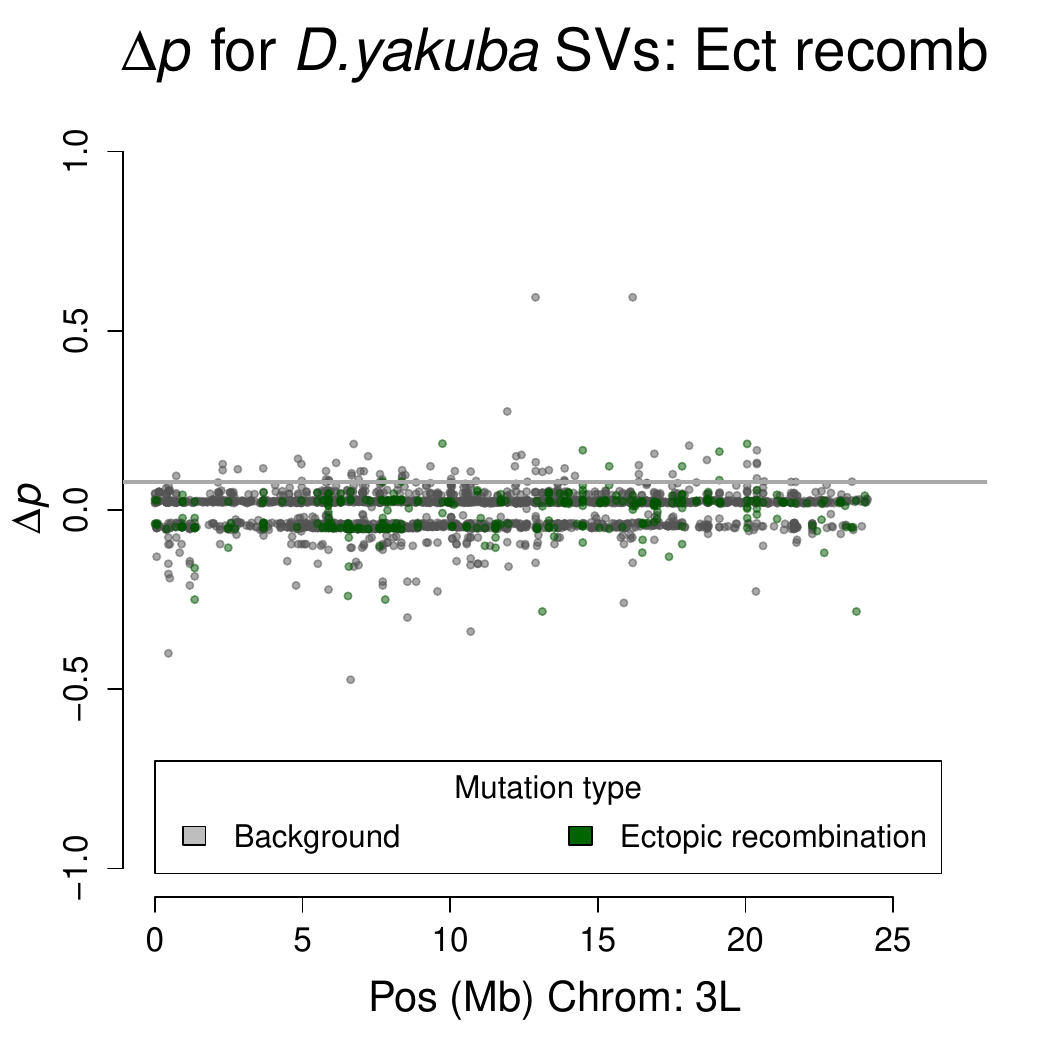}
    \medskip
    \includegraphics[page=16,width=.33\textwidth]{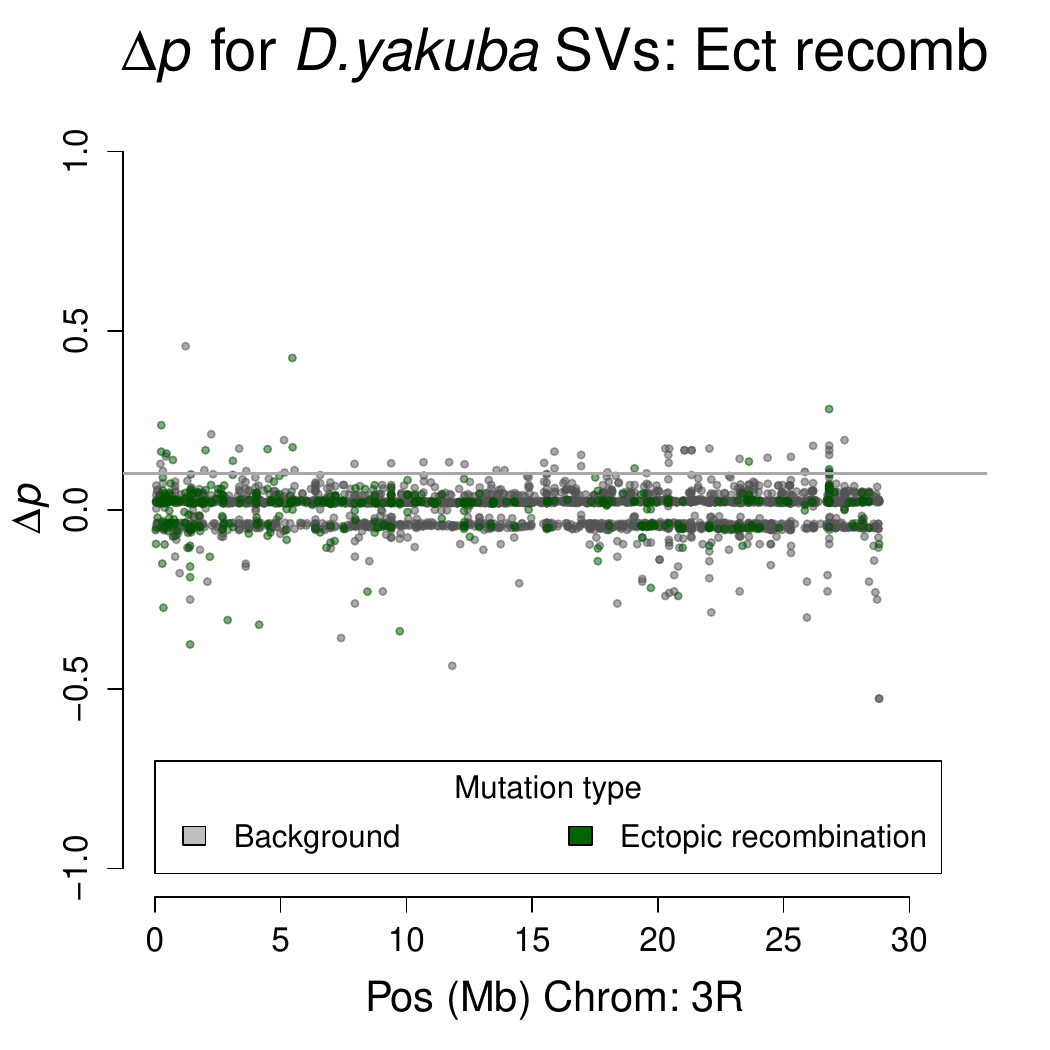}\quad
    \includegraphics[page=20,width=.33\textwidth]{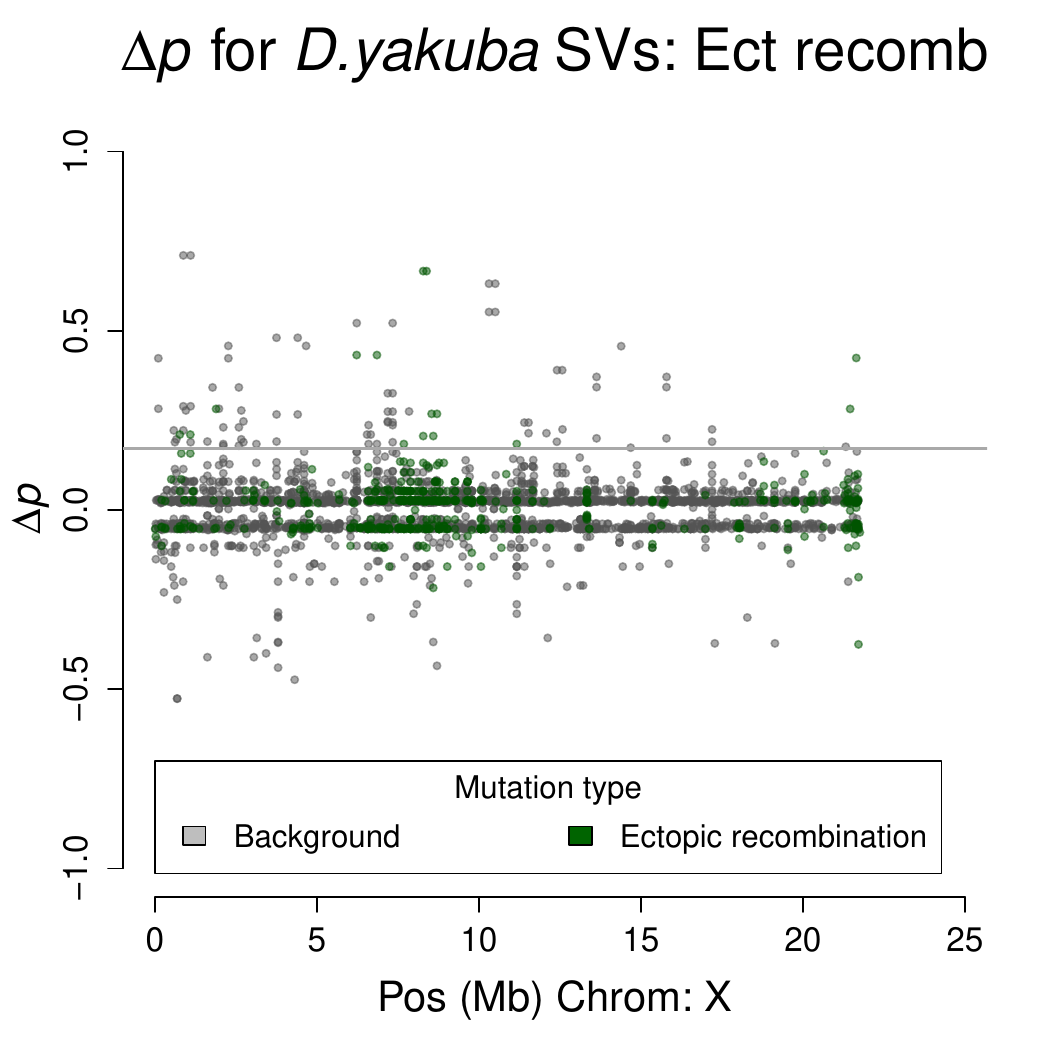}
    \caption{Population differentiation as shown via $\Delta$p between island \Dsant{} and mainland ancestral \Dyak{} on each chromosome. The x-axis shows genomic  position on the chromosome in megabases, and the y-axis is difference in allele frequencies between the island \Dsant{} and the mainland \Dyak{}. Points in dark green highlight the rearrangements associated with facilitating ectopic recombination. We observe greater levels of differentiation on the X compared to the autosomes, as well as a concentration of green points near the centromeres.}
    \label{deltap_ect_dyak_extra}
\end{figure}

% # of rearrangements TEs
% S11
\begin{figure}[h]
\centering
\subfloat[A]{\includegraphics[page=2,width=.49\linewidth]{Misc_Plots/TEBarPlot_sig.pdf}}
\subfloat[B]{\includegraphics[page=1,width=.49\linewidth]{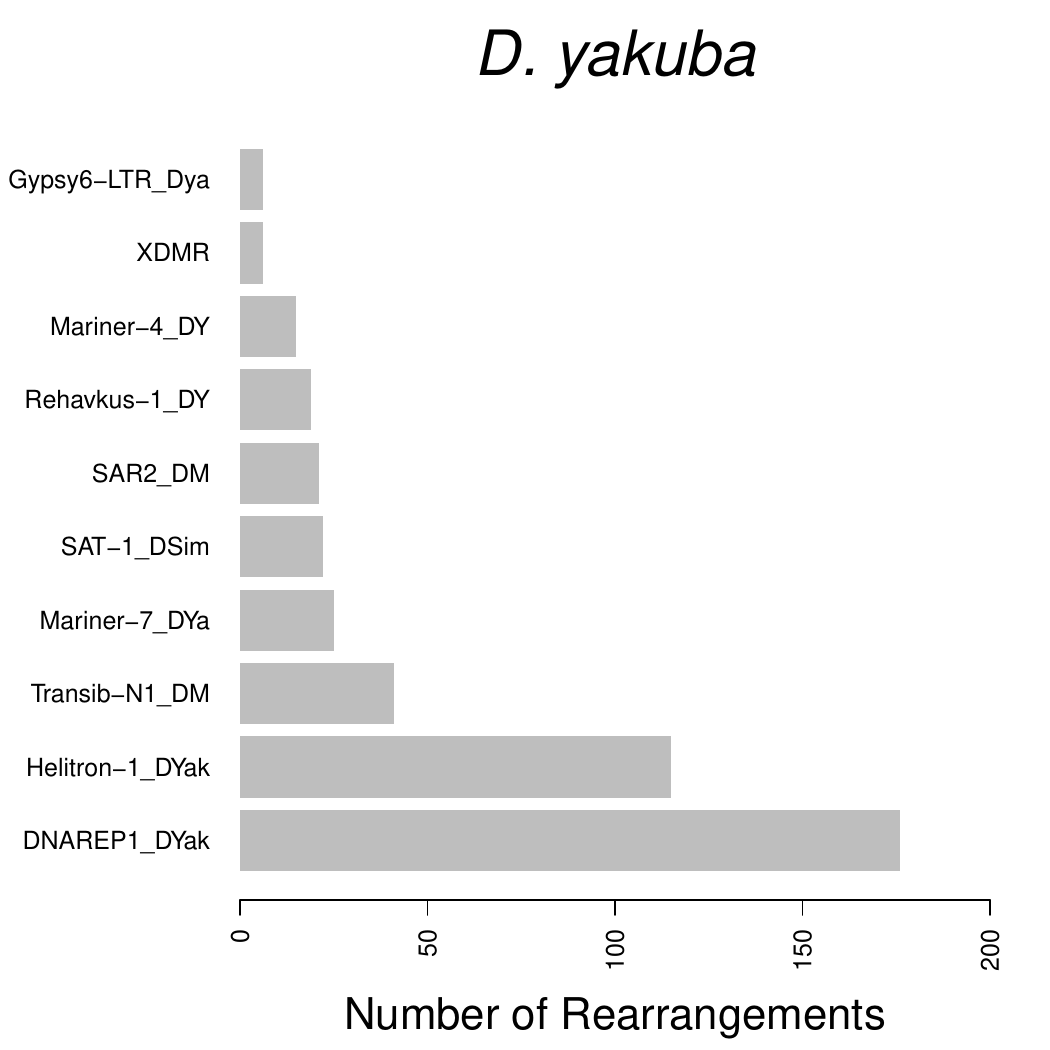}}
\caption{\label{AdaptiveTEs_sig} Number of rearrangements matching TE families associated with strong differentiation in A) \Dsant {} and B) \Dyak. The most common rearrangements come from \emph{DNAREP1} and \emph{Helitrons} in both species but minor elements constitute a greater proportion compared with background incidence rates.}
\end{figure}
\clearpage

%Theresa gene exp male 
\begin{figure}
    \centering
    \includegraphics[width=1\linewidth]{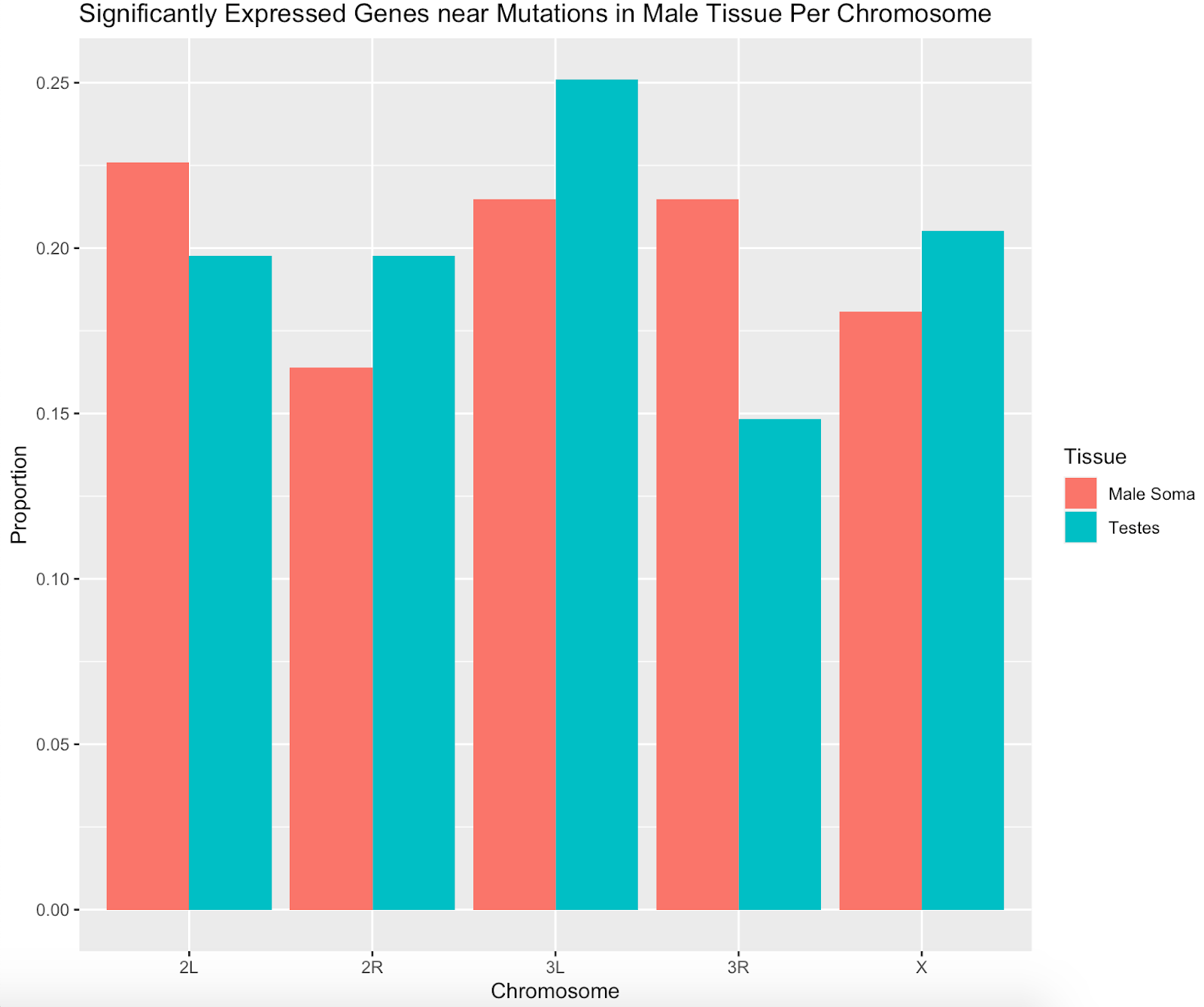}
    \caption{}
    \label{sigExp_male}
\end{figure}
\clearpage

%Theresa gene exp female
\begin{figure}
    \centering
    \includegraphics[width=1\linewidth]{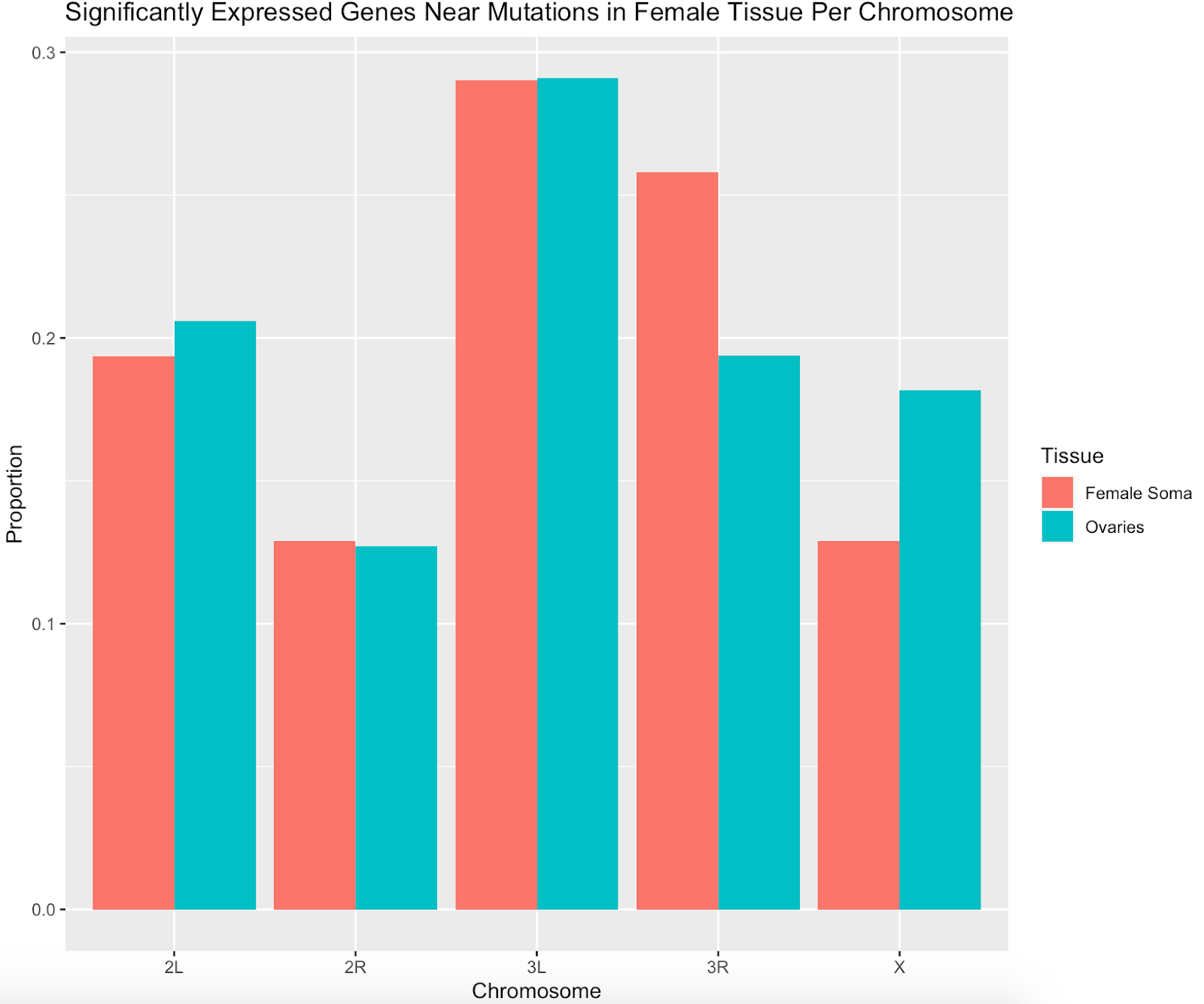}
    \caption{}
    \label{sigExp_female}
\end{figure}
\clearpage

% Extra deltap: gene express dsan
% S12
\begin{figure}[htp]
    \centering
    \includegraphics[page=1,width=.33\textwidth]{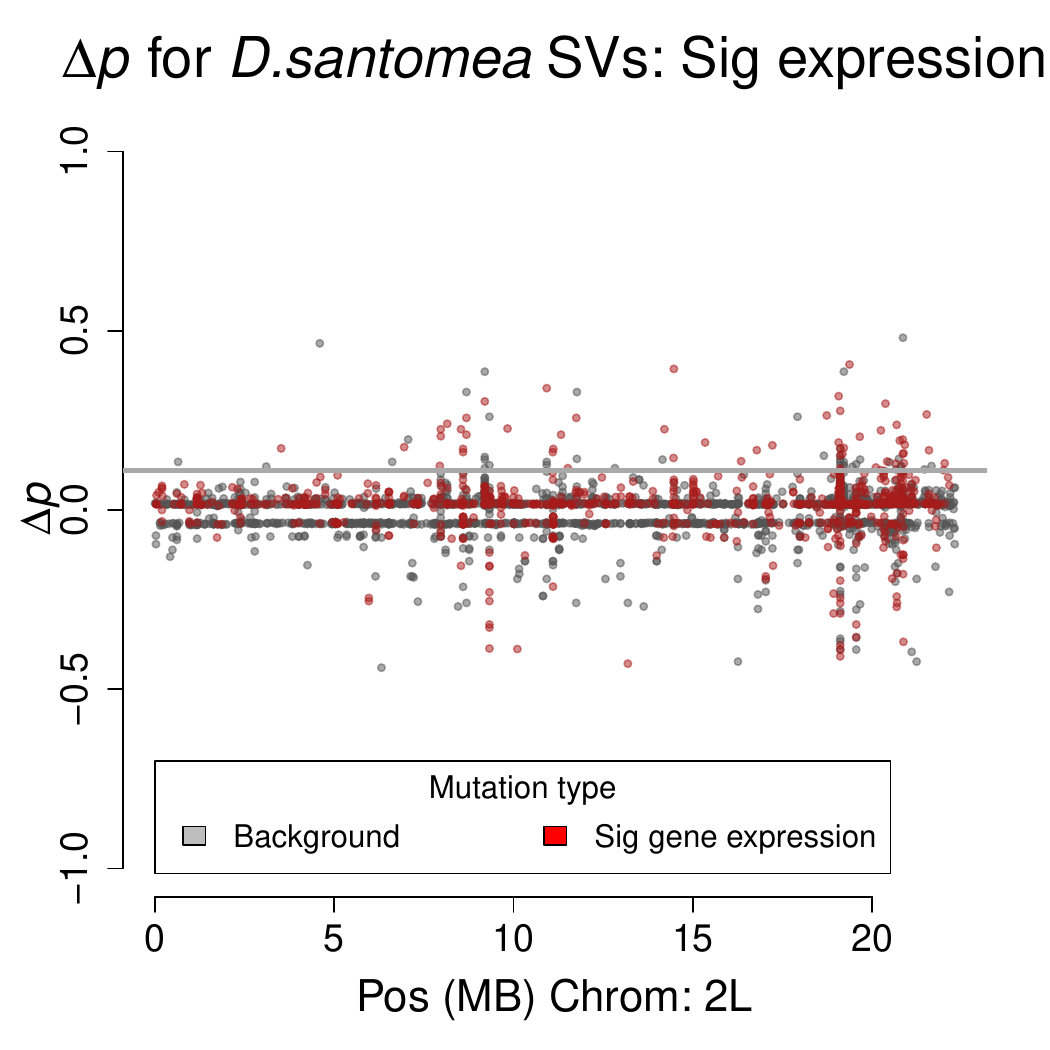}\quad
    \includegraphics[page=3,width=.33\textwidth]{Sig_exp_deltap/r_deltap_sig_thresholds.pdf}\quad
    \includegraphics[page=5,width=.33\textwidth]{Sig_exp_deltap/r_deltap_sig_thresholds.pdf}
    \medskip
    \includegraphics[page=7,width=.33\textwidth]{Sig_exp_deltap/r_deltap_sig_thresholds.pdf}\quad
    \includegraphics[page=9,width=.33\textwidth]{Sig_exp_deltap/r_deltap_sig_thresholds.pdf}
    \caption{Population differentiation as shown via $\Delta$p between island \Dsant{} and mainland ancestral \Dyak{} on each chromosome. The x-axis shows genomic  position on the chromosome in megabases, and the y-axis is difference in allele frequencies between the island \Dsant{} and the mainland \Dyak{}. Points in red highlight the rearrangements associated with significant gene expression.  We observe greater levels of differentiation on the X compared to the autosomes in both populations and rearrangements associated with significant expression changes on the X show strong differentiation.  These results suggest that gene expression changes induced by rearrangements are important for local adaptation.}
    \label{deltap_ge_dsan_extra}
\end{figure}

% Extra deltap: gene express dyak
% S13
\begin{figure}[htp]
    \centering
    \includegraphics[page=2,width=.33\textwidth]{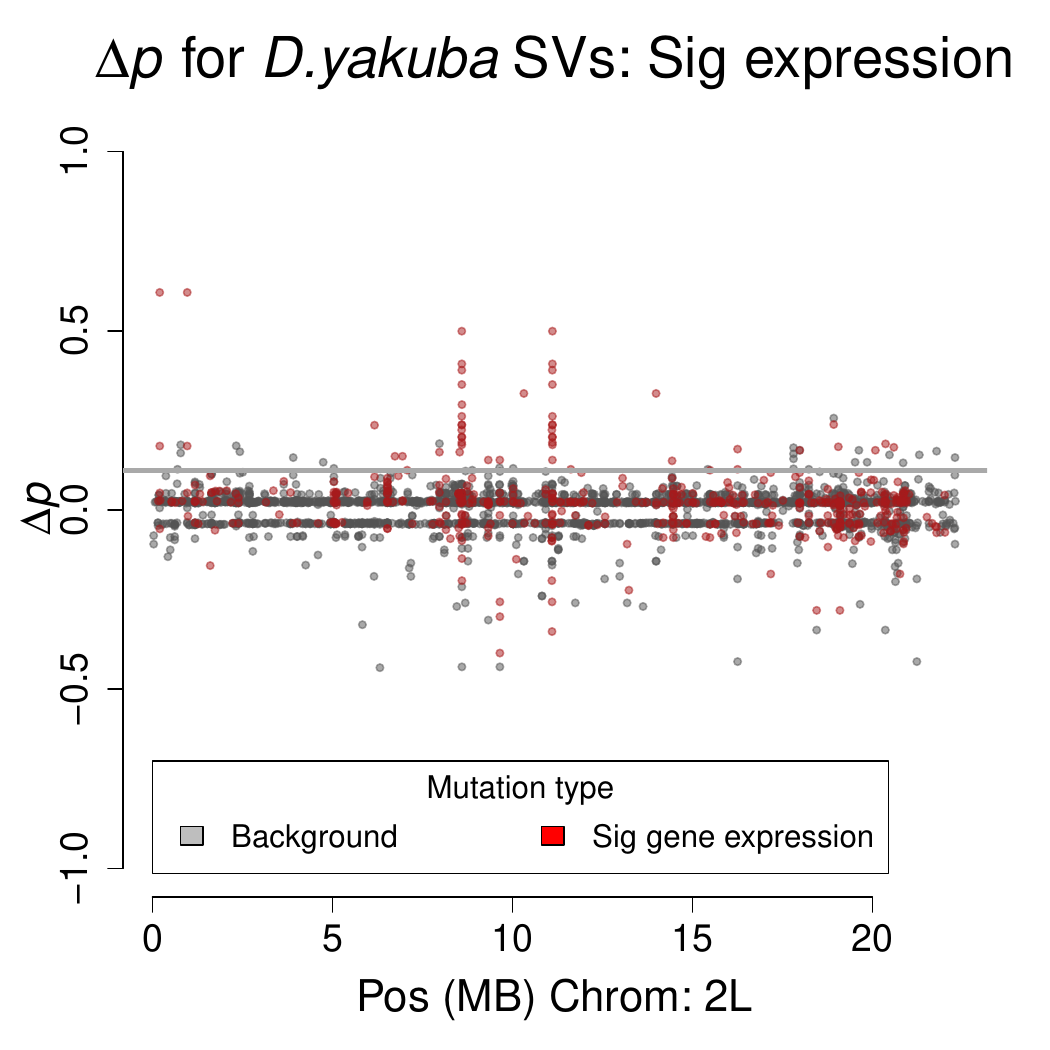}\quad
    \includegraphics[page=4,width=.33\textwidth]{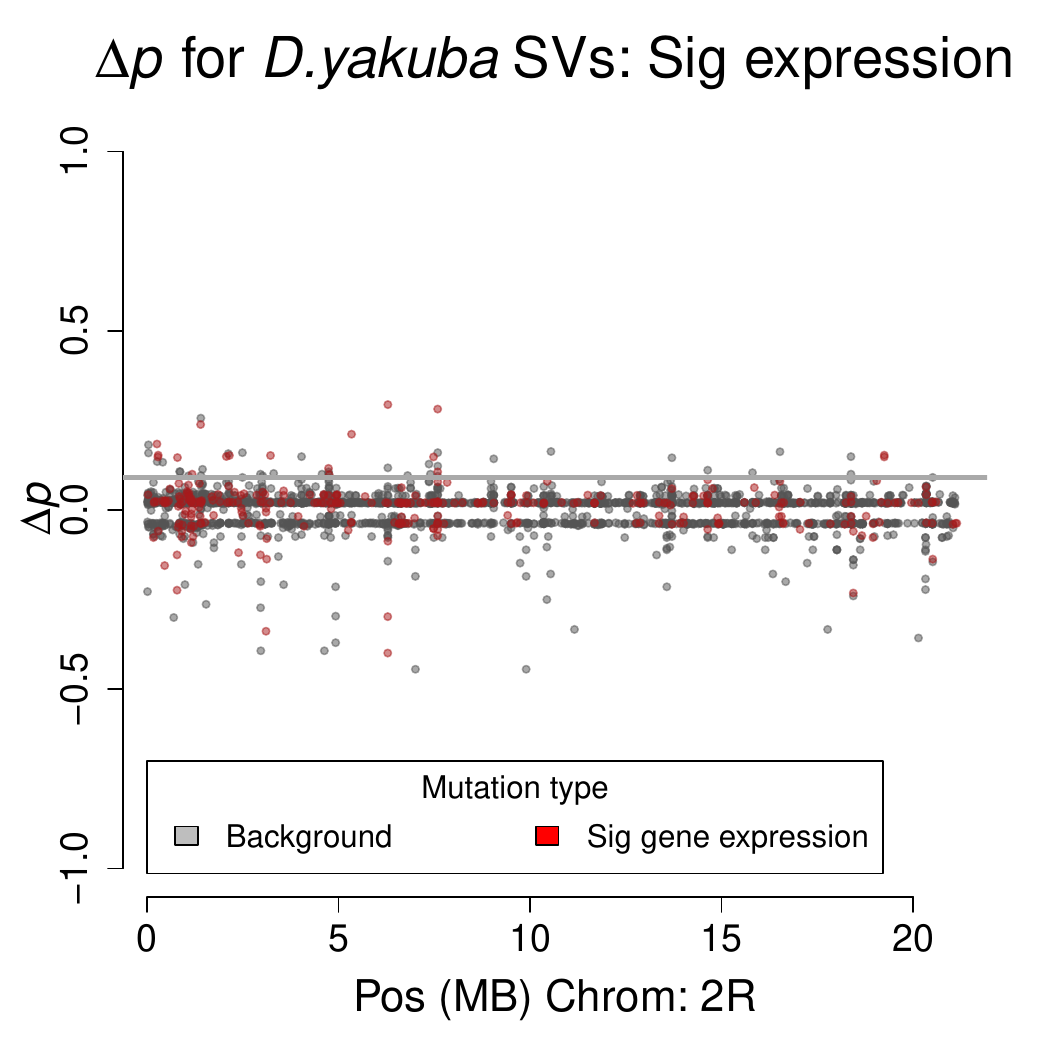}\quad
    \includegraphics[page=6,width=.33\textwidth]{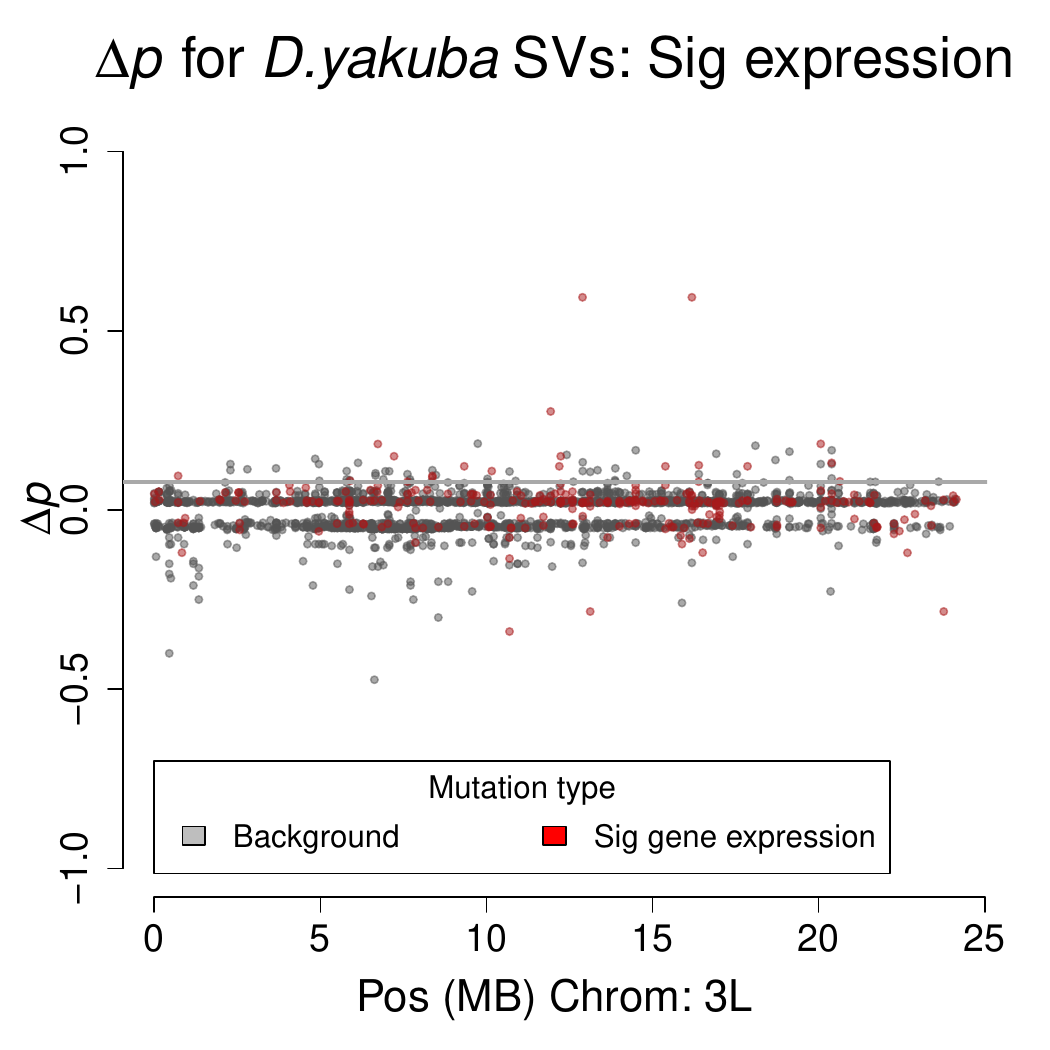}
    \medskip
    \includegraphics[page=8,width=.33\textwidth]{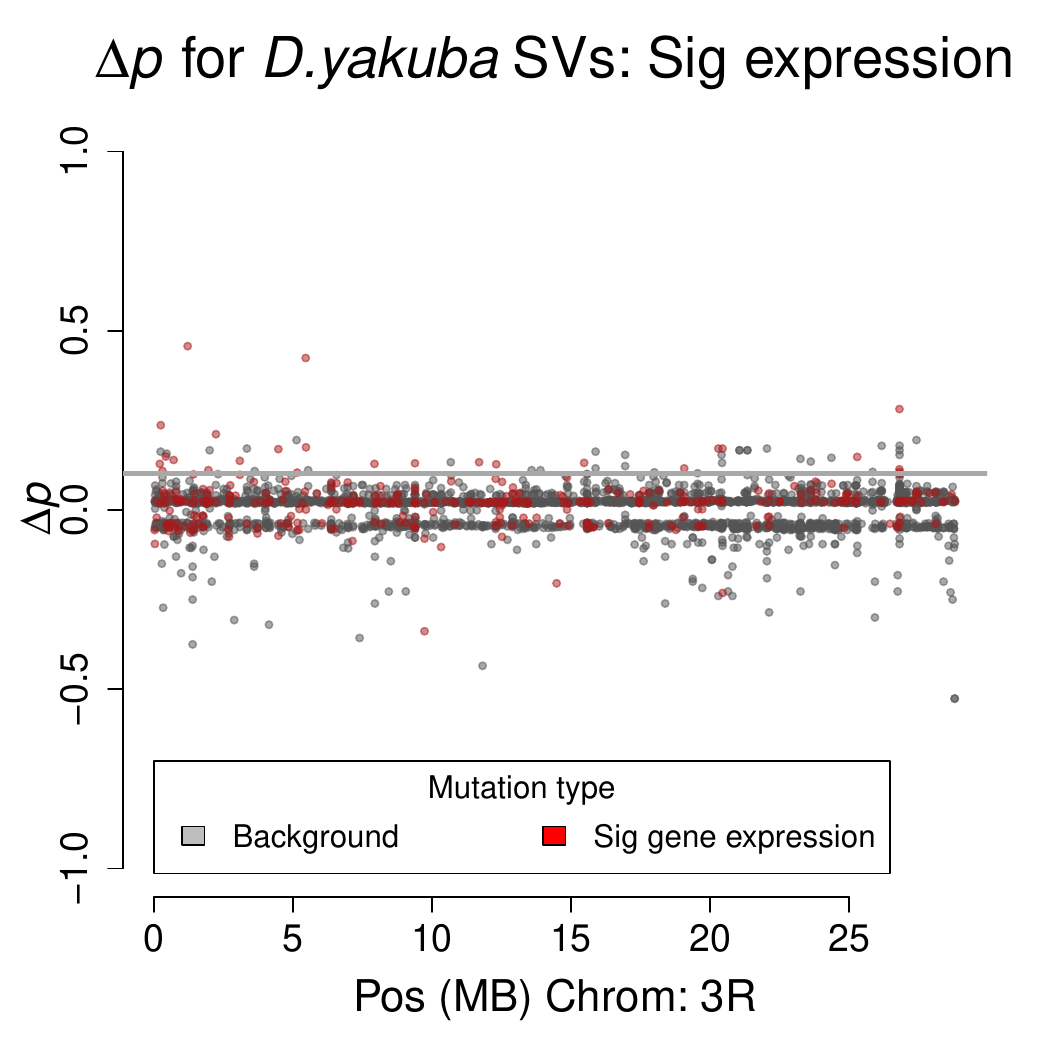}\quad
    \includegraphics[page=10,width=.33\textwidth]{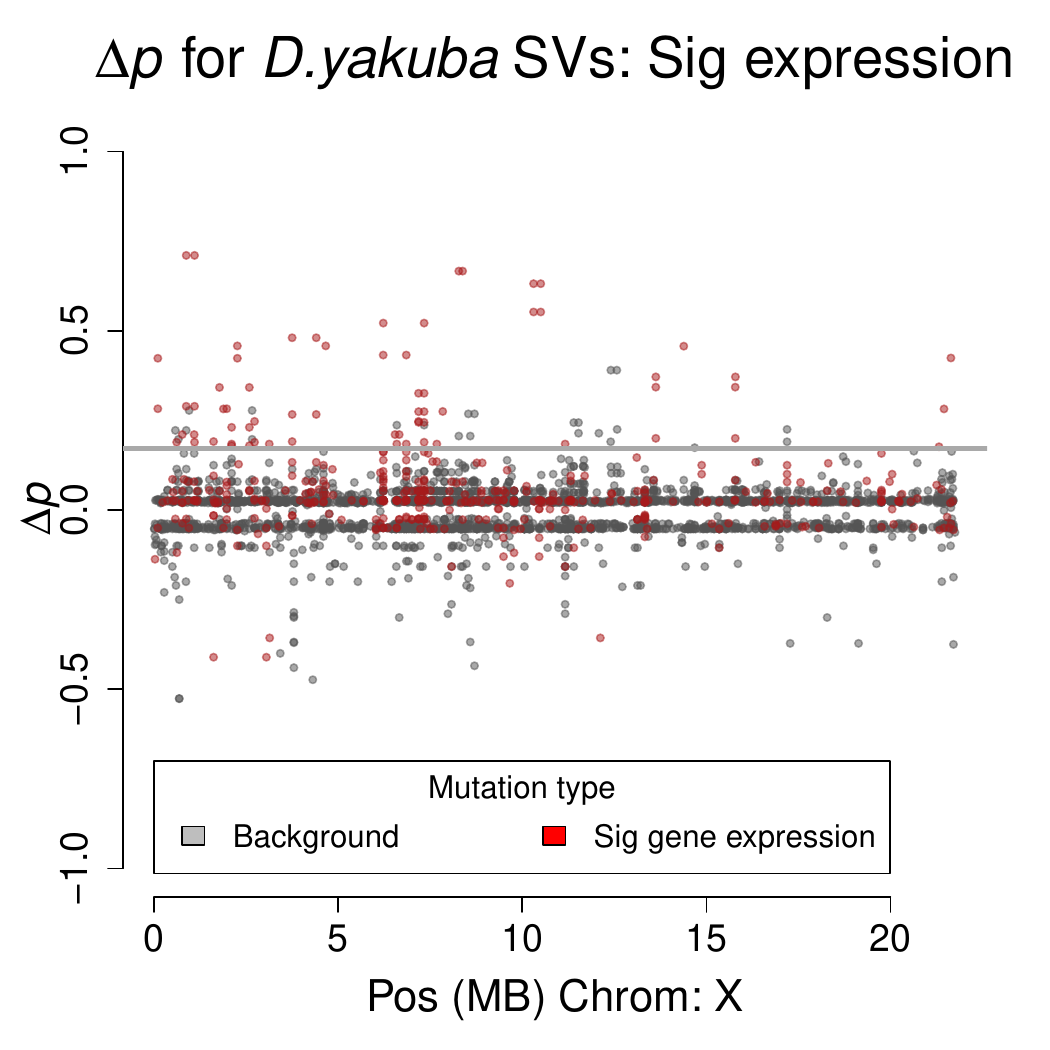}
    \caption{Population differentiation as shown via $\Delta$p between island \Dyak{} and mainland ancestral \Dyak{} on each chromosome. The x-axis shows genomic  position on the chromosome in megabases, and the y-axis is difference in allele frequencies between the island \Dyak{} and the mainland \Dyak{}. Points in red highlight the rearrangements associated with significant gene expression.  We observe greater levels of differentiation on the X compared to the autosomes in both populations and rearrangements associated with significant expression changes on the X show strong differentiation.  These results suggest that gene expression changes induced by rearrangements are important for local adaptation.}
    \label{deltap_ge_dyak_extra}
\end{figure}

% Extra sfs (vio)
% S5
\begin{figure}[htp]
    \centering
    \subfloat[A]{\includegraphics[page=1,width=.4\textwidth]{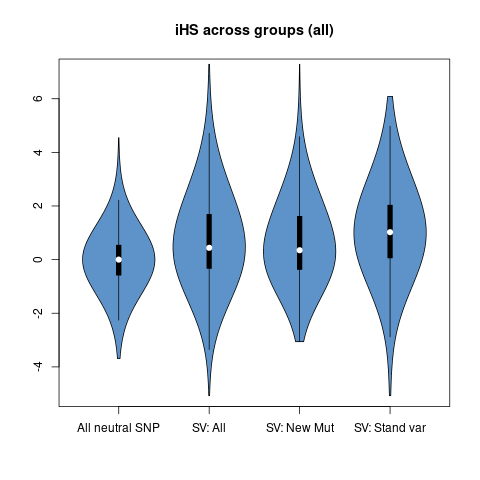}}\quad
    \subfloat[B]{\includegraphics[page=1,width=.4\textwidth]{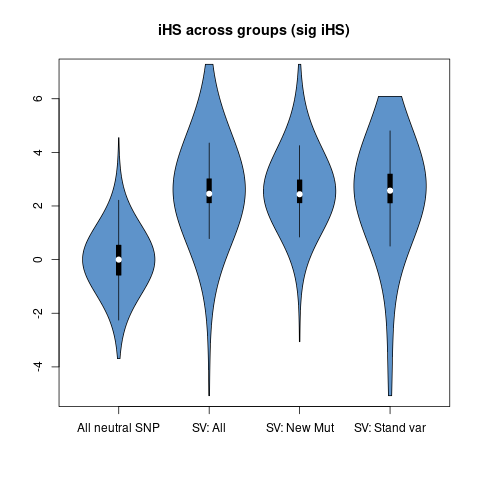}}\quad
    \subfloat[C]{\includegraphics[page=1,width=.4\textwidth]{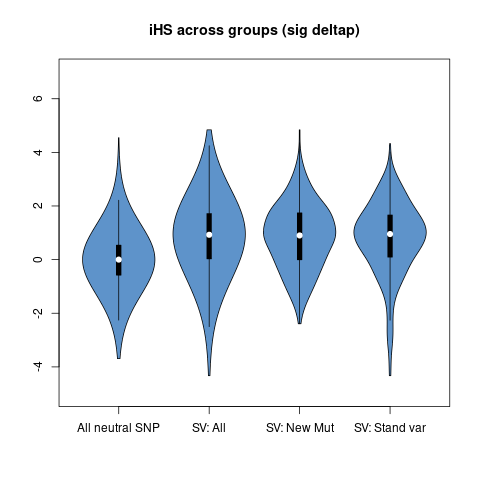}}
    \medskip
    \subfloat[D]{\includegraphics[page=1,width=.4\textwidth]{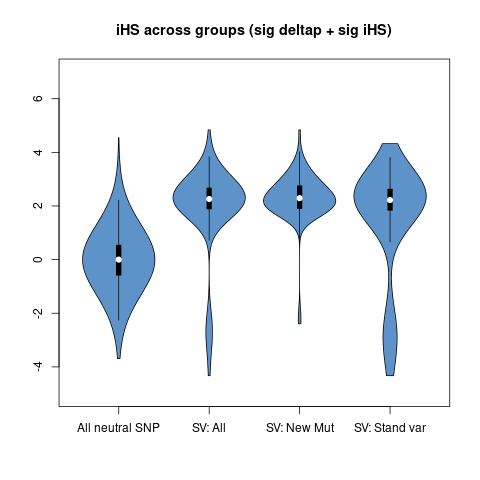}\quad}
    \caption{Distribution of iHS when comparing between all rearrangements, those from new mutations, and those from standing variation.}
    \label{vioPlots}
\end{figure}

%indv upset
% Upset S23
\begin{figure}
    \centering
    \includegraphics[page=2,width=.99\linewidth]{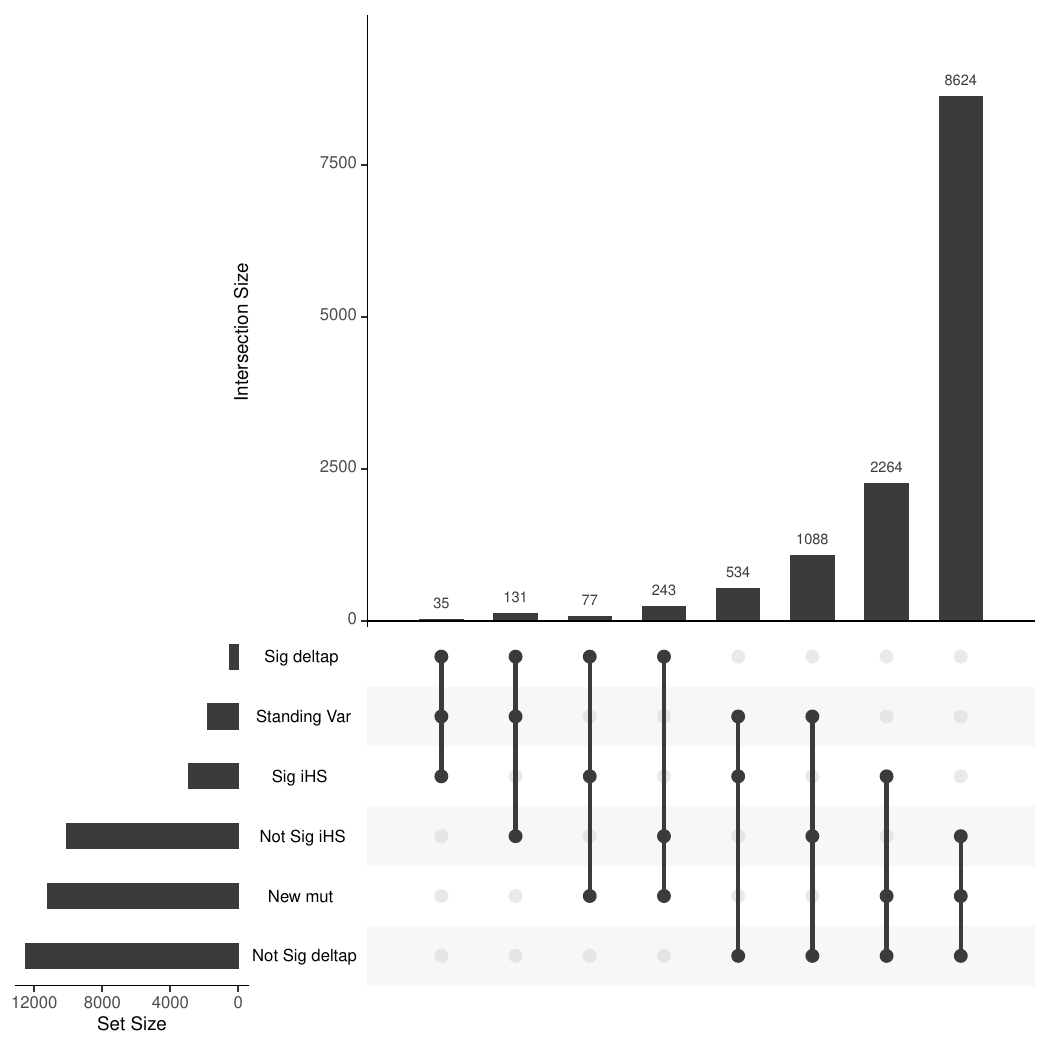}
    \caption{Upset plot comparing iHS and \deltap{} associated with rearrangements for \Dsant{}.  Novel TE insertions without population differentiation and without changes in gene expression are the most common type of mutation in \Dsant{}. New mutations induced by TE insertion with significant changes in gene expression are the most common type of variation that contributes to population differentiation.}
    \label{upsetihsPlot_dsan}
\end{figure}
\clearpage

% sim stuff S21
\begin{figure}[htp]
    \centering
    \includegraphics[page=1,width=.48\textwidth]{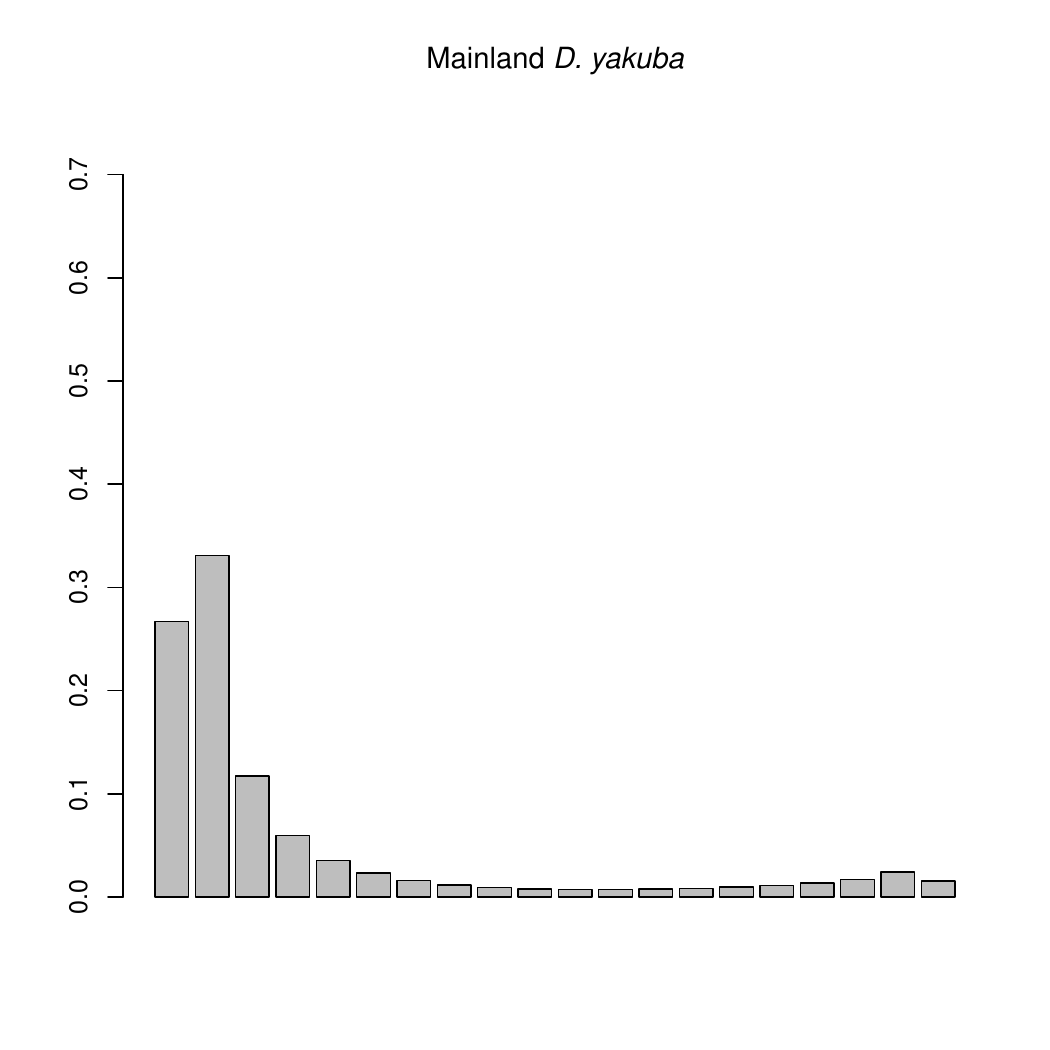}\quad
    \includegraphics[page=4,width=.49\textwidth]{Misc_Plots/simsfs.pdf}\quad
    \medskip
    \includegraphics[page=2,width=.49\textwidth]{Misc_Plots/simsfs.pdf}
    \includegraphics[page=5,width=.49\textwidth]{Misc_Plots/simsfs.pdf}\quad
    \caption{ Comparison of SFS between empirical distribution of SNPs in \Dsant{} and \Dyak{} and the simulated neutral distribution of the same subpopulations.}
    \label{sims}
\end{figure}

% sim stuff S22
\begin{figure}[htp]
    \centering
    \includegraphics[page=1,width=.48\textwidth]{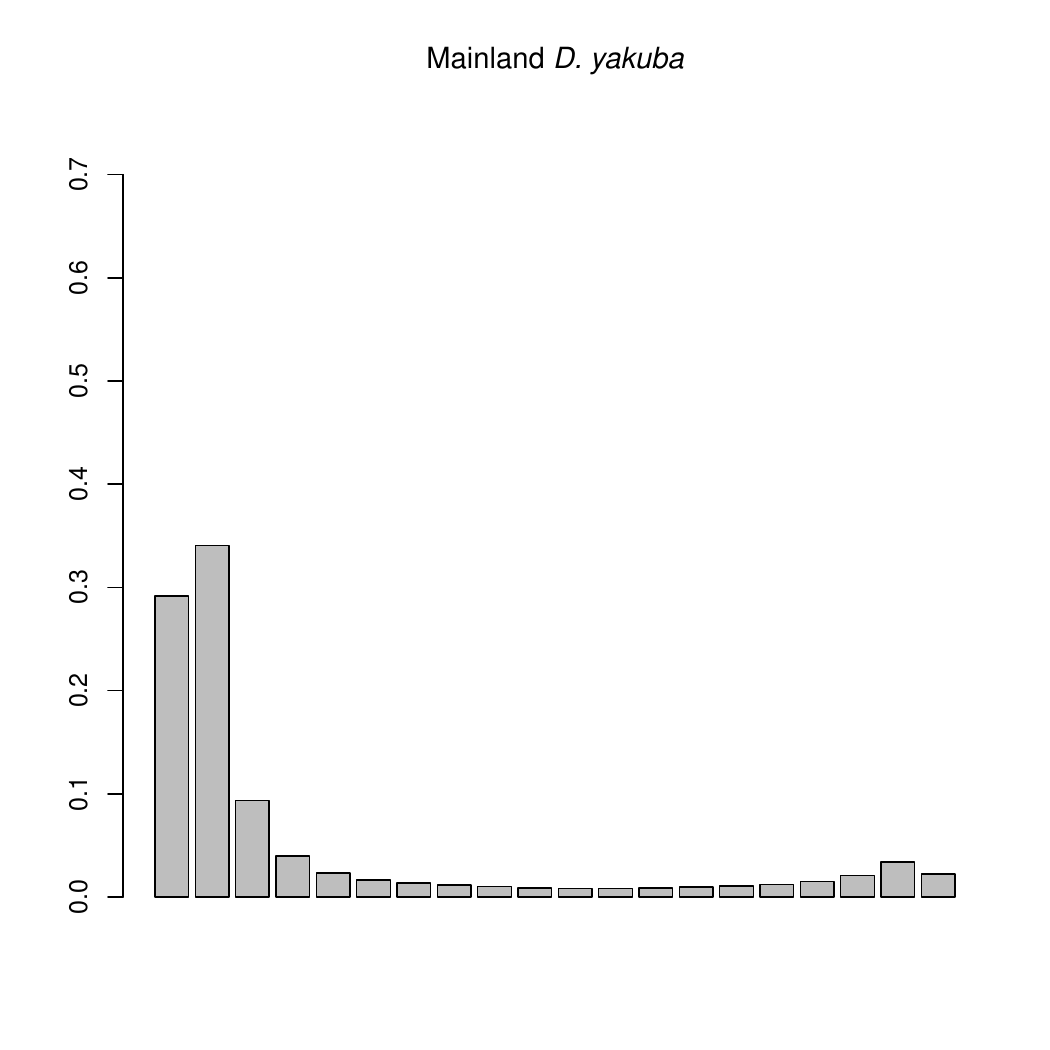}\quad
    \includegraphics[page=4,width=.49\textwidth]{Misc_Plots/simssfs_btl.pdf}\quad
    \medskip
    \includegraphics[page=2,width=.49\textwidth]{Misc_Plots/simssfs_btl.pdf}
    \includegraphics[page=5,width=.49\textwidth]{Misc_Plots/simssfs_btl.pdf}\quad
    \caption{ Comparison of SFS between empirical distribution of SNPs in \Dsant{} and \Dyak{} and the simulated neutral distribution of the same subpopulations, including a significant bottleneck.}
    \label{sims_btl}
\end{figure}

% Upset S24
\begin{figure}
    \centering
    \includegraphics[page=2,width=.99\linewidth]{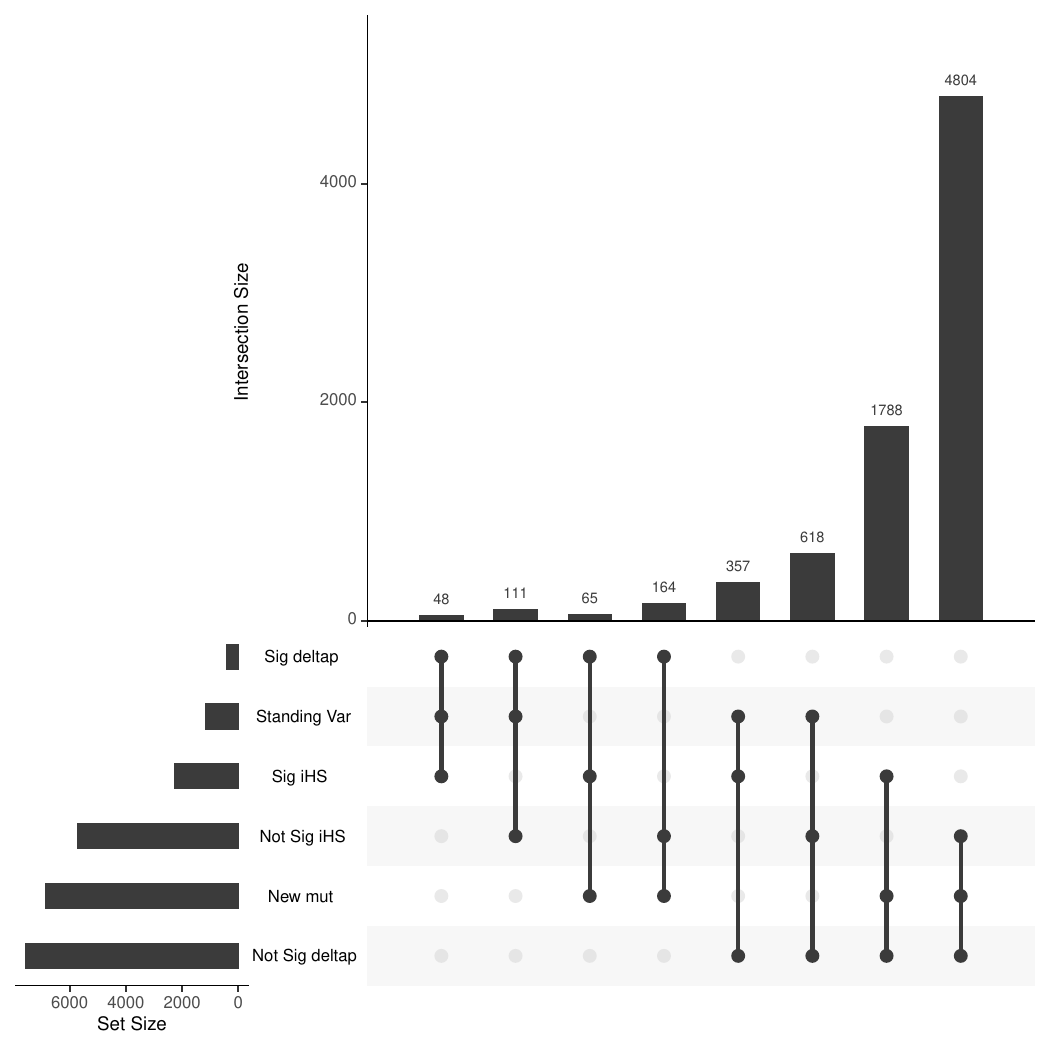}
    \caption{Upset plot comparing iHS and \deltap{} associated with rearrangements for \Dyak{}.  Novel TE insertions without population differentiation and without changes in gene expression are the most common type of mutation in \Dyak{}. New mutations induced by TE insertion with significant changes in gene expression are the most common type of variation that contributes to population differentiation.}
    \label{upsetihsPlot_oran}
\end{figure}
\clearpage

% Upset 
\begin{figure}[h]
    \centering
    \subfloat[A]{\label{upset_dsan}\includegraphics[page=2,width=.475\linewidth]{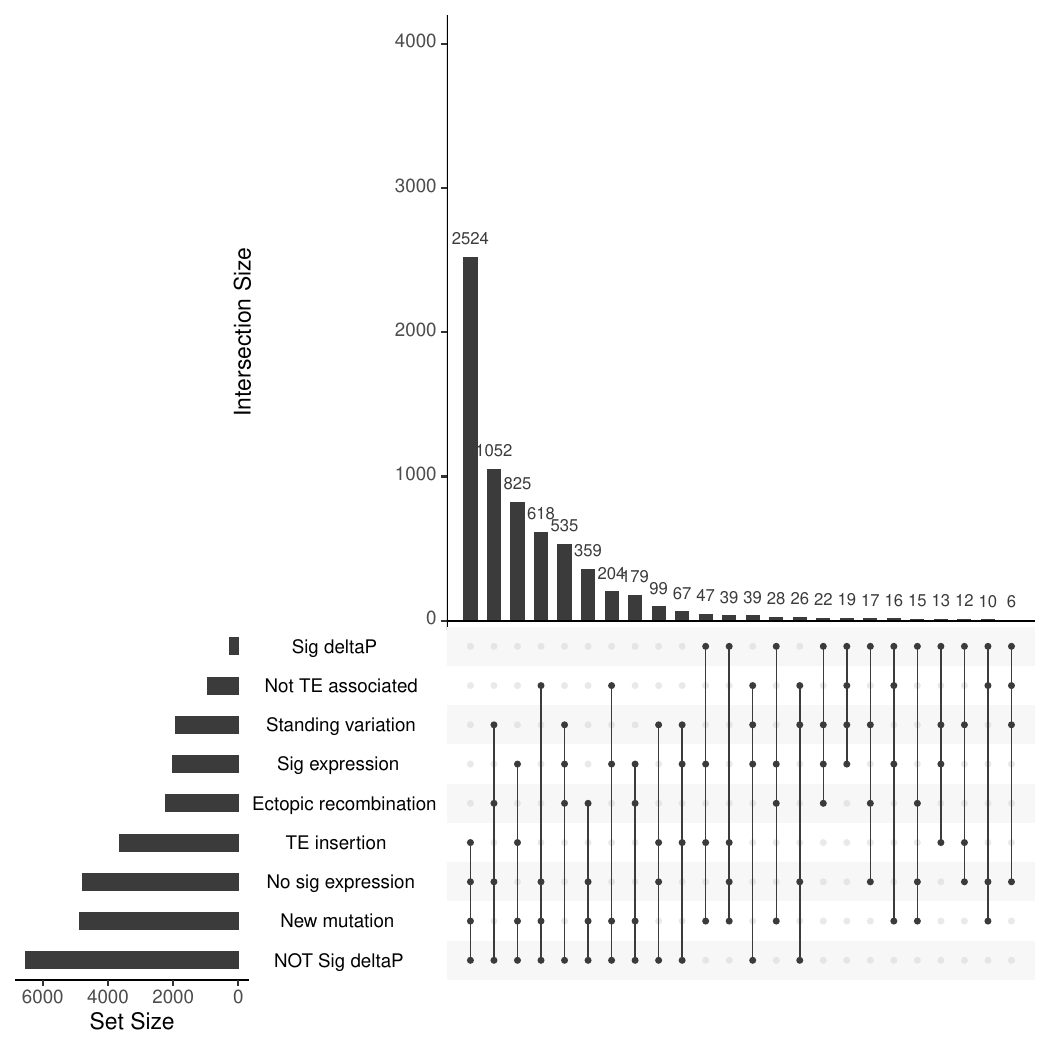}}
    \subfloat[B]{\label{upset_oran}\includegraphics[page=3,width=.475\linewidth]{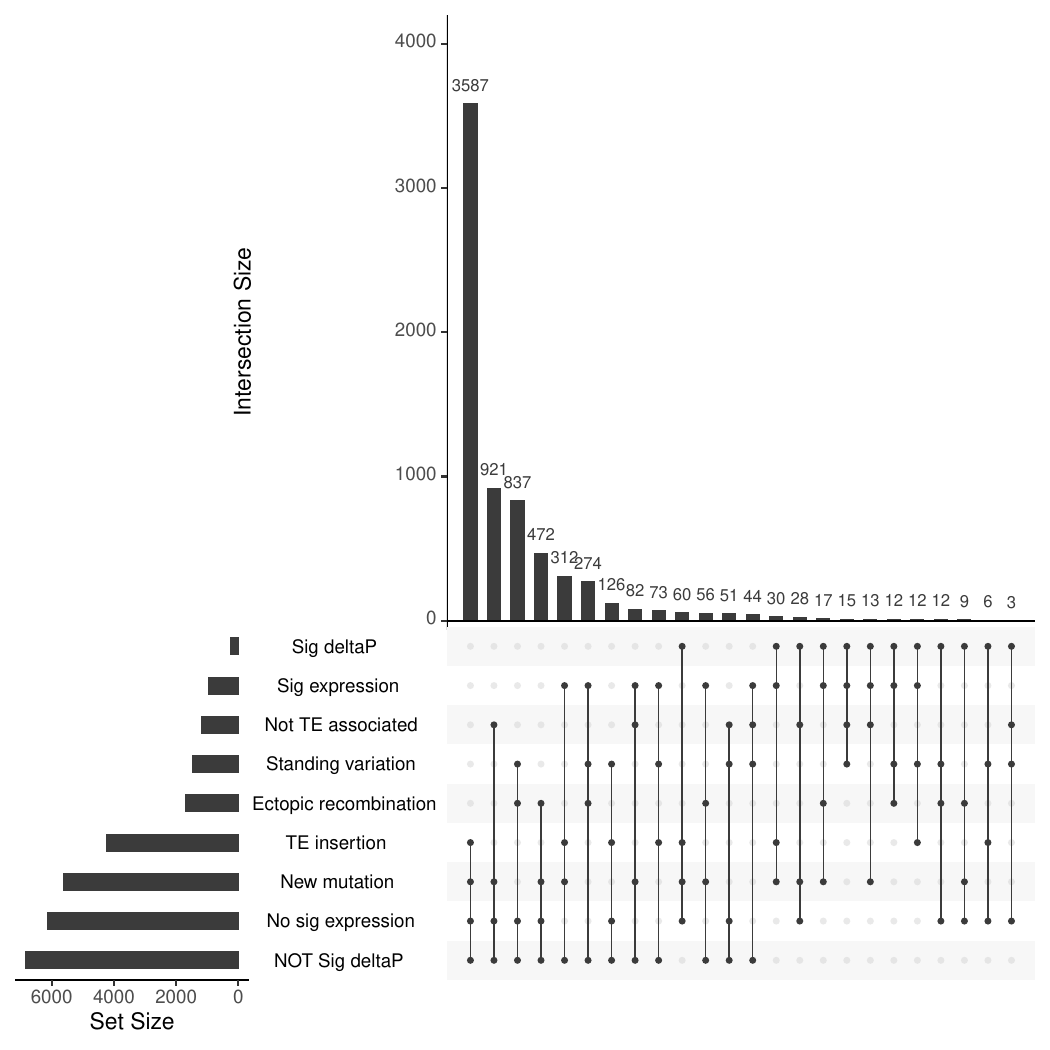}} \\ 
    \caption{Upset plot showing factors associated with rearrangements. Novel TE insertions without population differentiation and without changes in gene expression are the most common type of mutation in (A) \Dsant{} and (B) \Dyak{}. New mutations induced by TE insertion with significant changes in gene expression are the most common type of variation that contributes to population differentiation. }
    \label{upsetPlot}
\end{figure}

% hihglight 2 NON-randoms
% S18
\begin{figure}
    \centering
    \includegraphics[width=1\linewidth]{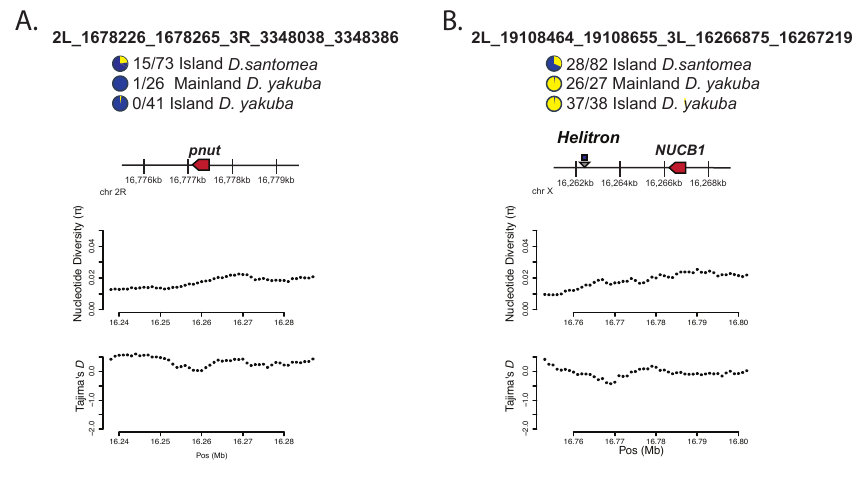}
    \caption{Two regions associated with chromosomal rearrangements that show significant changes in both population differentiation and signatures of selection and are within 5kb of genes  A) A rearrangement near the ortholog to a cell organization gene \emph{pnut}, and slight shows drops in \textpi{} and Tajima's (\emph{D}) (local minima \textpi{}=0.00966 and \emph{D}=-0.421).   B) A rearrangement mediated by a \emph{SAT} element, and shows minute drops in \textpi{} and Tajima's (\emph{D}) (local minima \textpi{}=0.0141 and \emph{D}=.0340). This rearrangement is heavily present in the \Dyak{} populations. Genomic averages for \textpi{} and Tajima's \emph{D} are \textpi{}=0.0159 and \emph{D}=-0.0445}
    \label{nonrandom_highlight}
\end{figure}
\clearpage

% hihglight 2 randoms
% S19
\begin{figure}
    \centering
    \includegraphics[width=1\linewidth]{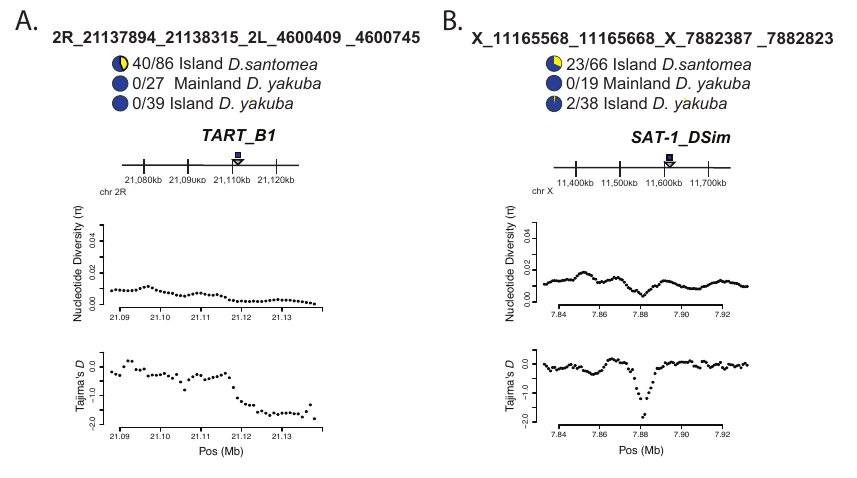}
    \caption{Two regions associated with chromosomal rearrangements that show significant changes in both population differentiation and signatures of selection A) A rearrangement mediated by a \emph{TART-B1} element that is only present in \Dsant{}, and shows drops in \textpi{} and Tajima's (\emph{D}) (local minima \textpi{}=0.000111 and \emph{D}=-1.80).    B) A rearrangement mediated by a \emph{SAT-1} element, and shows drops in \textpi{} and Tajima's (\emph{D}) (local minima \textpi{}=0.00176 and \emph{D}=-1.84). Genomic averages for \textpi{} and Tajima's \emph{D} are \textpi{}=0.0159 and \emph{D}=-0.0445}
    \label{random_highlight}
\end{figure}
\clearpage

% Extra deltap: te ins dsan
% S6
\begin{figure}[htp]
    \centering
    \includegraphics[page=1,width=.4\textwidth]{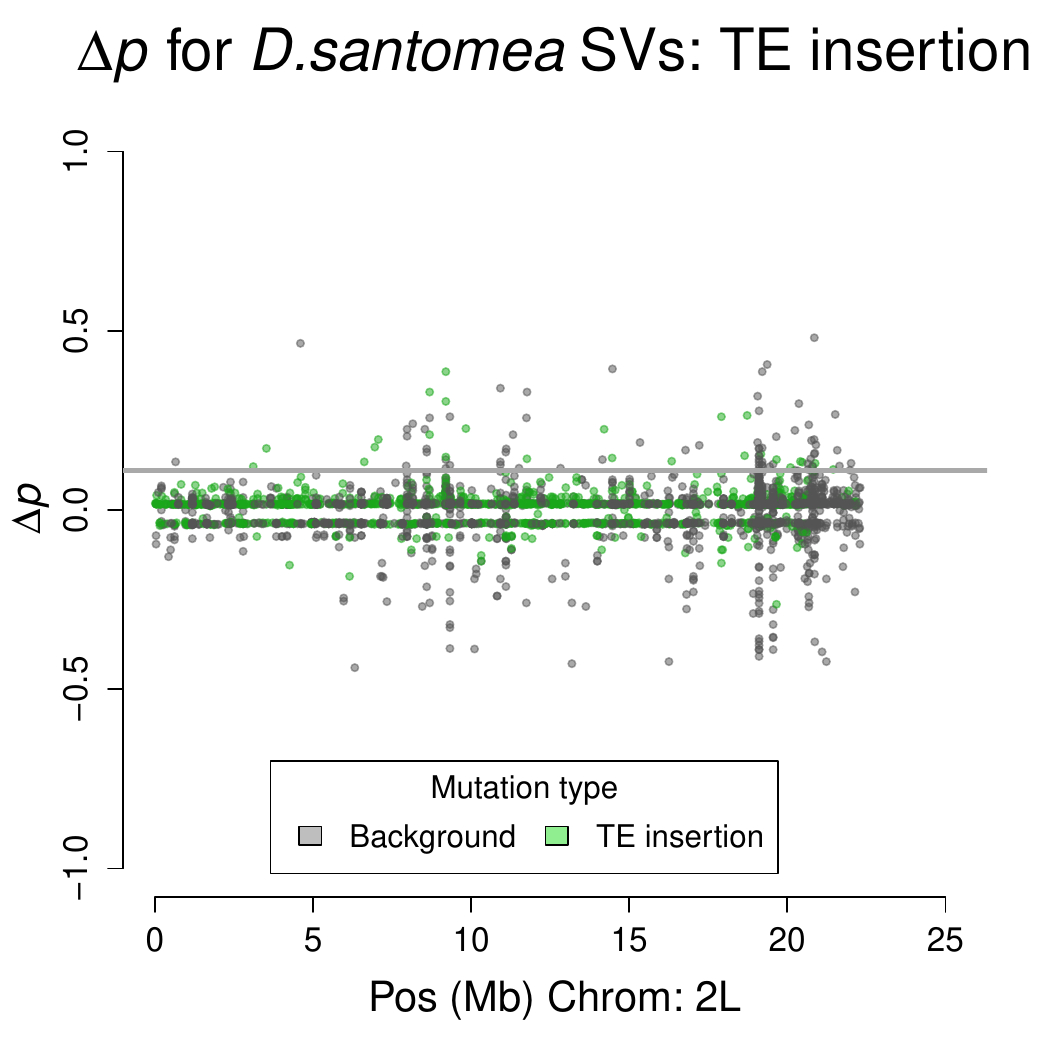}\quad
    \includegraphics[page=5,width=.4\textwidth]{Deltap/r_deltap_thresholds.pdf}\quad
    \includegraphics[page=9,width=.4\textwidth]{Deltap/r_deltap_thresholds.pdf}
    \medskip
    \includegraphics[page=13,width=.4\textwidth]{Deltap/r_deltap_thresholds.pdf}\quad
    \includegraphics[page=17,width=.4\textwidth]{Deltap/r_deltap_thresholds.pdf}
    \caption{Population differentiation as shown via $\Delta$p between island \Dsant{} and mainland ancestral \Dyak{} on each chromosome. The x-axis shows genomic  position on the chromosome in megabases, and the y-axis is difference in allele frequencies between the island \Dsant{} and the mainland. Points in green highlight the rearrangements associated with TE insertions.  We observe greater levels of differentiation on the X compared to the autosomes.}
    \label{deltap_teins_dsan_extra}
\end{figure}
\clearpage

% Extra deltap: te ins dyak
% S7
\begin{figure}[htp]
    \centering
    \includegraphics[page=3,width=.4\textwidth]{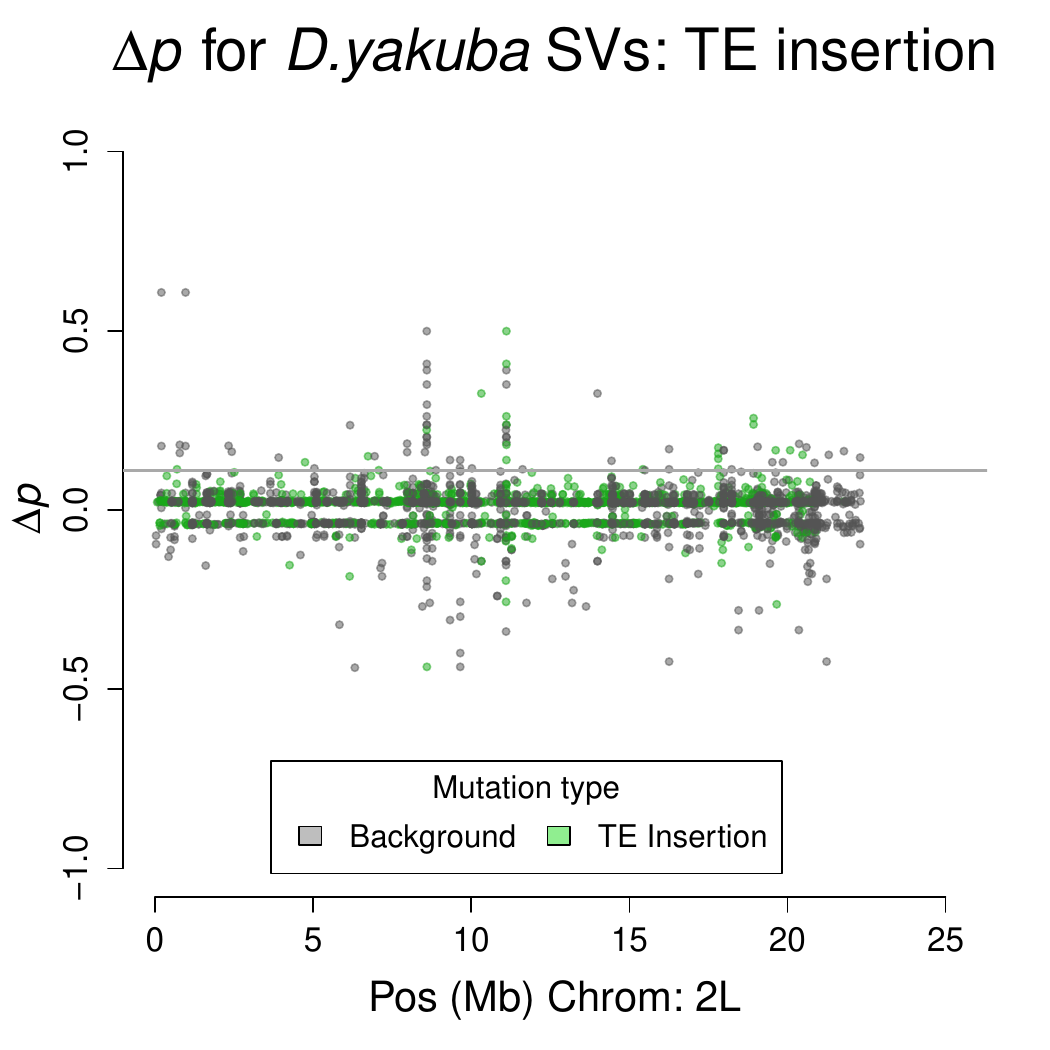}\quad
    \includegraphics[page=7,width=.4\textwidth]{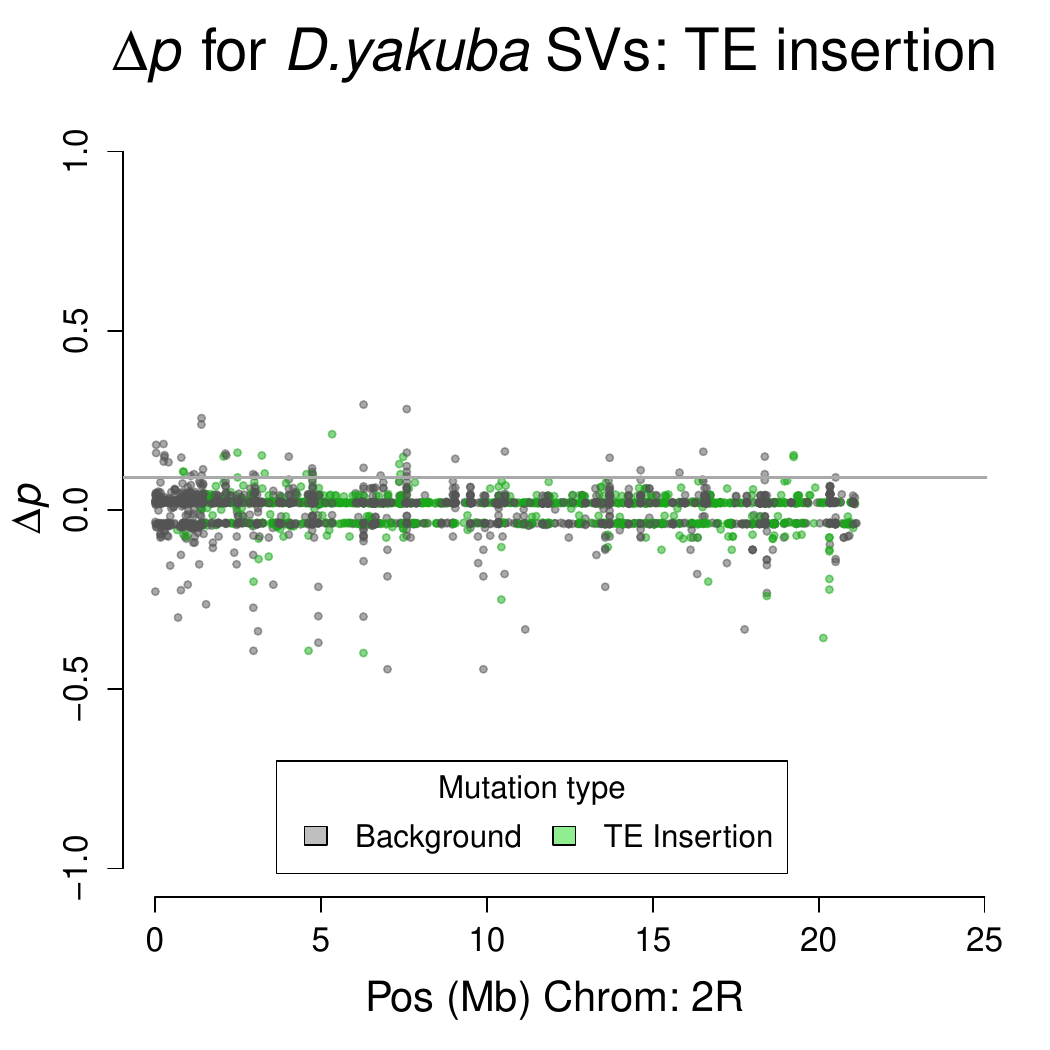}\quad
    \includegraphics[page=11,width=.4\textwidth]{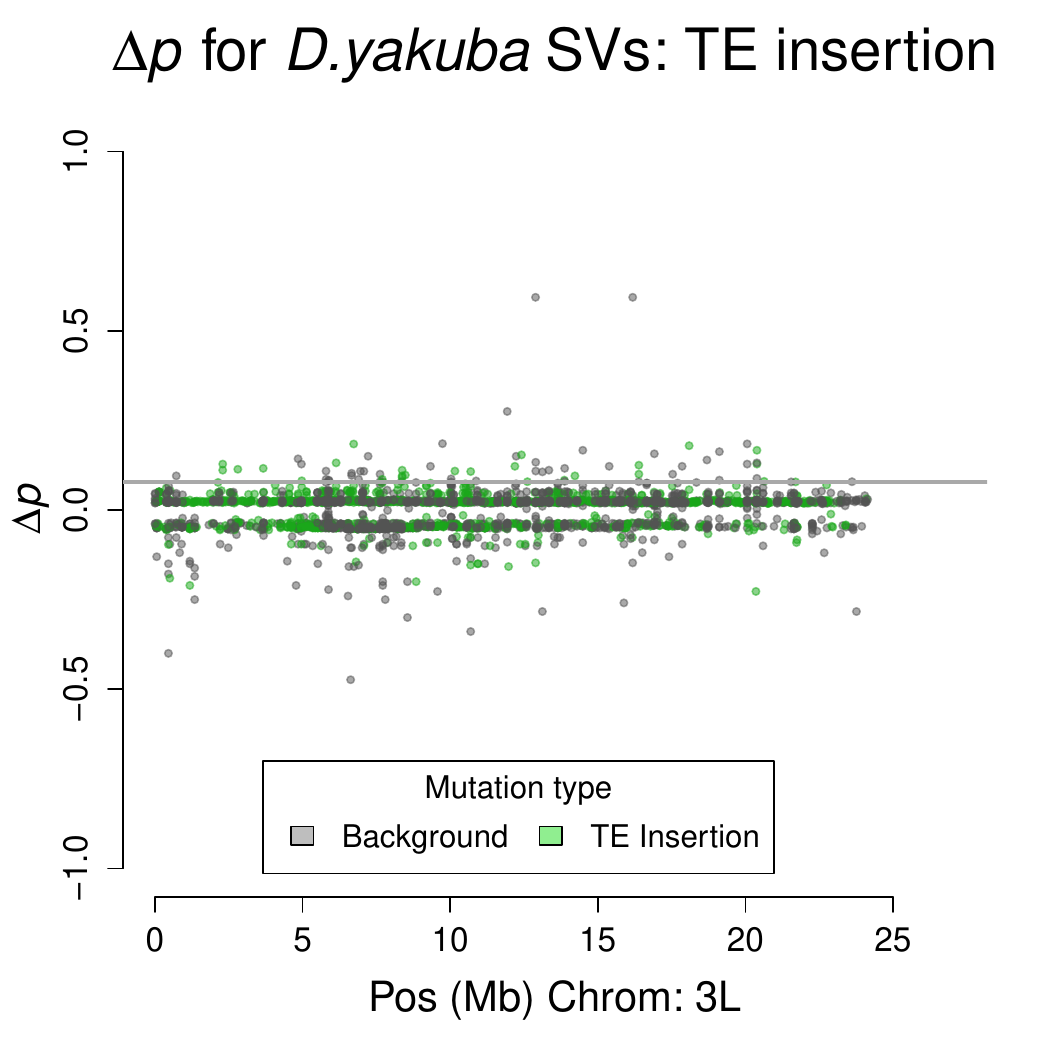}
    \medskip
    \includegraphics[page=15,width=.4\textwidth]{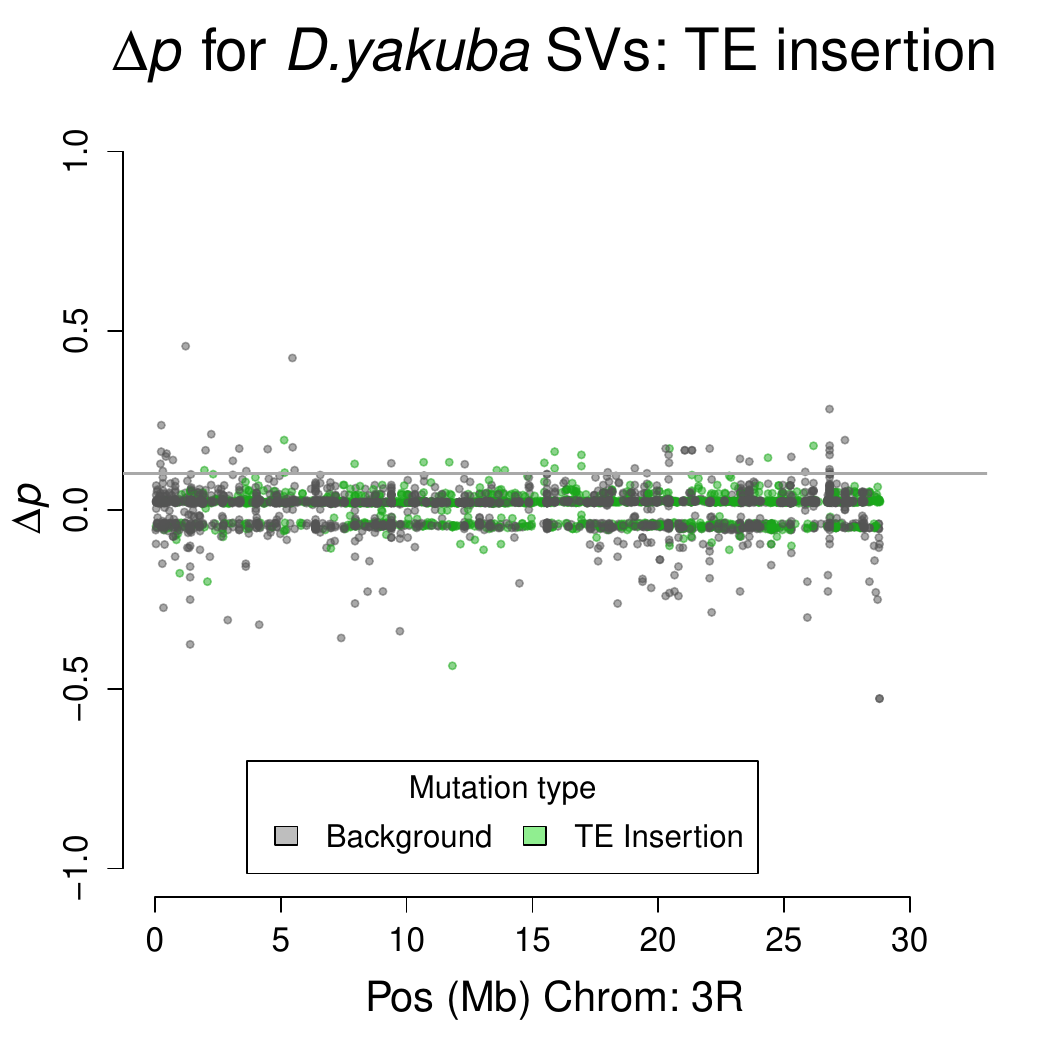}\quad
    \includegraphics[page=19,width=.4\textwidth]{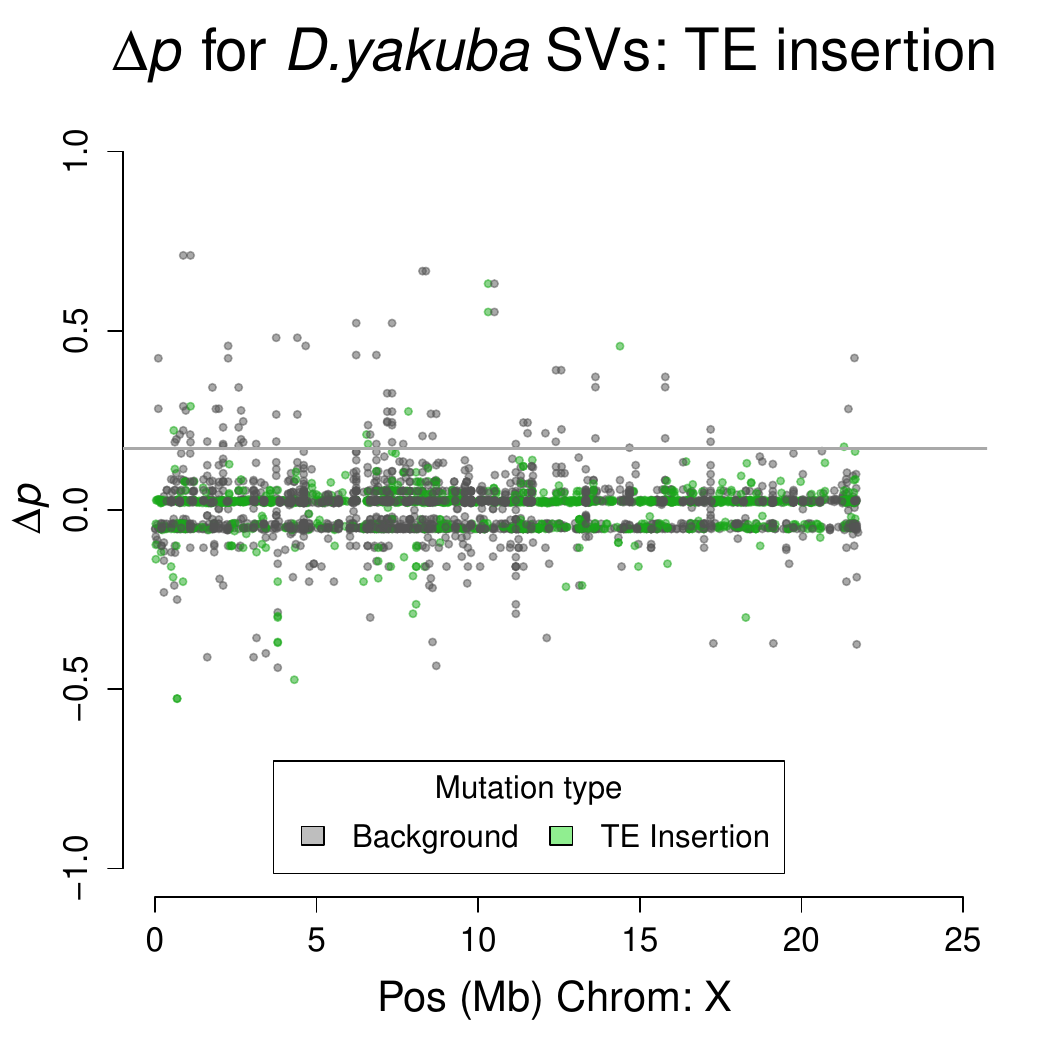}
    \caption{Population differentiation as shown via $\Delta$p between island \Dyak{} and mainland ancestral \Dyak{} on each chromosome. The x-axis shows genomic  position on the chromosome in megabases, and the y-axis is difference in allele frequencies between the island \Dyak{} and the mainland \Dyak{}. Points in green highlight the rearrangements associated with TE insertions.  We observe greater levels of differentiation on the X compared to the autosomes.}
    \label{deltap_teins_dyak_extra}
\end{figure}
\clearpage

 % Extra ect 
 % S3
\begin{figure}[h]\centering
    \subfloat[A]{\label{deltap_ect_3L_sub1A}\includegraphics[width=.48\linewidth]{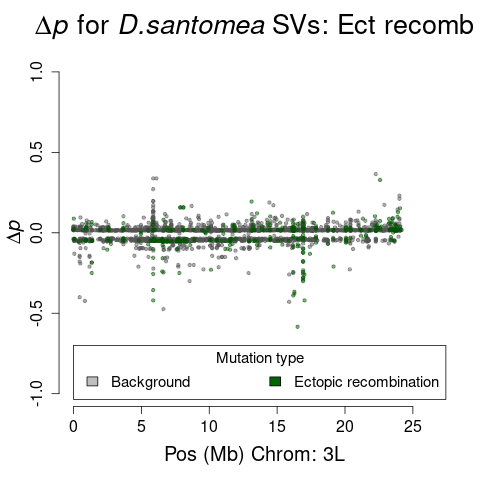}} \hfill
    \subfloat[B]{\label{deltap_ect_X_sub1B}\includegraphics[width=.48\linewidth]{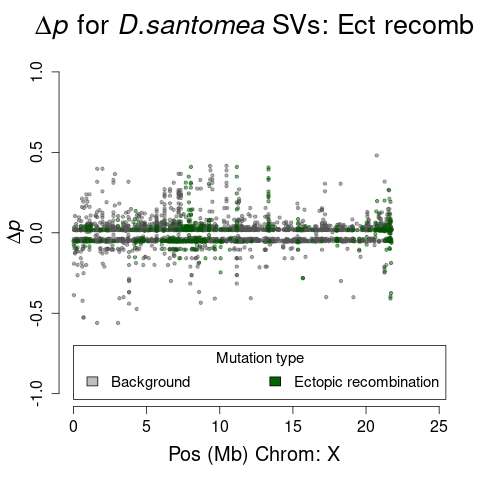}} \\
    \subfloat[C]{\label{deltap_ect_3L_sub1C}\includegraphics[width=.48\linewidth]{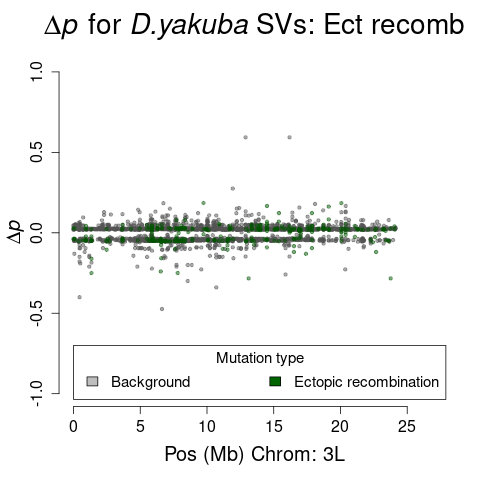}} \hfill
    \subfloat[D]{\label{deltap_ect_X_sub1D}\includegraphics[width=.48\linewidth]{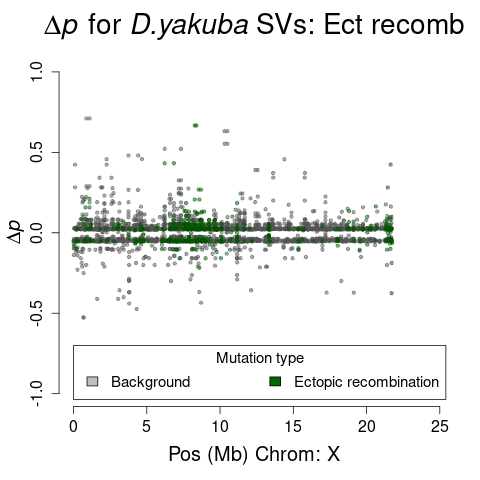}} \\
\caption{Population differentiation as shown via $\Delta$p A) Between \Dsant{} and mainland ancestral \Dyak{} on autosome 3L.  B)  Between \Dsant{} and mainland ancestral \Dyak{} on the X chromosome  C)  Between Island \Dyak{} and ancestral mainland \Dyak{} on the autosome 3L.  D)   Between Island \Dyak{} and ancestral mainland \Dyak{} on the X chromosome.  The x-axis shows genomic  position on the chromosome in megabases, and the y-axis is difference in allele frequencies between the test population and the mainland. Points in dark green highlight the rearrangements associated facilitating ectopic recombination. We observe greater levels of differentiation on the X compared to the autosomes in both populations.}
\label{dsanVsDyak_ect}
\end{figure}
\clearpage

% Figure 4 
\begin{figure}[ht]
    \centering
    \includegraphics[width=1\textwidth]{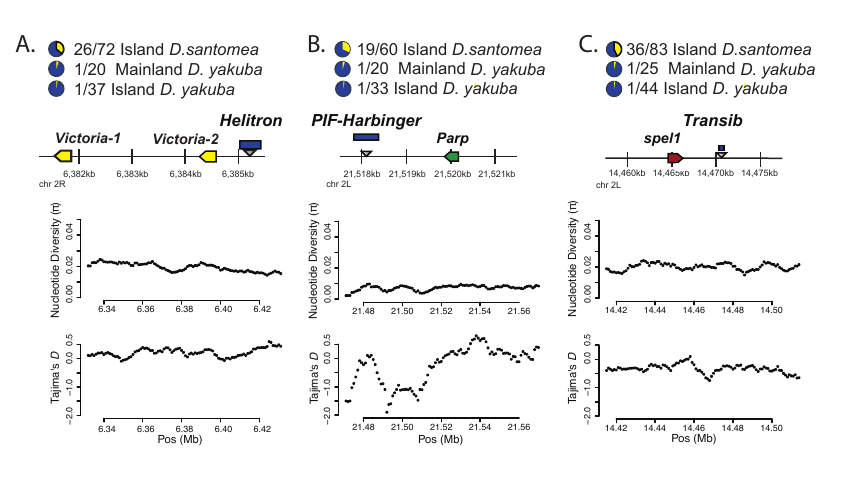}
    \caption{In \Dsant{}, three genes show rearrangements associated with significant enriched stress related genes and TEs, as shown via A) Rearrangements associated with \emph{Victoria}, which are mediated by a \emph{Helitron} element, and drops in \textpi{} and Tajima's (\emph{D}) (local minima \textpi{}=0.0160 and \emph{D}=-0.0917)    B) Rearrangements associated with \emph{Parp}, which are mediated by a \emph{Harbinger} element, and drops in \textpi{} and Tajima's (\emph{D}) (local minima \textpi{}=0.00134 and \emph{D}=-1.89). C) Rearrangements associated with \emph{spel1}, which are mediated by a \emph{Transib} element, and drops in \textpi{} and Tajima's (\emph{D}) (local minima \textpi{}=0.0135 and \emph{D}=-0.748). Rearrangement show a large shift in allele frequency and population differentiation. \emph{Parp} and \emph{spel1} also show significant differential gene expression via Fisher's combined p-values. Genomic averages for \textpi{} and Tajima's \emph{D} are \textpi{}=0.0159 and \emph{D}=-0.0445}
    \label{bigHighlight}
\end{figure}

% S10
\begin{figure}[h]
\centering
\subfloat[A]{\includegraphics[page=2,width=.49\linewidth]{Misc_Plots/TEBarPlot.pdf}}
\subfloat[B]{\includegraphics[page=1,width=.49\linewidth]{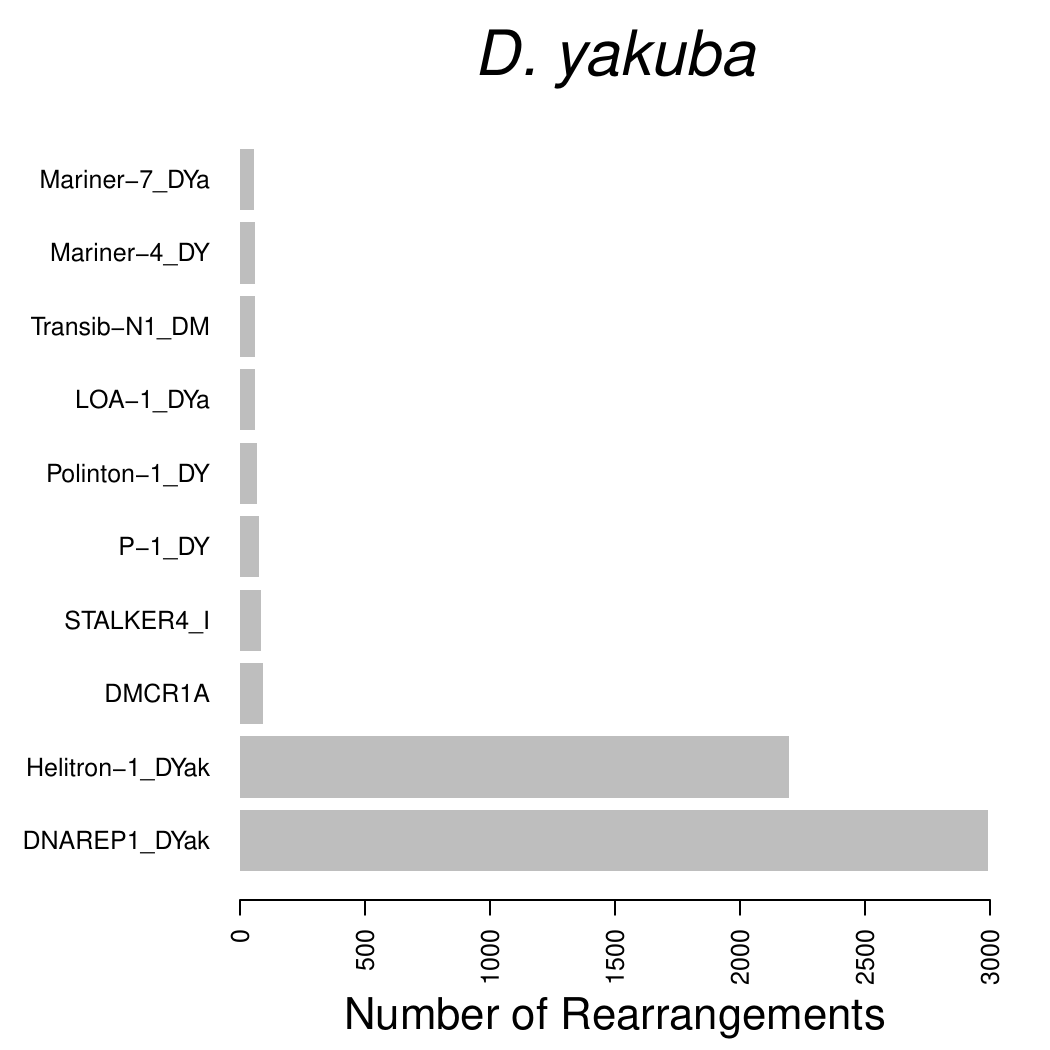}}
\caption{\label{AdaptiveTEs_all} Number of rearrangements matching TE families by population A) \Dsant {} and B) \Dyak. The most common rearrangements come from \emph{DNAREP1} and \emph{Helitrons} in both species.}
\end{figure}
\clearpage

% S Table 1 for ihs and dp thresholds
\clearpage
\begin{table}[ht!]
  \begin{center}
    \caption{Thresholds for significance for iHS and \deltap{}.}
    \label{threshTable}
    \begin{tabular}{c|c|c|c} % <-- Alignments: 1st column left, 2nd middle and 3rd right, with vertical lines in between
      \textbf{Chrom} & \textbf{\deltap{}} & \textbf{iHS lower} & \textbf{iHS upper}\\
      \hline
       2L & 0.0936917 & -2.05632375643193 & 1.86580699048154\\
       2R & 0.09405691 & -2.10724653850541 & 1.8012844093028\\
       3L & 0.08218144 & -1.97339364040109 & 2.00827097495942\\
       3R & 0.08330457 & -2.09598808898755 & 1.99649354809924\\
       X & 0.1462702 & -2.20695492241181 & 1.55353268541448\\     

      %& 1,190 & 590 & 600\\
    \end{tabular}
  \end{center}
\end{table}
\clearpage

% FST sant
% S14
\begin{figure}[htp]
    \centering
    \includegraphics[page=2,width=.4\textwidth]{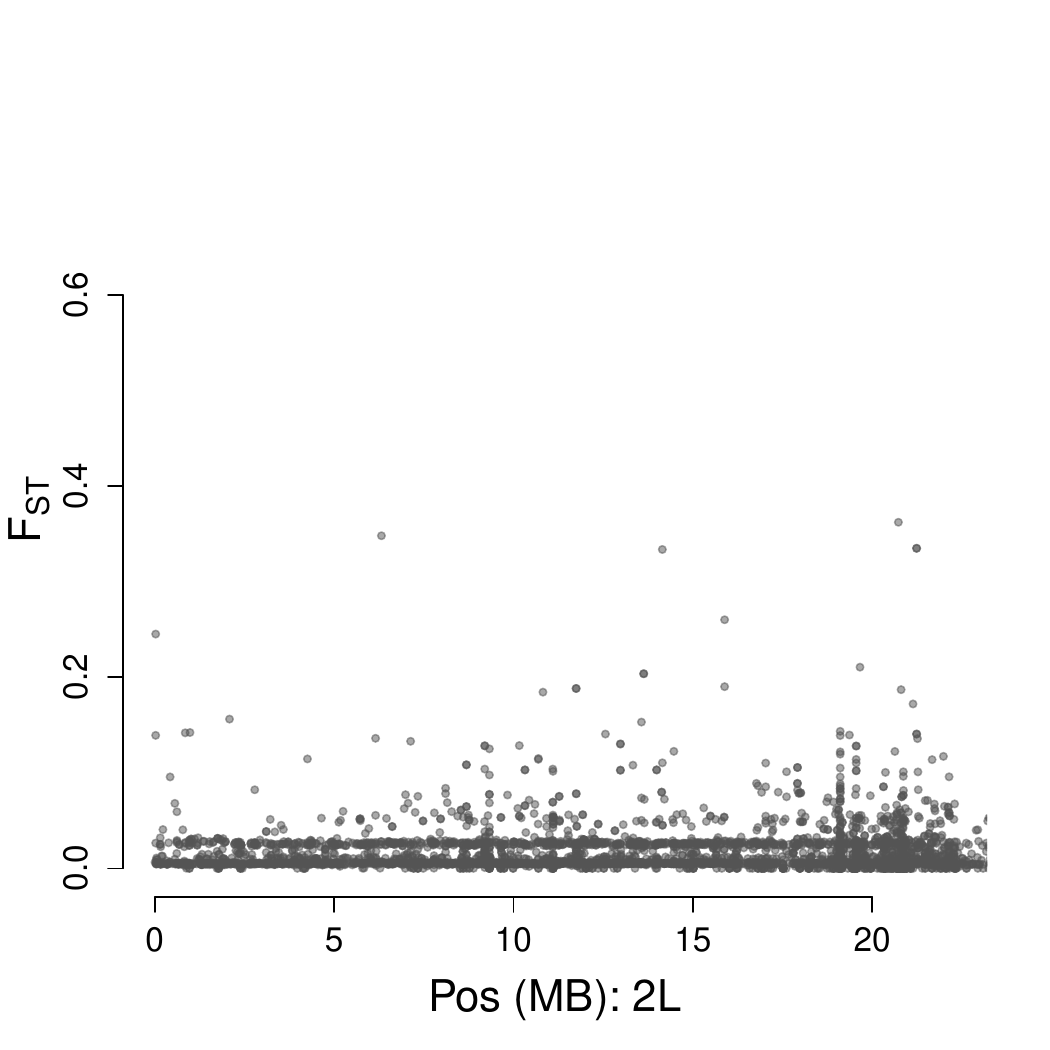}\quad
    \includegraphics[page=4,width=.4\textwidth]{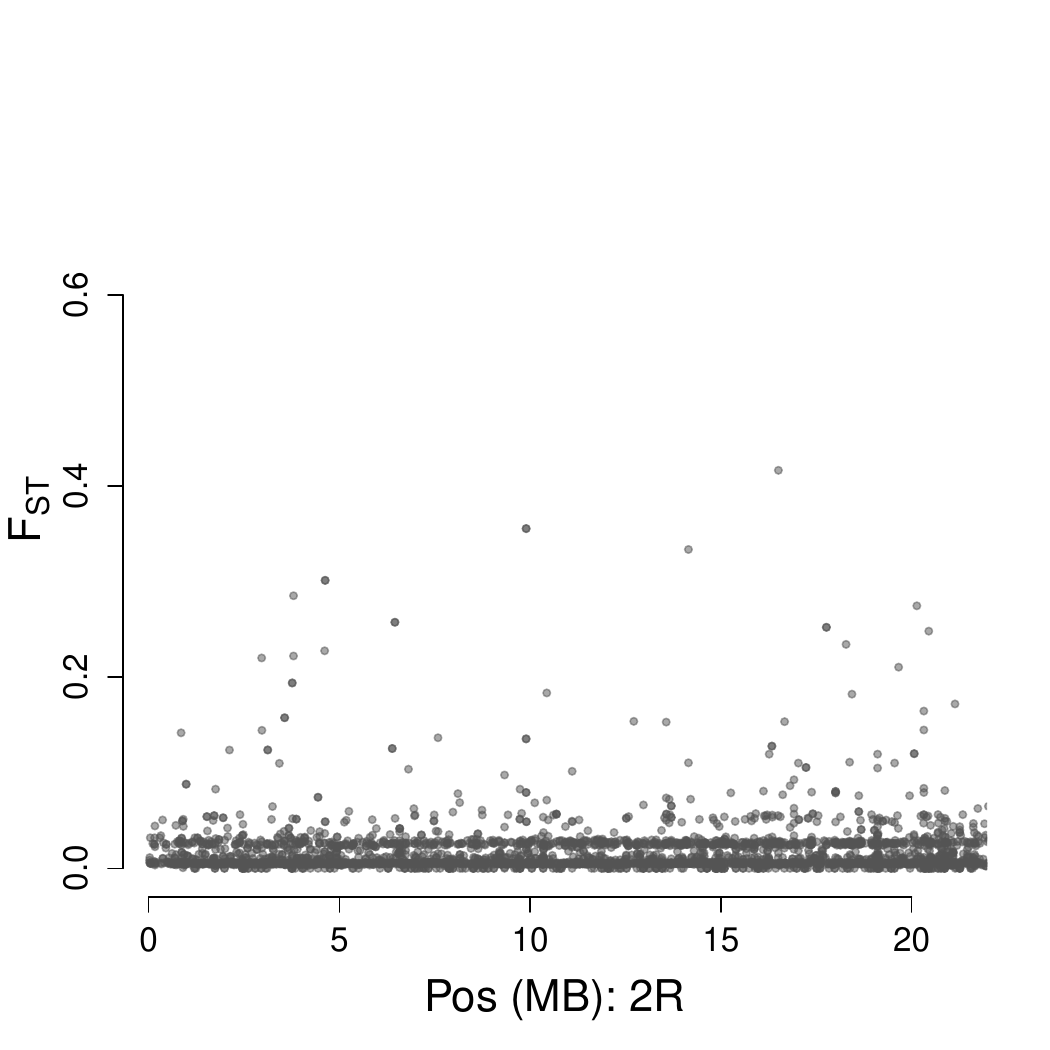}\quad
    \includegraphics[page=6,width=.4\textwidth]{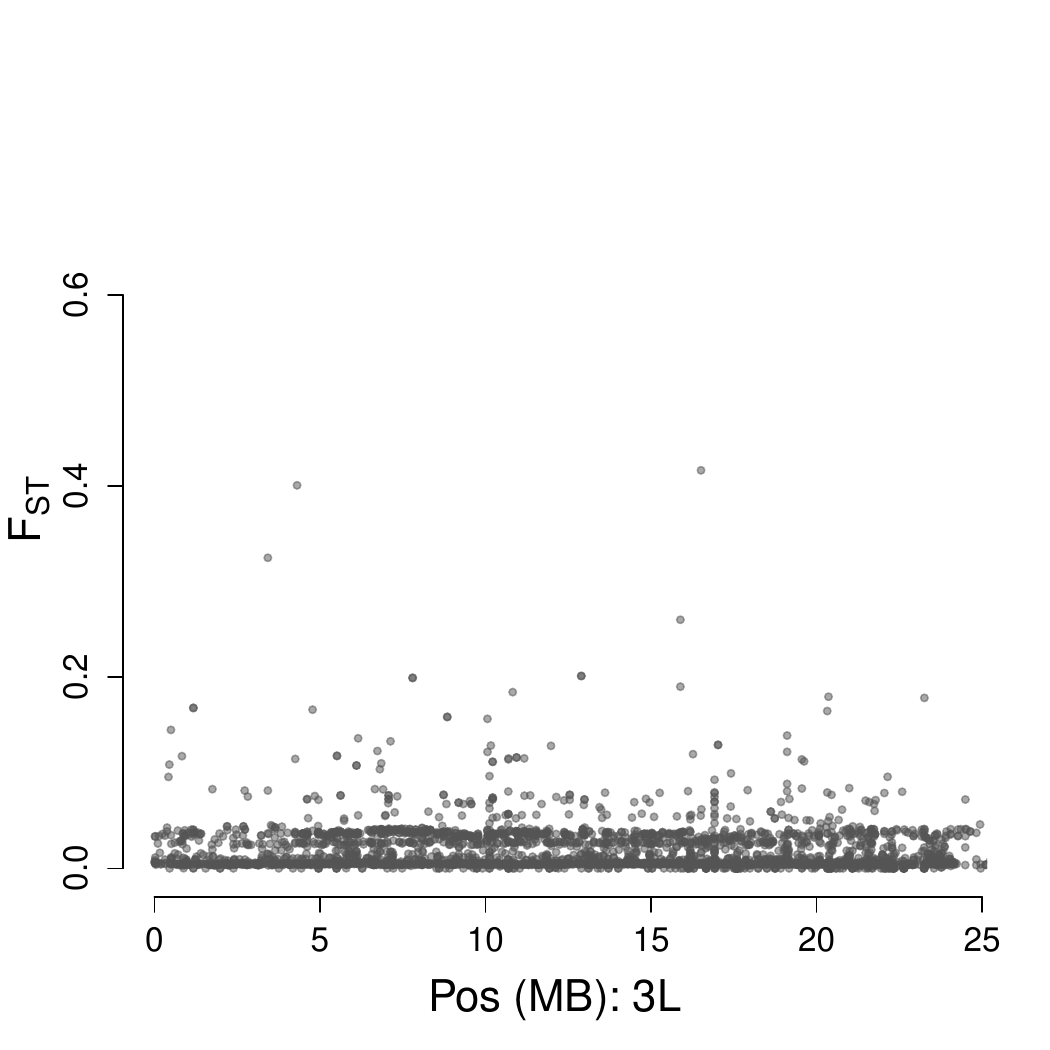}
    \medskip
    \includegraphics[page=8,width=.4\textwidth]{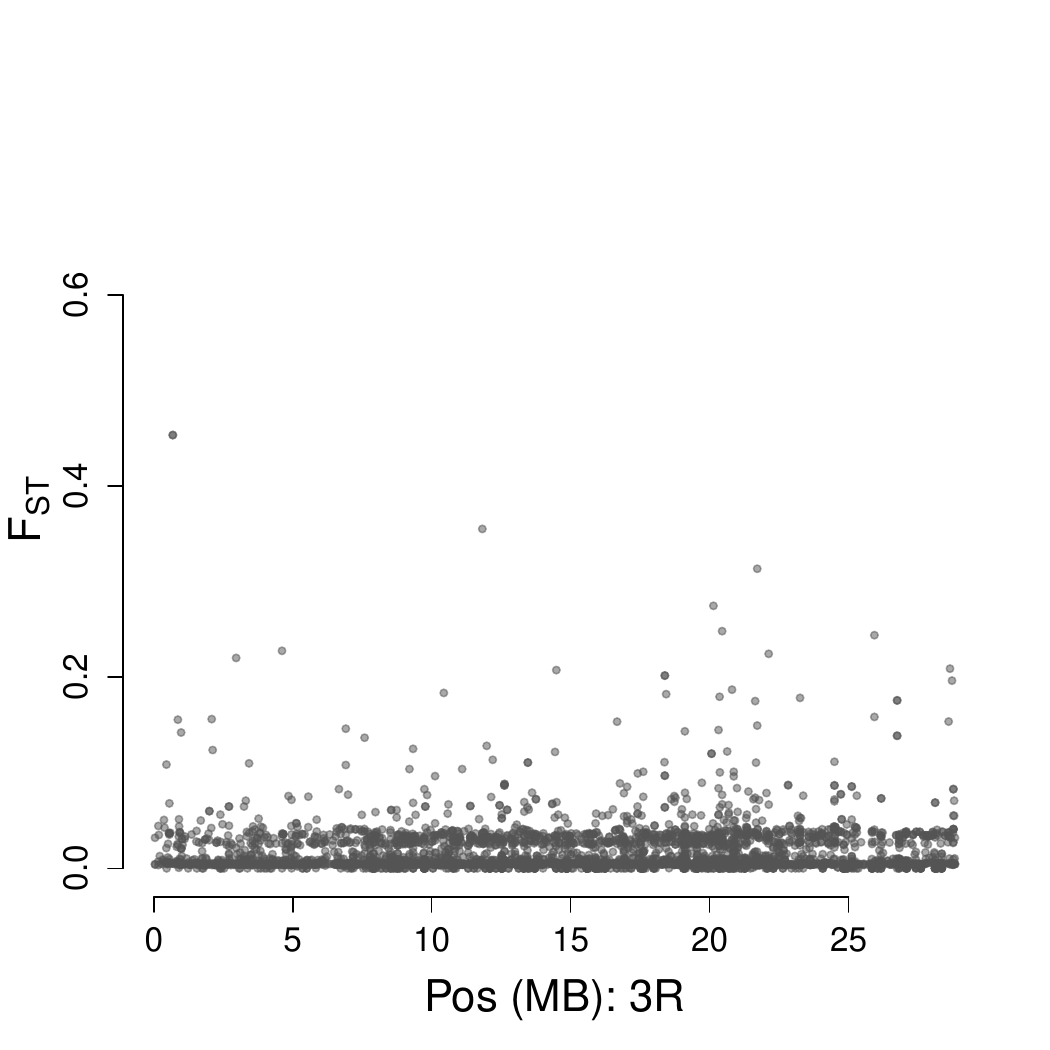}\quad
    \includegraphics[page=10,width=.4\textwidth]{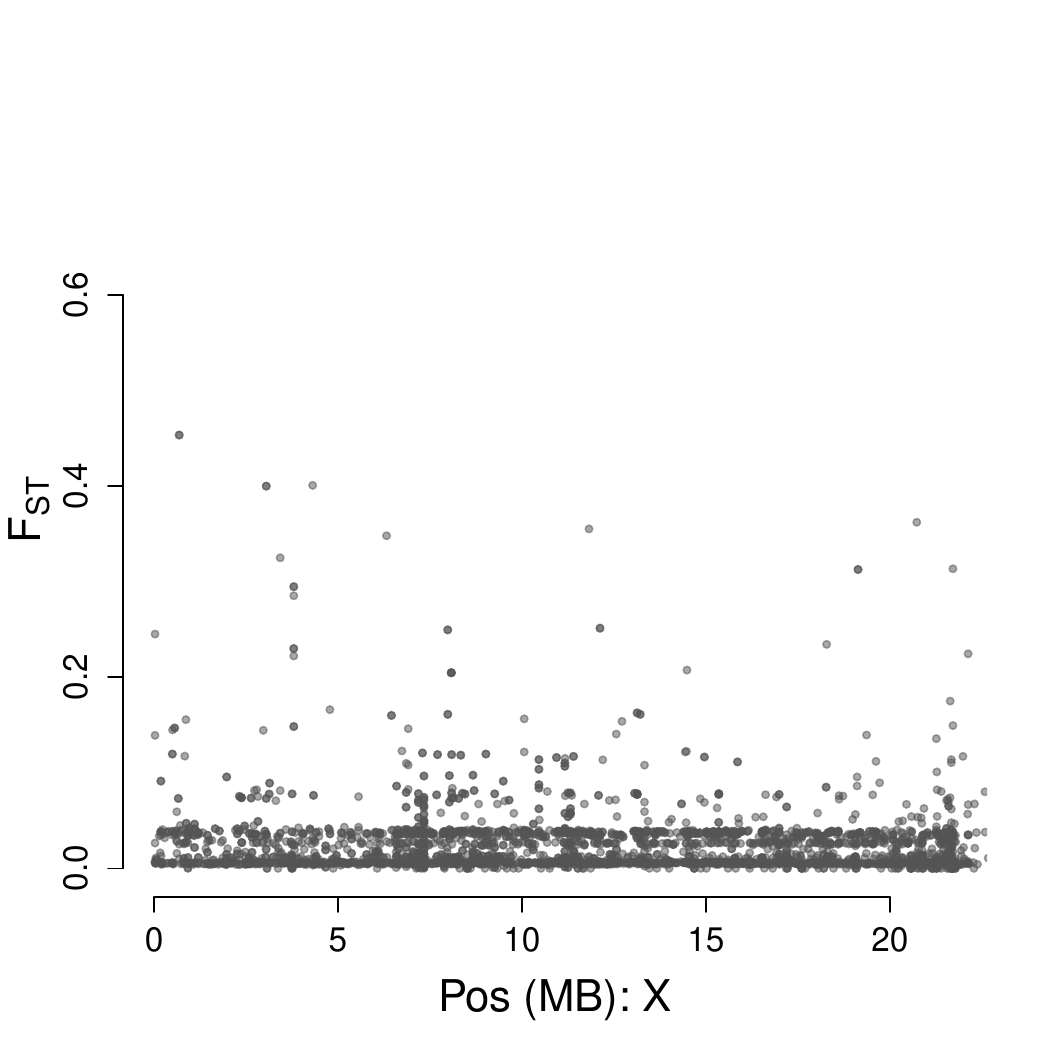}
    \caption{\FST {} along the autosomes and the X chromosome shows high rates of population differentiation on the X and autosomes in \Dsant {} compared with the ancestral mainland population of \Dyak.  Most variation shows minimal divergence, with peaks associated with genome outliers. This complementary measure of differentiation to \deltap {} also shows that structural variation is associated with population divergence on the island of \STome.}
    \label{fst_dsan}
\end{figure}
\clearpage

% FST yak
% S15
\begin{figure}[htp]
    \centering
    \includegraphics[page=1,width=.4\textwidth]{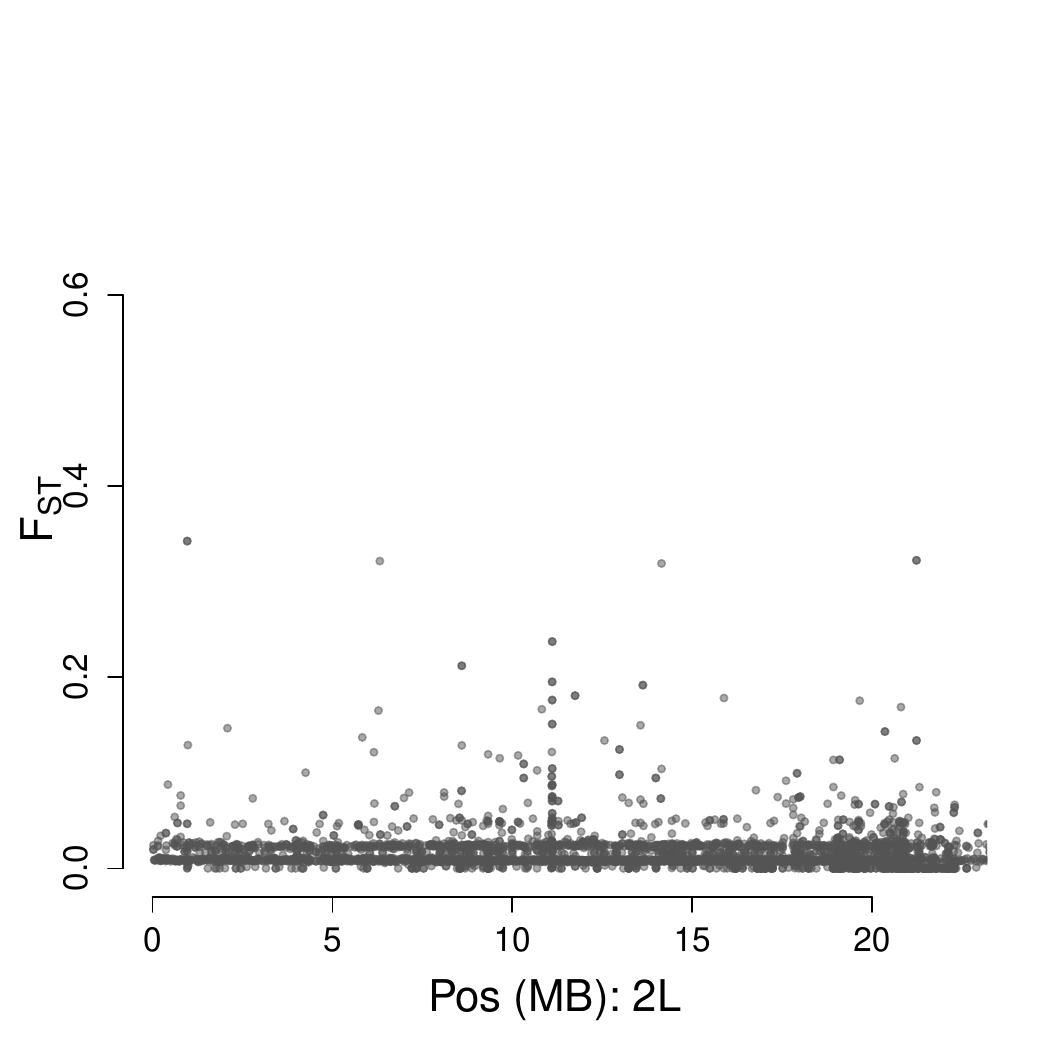}\quad
    \includegraphics[page=3,width=.4\textwidth]{FST/r_fst.pdf}\quad
    \includegraphics[page=5,width=.4\textwidth]{FST/r_fst.pdf}
    \medskip
    \includegraphics[page=7,width=.4\textwidth]{FST/r_fst.pdf}\quad
    \includegraphics[page=9,width=.4\textwidth]{FST/r_fst.pdf}
    \caption{\FST {} along the autosomes and the X chromosome shows high rates of population differentiation on the X and autosomes in island \Dyak{} compared with the ancestral mainland population.  Most variation shows minimal divergence, with peaks associated with genome outliers. This complementary measure of differentiation to \deltap {} also shows that structural variation is associated with population divergence on the island of \STome.}
    \label{fst_dyak}
\end{figure}
\clearpage

%indv upset
% Upset S16
\begin{figure}
    \centering
    \includegraphics[page=2,width=.99\linewidth]{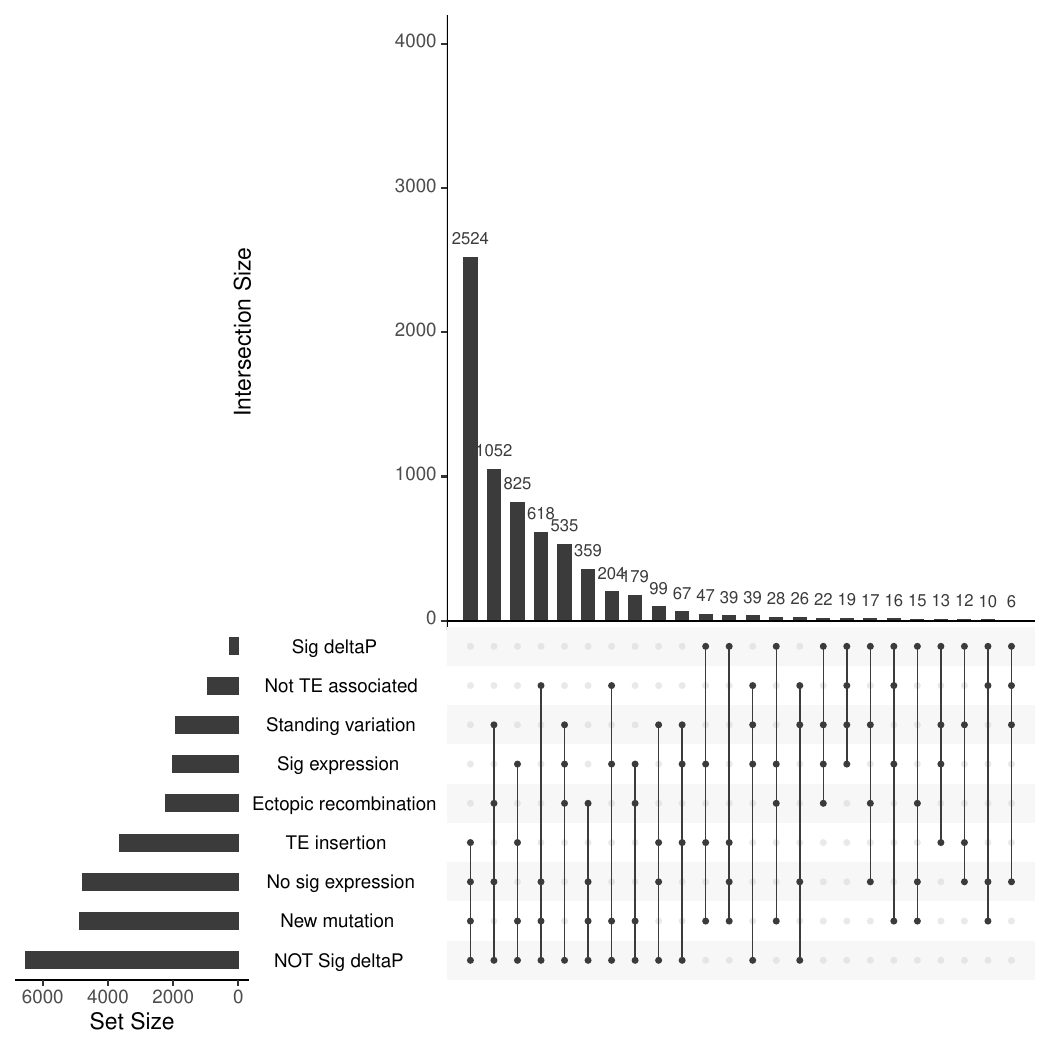}
    \caption{Upset plot showing factors associated with rearrangements for \Dsant{}.  Novel TE insertions without population differentiation and without changes in gene expression are the most common type of mutation in \Dsant{}. New mutations induced by TE insertion with significant changes in gene expression are the most common type of variation that contributes to population differentiation.}
    \label{upsetPlot_dsan}
\end{figure}
\clearpage

% Upset S17
\begin{figure}
    \centering
    \includegraphics[page=3,width=.99\linewidth]{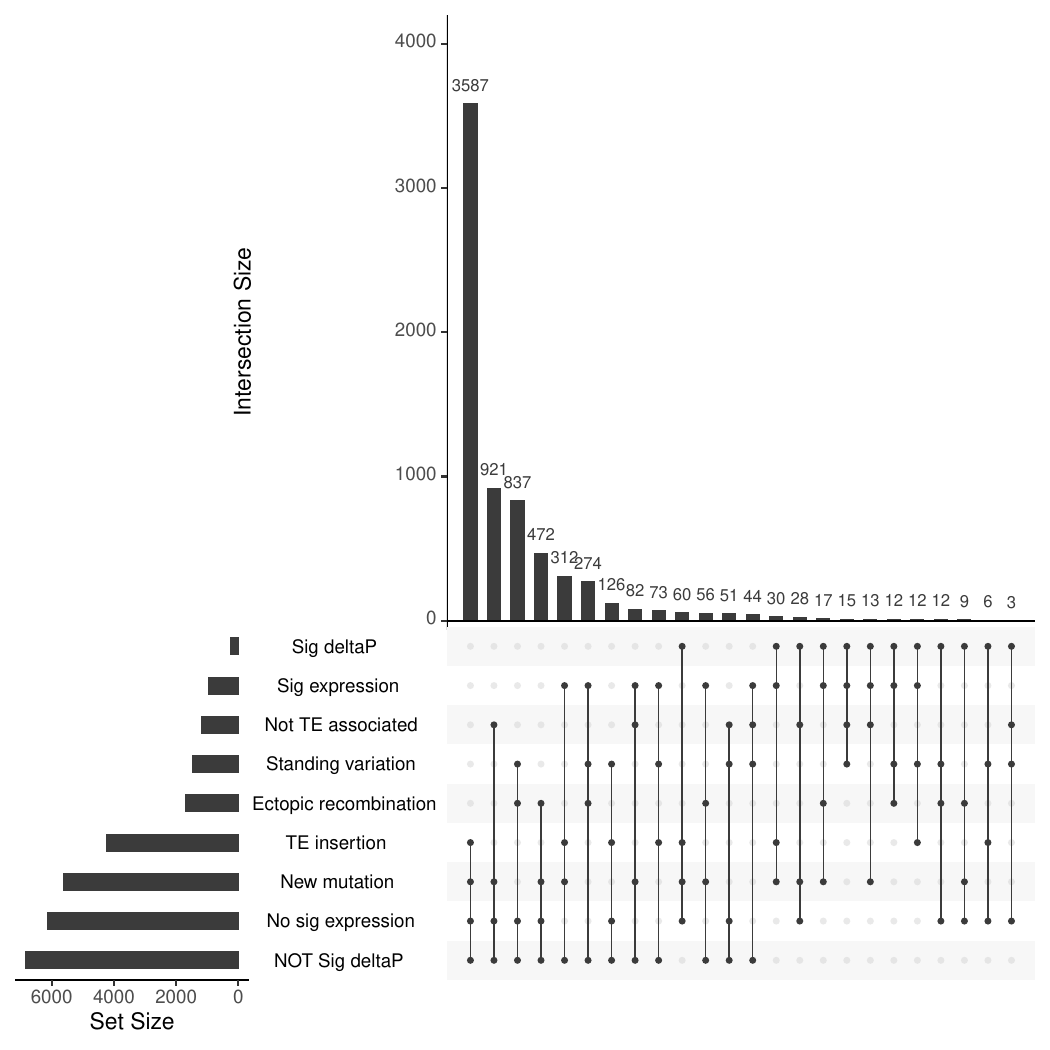}
    \caption{Upset plot showing factors associated with rearrangements for \Dyak{}.  Novel TE insertions without population differentiation and without changes in gene expression are the most common type of mutation in \Dyak{}. New mutations induced by TE insertion with significant changes in gene expression are the most common type of variation that contributes to population differentiation.}
    \label{upsetPlot_oran}
\end{figure}
\clearpage

% %Theresa gene exp by chrom
% \begin{figure}
%     \centering
%     \includegraphics[width=1\linewidth]{Expression/sig_express_by_chrom.png}
%     \caption{}
%     \label{sigExp_byChrom}
% \end{figure}
% \clearpage

% %Theresa gene exp by tissue
% \begin{figure}
%     \centering
%     \includegraphics[width=1\linewidth]{Expression/sig_genes_by_tissue.png}
%     \caption{}
%     \label{sigExp_byTissue}
% \end{figure}
% \clearpage

% hihglight 2 randoms
% S19
\begin{figure}
    \centering
    \includegraphics[width=1\linewidth]{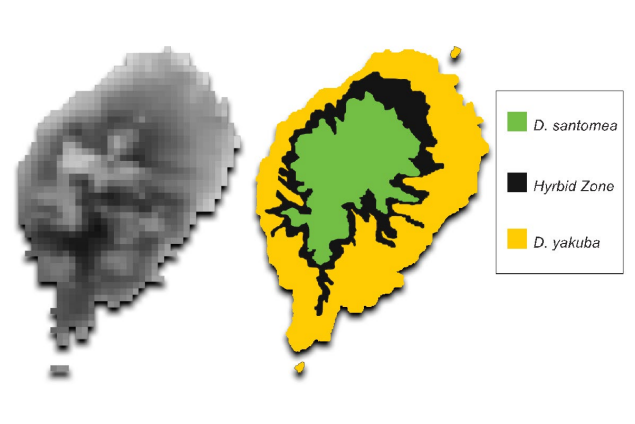}
    \caption{Maps of the island of São Tomé. The left map is solar radiation data for WorldClim 2.1 historical climate data. The left map is an overlay among all months for a holistic annual solar radiation São Tomé is exposed to with lighter colors signifying higher solar radiation (kJ m-2 day1) and darker colors delineating less solar radiation. The right map gives the estimated ranges between D. santomea and D. yakuba adapted from Lachaise et al. 2000).}
    \label{james_exposed}
\end{figure}
\clearpage

% Confirm rate 
% S20
\clearpage

\end{document}